\documentclass[]{hdsr}

\graphicspath{{figs/}}
\usepackage{adjustbox}
\usepackage{multirow}
\usepackage{booktabs}
\usepackage{tabularx}

\newcommand{\PP}{\mathbb{P}}
\newcommand{\err}{\Delta}
\newcommand{\pint}{\widehat{\textrm{PI}}}

\newcommand{\tidx}{t}
\newcommand{\period}{\ensuremath{\mathcal{T}}}
\newcommand{\ytk}[1][\tidx]{\widehat{y}_{#1}}

\begin{document}

% Larger bottom margin for the first page
\newgeometry{bottom=1.5in}

% Editorial staff will replace the following values:
% 1. Volume number
% 2. Issue number
% 3. Article DOI
% e.g. for Volume 2, Issue 3, DOI 12.345:
% \volumeheader{2}{3}{12.345}
% \volumeheader{0}{0}{00.000}

\begin{center}
  \title[COVID-19 Data Repository and County-Level Forecasting]{Curating a COVID-19 data repository and forecasting county-level death counts in the United States}
  \maketitle

  % Start page numbering on second page. Must appear *after* \maketitle
  \thispagestyle{empty}

  \vspace*{.2in}

  % Authors and Affiliations
  \begin{tabular}{cc}
    Nick Altieri\upstairs{\affilone}, 
    Rebecca L Barter\upstairs{\affilone}, 
    James Duncan\upstairs{\affilfour},
    Raaz Dwivedi\upstairs{\affiltwo},
    Karl Kumbier\upstairs{\affilsix},\\
    Xiao Li\upstairs{\affilone},
    Robert Netzorg\upstairs{\affiltwo}, 
    Briton Park\upstairs{\affilone}, 
    Chandan Singh\upstairs{\affiltwo, *}, 
    Yan Shuo Tan\upstairs{\affilone},\\ 
    Tiffany Tang\upstairs{\affilone}, 
    Yu Wang\upstairs{\affilone},
    Chao Zhang\upstairs{\affilthree}, 
    Bin Yu\upstairs{\affilone, \affiltwo, \affilfour, \affilfive, \affilseven, *}
   \\[1ex]
   {\small \upstairs{\affilone} Department of Statistics,
   \upstairs{\affiltwo} Department of EECS,
   \upstairs{\affilthree} Department of IEOR} \\[-0.25ex]
   {\small \upstairs{\affilfour} Division of Biostatistics,
   \upstairs{\affilfive} Center for Computational Biology
   } 
   \\[-0.25ex]
   {\small University of California, Berkeley} \\[2ex]
   {\small \upstairs{\affilsix} Department of Pharmaceutical Chemistry}\\[-0.25ex]
   {\small  University of California, San Francisco} \\[2ex]
   {\small \upstairs{\affilseven} Chan Zuckerberg Biohub, San Francisco} \\
  \end{tabular}
  
  \emails{
    Authors ordered alphabetically. Corresponding authors' emails: \upstairs{*}\{chandan\_singh, binyu\}@berkeley.edu
    }
  \vspace*{0.4in}

\begin{abstract}
As the COVID-19 outbreak evolves, accurate forecasting continues to play an extremely important role in informing policy decisions. In this paper, we present our continuous curation of a large data repository containing COVID-19 information from a range of sources. 
We use this data to develop predictions and corresponding prediction intervals for the short-term trajectory of COVID-19 cumulative death counts at the county-level in the United States up to two weeks ahead. Using data from January 22 to June 20, 2020, we develop and combine multiple forecasts using ensembling techniques, resulting in an ensemble we refer to as Combined Linear and Exponential Predictors (CLEP). 
Our individual predictors include county-specific exponential and linear predictors, a shared exponential predictor that pools data together across counties, an expanded shared exponential predictor that uses data from neighboring counties, and a demographics-based shared exponential predictor. We use prediction errors from the past five days to assess the uncertainty of our death predictions, resulting in generally-applicable prediction intervals, Maximum (absolute) Error Prediction Intervals (MEPI). MEPI achieves a coverage rate of more than 94\% when averaged across counties for predicting cumulative recorded death counts two weeks in the future.
Our forecasts are currently being used by the non-profit organization, Response4Life, to determine the medical supply need for individual hospitals and have directly contributed to the distribution of medical supplies across the country. We hope that our forecasts and data repository at \url{https://covidseverity.com} can help guide necessary county-specific decision-making and help counties prepare for their continued fight against COVID-19. 
\end{abstract}
\end{center}

\hspace{10pt}
  
% \copyrightnotice
\restoregeometry

\section*{Media Summary}
Accurate short-term forecasts for COVID-19 fatalities (e.g., over the next two weeks) are critical for making immediate policy decisions such as whether or not counties should re-open.
This paper presents: (i) a large publicly available data repository that continuously scrapes, combines, and updates data from a range of different public sources, and (ii) a predictive algorithm CLEP along with a prediction interval MEPI, for forecasting short-term county-level COVID-19 mortality in the US. 
By combining different trends in the death count data, our county-level CLEP forecasts for cumulative deaths due to COVID-19 are accurate for 7-day into the future and decent for 14-day into the future. The MEPI prediction intervals exhibit high coverage for both 7-day and 14-day forecasts. Our approach was the first to develop forecasts for individual counties (rather than for entire countries or states). Our predictions, along with data and code, are open-source at \url{https://covidseverity.com}. They are currently being used by the non-profit organization, Response4Life, to determine the medical supply need for individual hospitals, and have directly contributed to the distribution of medical supplies across the country. 

{\footnotesize{
\tableofcontents
\listoffigures
\listoftables
}
}
\section{Introduction}
\label{sec:intro}

In recent months, the COVID-19 pandemic has dramatically changed the shape of our global society and economy to an extent modern civilization has never experienced. Unfortunately, the vast majority of countries, the United States included, were thoroughly unprepared for the situation we now find ourselves in. There are currently many new efforts aimed at understanding and managing this evolving global pandemic. This paper, together with the data we have collated (and are collating continuously), represents one such effort. 

Our goals are to provide access to a large data repository combining data from a range of different sources and to forecast short-term (up to two weeks) COVID-19 mortality at the county level in the United States. We also provide uncertainty assessments of our forecasts in the form of prediction intervals based on conformal inference~\cite{vovk2005algorithmic}.

Predicting the short-term impact (e.g., over the next week) of the virus in terms of the number of deaths is critical for many reasons. Not only can it help elucidate the overall impacts of the virus, but it can also help guide difficult policy decisions, such as whether or not to impose or ease lock-downs (or whether to re-open). While many other studies focus on predicting the long-term trajectory of COVID-19, these approaches are currently difficult to verify due to a lack of long-term COVID-19 data. On the other hand, predictions for immediate short-term trajectories are much easier to verify and are likely to be much more accurate than long-term forecasts due to comparatively fewer uncertainties involved, e.g., due to policy
changes, or behavioral changes in the society.
Short-term predictions are also necessary for PPE distribution planning
and policy decisions such as safe re-opening of the counties and states.
So far, a vast majority of predictive efforts have focused on modeling 
\mbox{COVID-19} case-counts or death-counts at the national or state-level~\cite{ferguson2020impact}, rather than the more fine-grained county-level
that we consider in this paper. To the best of our knowledge, ours was the
first work on county-level forecasts.\footnote{By the time of our first
submission to arXiv on May 16, 2020, we were not aware of any concurrent
work on county-level forecasts. See Section~\ref{sec:comparison} for discussion
on recent work~\cite{chiang2020hawkes} on county-level
forecasts (time stamp of June 8) which we became aware of in
mid-June while
revising the manuscript.}

The predictions we produce in this paper focus on recorded cumulative death counts, rather than recorded cases since recorded cases fail to accurately capture the true prevalence of the virus due to previously limited testing availability. Moreover, comparing different counties based on \textit{the number of} recorded cases is difficult since some counties have performed many more tests than others: the number of positive tests does not equal the number of actual cases. While the \textit{proportion of positive tests} is more comparable across different counties, our modeling approach focuses on recorded death counts rather than proportions. The original motivation to predict death counts was to provide a proxy for the severe case counts, where individuals would need intense care in a hospital (see Section~\ref{sec:impact} for further discussion). We note that the recorded death count is also likely to be an under-count of the number of true COVID-19 deaths (since it seems as though in many cases only deaths occurring in hospitals are being counted).\footnote{\url{https://www.nytimes.com/interactive/2020/04/28/us/coronavirus-death-toll-total.html}}
However, more recently, several efforts are being made to obtain better recorded (not using any algorithms or models) COVID-19 death counts, e.g., by including probable deaths
and deaths occurring at home.\footnote{\url{https://covidtracking.com/blog/confirmed-and-probable-covid-19-deaths-counted-two-ways}} 
Nonetheless, the recorded death count is generally believed to be more reliable than the recorded case count.\footnote{\url{https://coronavirus.jhu.edu/data/cumulative-cases}}

In Section~\ref{sec:data}, we introduce our data repository and summarize the data sources contained within. 
This data repository is being updated continuously (as of July 2020) and includes a wide variety of COVID-19 related information in addition to the county-level case-counts and death-counts
(see Tables~\ref{tab:data_features_county}--\ref{tab:data_hospital} for an overview).
Given the rapidly evolving and dynamic nature of COVID-19, several biases arise in the COVID-19 infection data.
We provide a detailed discussion on these biases in the context of our forecasts in Section~\ref{sec:data_biases}.

In Section~\ref{sec:predictor}, we introduce our predictive approach, wherein
we fit a range of different exponential and linear predictor models using
our curated data. Each predictor captures a different aspect of the behaviors
exhibited by COVID-19, both spatially and temporally, i.e., across regions
and time. The predictions generated by the different methods are combined
using an ensembling technique by Schueller et al.~\cite{schuller_2002},
which we refer to as Combined Linear and Exponential Predictors (CLEP). Additional predictive approaches, including those using social distancing information, are presented in Appendix~\ref{sec:additional_models} (even though they do not outperform the CLEP predictors in the main text based on the COVID-19  case and death counts in the past and neighboring counties).

In Section~\ref{sec:prediction_intervals}, we develop uncertainty estimates for our predictors in the form of prediction intervals, which we call Maximum (absolute) Error Prediction Intervals (MEPI). The ideas behind these intervals come from conformal inference~\cite{vovk2005algorithmic} where the prediction interval coverage is well defined by observing the empirical proportion of time that the observed (cumulative) death counts fall into the prediction intervals over a period of time.
Moreover, their guarantees rely on an exchangeability property of the prediction errors in the past several days, which we also examine in the context our prediction tasks.

Section~\ref{sec:results} details the evaluation of the predictors and the prediction intervals for the $3$, $5$, $7$, and $14$-days-ahead forecasts. We use the data from January 22, 2020 (the day of the first COVID-19 death in the US\footnote{\url{https://www.cdc.gov/mmwr/volumes/69/wr/mm6924e2.htm}})  and report the prediction performance over the period March 22, 2020 to June 20, 2020.
Overall, we find that CLEP predictions are adaptive to the exponential and sub-exponential nature of COVID-19 outbreak (with about 15\% error for 7-day-ahead and 30\% error for 14-day-ahead predictions; e.g., see Table~\ref{table:ensembleresults}).
We also provide detailed results for our prediction intervals MEPI from April 11, 2020 to June 20, 2020. And, we observe that MEPIs are reasonably narrow and cover the recorded number of deaths for more than 90\% of days for most of the counties in the US (e.g., see Figures~\ref{fig:mepi_coverage_and_length} and \ref{fig:mepi_coverage_and_length_14_days}). 

Finally, we describe related work by other authors in Section~\ref{sec:comparison}, discuss the impact of our work in distributing medical supplies across the country in Section~\ref{sec:impact}, and conclude in Section~\ref{sec:conclusion}.

\begin{figure}
    \centering
    \includegraphics[width=0.85\textwidth]{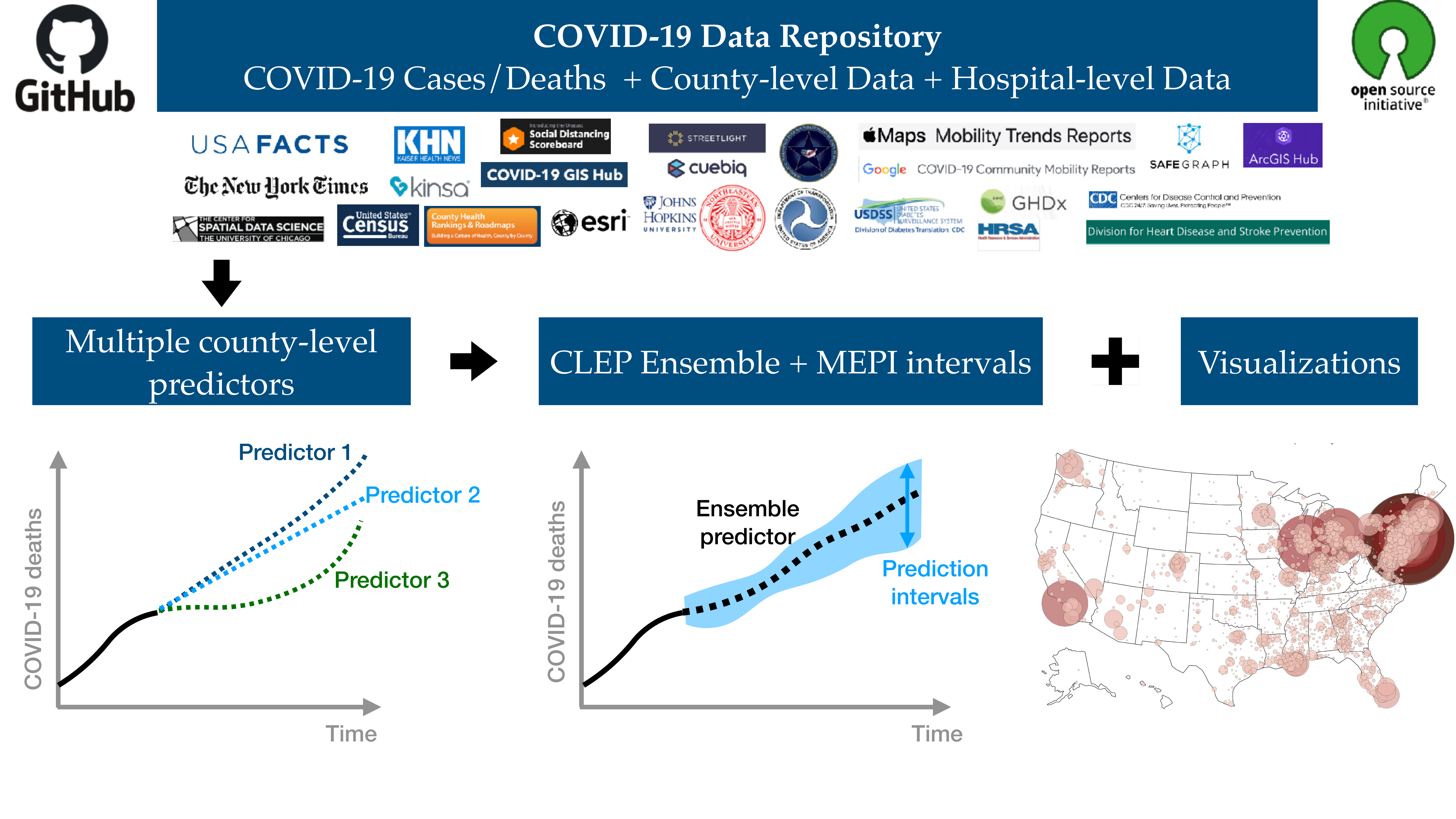}
    \vspace{-20pt}
    \caption[An overview of the paper]{An overview of the paper. We curate an extensive data repository combining data from multiple data sources.
    We then build several predictors for county-level predictions of cumulative COVID-19 death counts, and develop an ensembling procedure (CLEP) and a prediction interval scheme (MEPI) for these predictions.
    Both CLEP and MEPI are generic machine learning methods and can be of independent interest (see Sections~\ref{sec:combined-predictor}
    and \ref{sub:mepi} respectively). All the data, and predictions are publicly available at GitHub repo (link in footnote). Visualizations are available at \url{https://covidseverity.com/} and \url{https://geodacenter.github.io/covid/map.html}, in collaboration with the Center for Spatial Data Science at the University of Chicago.}
    \label{fig:summary}
\end{figure}

Making both data and the predictive algorithms used in this paper accessible to others is key to ensuring the usefulness of these resources. Thus the data, code, and predictors we discuss in this paper are open-source on GitHub (\url{https://github.com/Yu-Group/covid19-severity-prediction}) and are also updated daily with several visualizations at \url{https://covidseverity.com}.
The results in this paper contain case and death information at county level in the U.S. from January 22, 2020 to June 20, 2020 but the data, forecasts, and visualizations in the GitHub repository and at website are updated daily. 
See Figure~\ref{fig:summary} for a high-level summary of the contributions made in this work.

\section{COVID-19 data repository}
\label{sec:data}
One of our primary contributions is the curation of a COVID-19 data repository that we have made publicly available on GitHub. It is updated daily with new information. Specifically, we have compiled and cleaned a large corpus of hospital-level and county-level data from 20+ public sources to aid data science efforts to combat COVID-19. 

\subsection{Overview of the datasets on June 20, 2020}
\label{sub:overview_datasets}
At the \emph{hospital-level}, our dataset covers over 7000 US hospitals and over 30 features including the hospital's CMS certification number (a unique ID of each hospital used by Centers for Medicare and Medicaid Services), the hospital's location, the number of ICU beds, the hospital type (e.g., short-term acute care, critical access), and numerous other hospital statistics. 

There are more than 3,100 counties in the US. At the \emph{county-level}, our repository includes data on 
\begin{enumerate}[label=(\roman*)]
\itemsep0em
    \item daily recorded COVID-19-related case count and (recorded) death
    count by NY Times~\cite{nytimes_data} and USA Facts~\cite{usa_facts_data};
    \item demographic features such as age-wise population, and population density;
    \item socioeconomic factors including poverty levels, unemployment, education, and social vulnerability measures;
    \item health resource availability such as the number of hospitals, ICU beds, and medical staff;
    \item health risk indicators including heart disease, chronic respiratory disease, smoking, obesity, and diabetes prevalence;
    \item social mobility measures such as the percent change in mobility from a pre-COVID-19 baseline; and
    \item other relevant information such as county-level presidential election results from 2000 to 2016, county-level commute data that includes the number of workers in the commuting flow, and airline ticket survey data that includes origin, destination, and other itinerary details.
\end{enumerate}
In total, there are over 8000 features in the county-level dataset. We provide a feature-level snapshot of the different types of data available in our repository, highlighting features in the county-level datasets in Table~\ref{tab:data_features_county} and the hospital-level datasets in Table~\ref{tab:data_features_hospital}. Alternatively, in Tables~\ref{tab:data_county} and ~\ref{tab:data_hospital}, we provide an overview of the county-level and hospital-level data sources in our repository, respectively, organized by the dataset. 

The full corpus of data, along with further details and extensive documentation, are available on GitHub. In particular, we have created a comprehensive data dictionary with the available data features, their descriptions, and source dataset for ease of navigation on \href{https://github.com/Yu-Group/covid19-severity-prediction/blob/master/data/list_of_columns.md}{our github}. We have also provided a quick-start guide for accessing the unabridged county-level and hospital-level datasets with a single Python code line. \\

\paragraph{\emph{Datasets used by our predictors:}}
In this paper, we focus on predicting the number of recorded COVID-19-related cumulative death counts in each county. For our analysis, we primarily use the county-level case and death reports provided by USAFacts from January 22, 2020 to June 20, 2020 (pulled on June 21, 2020) 
along with some county-level demographics and health data. We have marked these datasets with an asterisk (*) in Table~\ref{tab:data_county}.
We discuss our prediction algorithms in detail in Sections \ref{sec:predictor}, \ref{sec:prediction_intervals} and \ref{sec:results}.\\

\paragraph{\emph{Other potential use-cases for our repository:}}
The original intent of our data repository was indeed to facilitate our work with Response4Life and aid medical supply allocation efforts. However, with time the data repository has grown to encompass a much larger audience and now supports investigations into a wide range of COVID-19 related problems.
For instance, using the breadth of travel information in our repository, including (aggregated) air travel, work commutes, and social mobility data, researchers can investigate the impact of both local and between-city travel patterns on the spread of COVID-19. 
Our data repository also includes data on the prevalence of various COVID-19 health risk factors, including diabetes, heart disease, and chronic respiratory disease, which can be used to stratify counties. 
Furthermore, one can also potentially leverage socioeconomic and demographic information, as well as health resource data (e.g., number of ICU beds, medical staff) to gain a better understanding of the severity of the pandemic in a county. 
 Stratification using these covariates is particularly crucial for assessing the COVID-19 status of rural communities, which are not directly comparable, both in terms of people and resources, to larger cities and counties that have received the most attention. \\

\begin{table}
\centering
\rowcolors{2}{gray!20}{white}
\begin{adjustbox}{width=1.01\textwidth,center=\textwidth} 
\begin{tabular}[t]{p{3.8in}p{3in}}
\toprule
{\normalsize \textsc{\textbf{Description of County-level Features}}} & {\normalsize \textsc{\textbf{Data Source(s)}}}\\
\midrule
\addlinespace[0.1em]
\multicolumn{2}{l}{\textbf{COVID-19 Cases/Deaths}}\\
\addlinespace[0.1em]
\hspace{1.5em}\hangindent=1.5em Daily $\#$ of COVID-19-related recorded cases by US county& USAFacts \cite{usa_facts_data}; The New York Times \cite{nytimes_data} \\
\addlinespace[0.1em]
\hspace{1.5em}\hangindent=1.5em Daily $\#$ of COVID-19-related deaths by US county & USAFacts \cite{usa_facts_data}; The New York Times \cite{nytimes_data} \\
\addlinespace[.1em]
\midrule
\addlinespace[.1em]
\multicolumn{2}{l}{\textbf{Demographics}}\\
\addlinespace[0.1em]
\hspace{1.5em}\hangindent=1.5em Population estimate by county (2018) & \hangindent=1.5em Health Resources and Services Administration \cite{ahrf_data} (Area Health Resources Files) \\
\addlinespace[0.1em]
\hspace{1.5em}\hangindent=1.5em Census population by county (2010) & \hangindent=1.5em Health Resources and Services Administration \cite{ahrf_data} (Area Health Resources Files) \\
\addlinespace[0.1em]
\hspace{1.5em}\hangindent=1.5em Age 65+ population estimate by county (2017) & \hangindent=1.5em Health Resources and Services Administration \cite{ahrf_data} (Area Health Resources Files) \\
\addlinespace[0.1em]
\hspace{1.5em}\hangindent=1.5em Median age by county (2010) & \hangindent=1.5em Health Resources and Services Administration \cite{ahrf_data} (Area Health Resources Files) \\
\addlinespace[0.1em]
\hspace{1.5em}\hangindent=1.5em Population density per square mile by county (2010) & \hangindent=1.5em Health Resources and Services Administration \cite{ahrf_data} (Area Health Resources Files) \\
\addlinespace[0.1em]
\midrule
\addlinespace[.1em]
\multicolumn{2}{l}{\textbf{Socioeconomic Factors}}\\
\addlinespace[0.1em]
\hspace{1.5em}\hangindent=1.5em \% uninsured by county (2017) & \hangindent=1.5em County Health Rankings \& Roadmaps \cite{smoking_data} \\
\addlinespace[0.1em]
\hspace{1.5em}\hangindent=1.5em High school graduation rate by county (2016-17) & \hangindent=1.5em County Health Rankings \& Roadmaps \cite{smoking_data} \\
\addlinespace[0.1em]
\hspace{1.5em}\hangindent=1.5em Unemployment rate by county (2018) & \hangindent=1.5em County Health Rankings \& Roadmaps \cite{smoking_data} \\
\addlinespace[0.1em]
\hspace{1.5em}\hangindent=1.5em \% with severe housing problems in each county (2012-16) & \hangindent=1.5em County Health Rankings \& Roadmaps \cite{smoking_data} \\
\addlinespace[0.1em]
\hspace{1.5em}\hangindent=1.5em Poverty rate by county (2018) & \hangindent=1.5em United States Department of Agriculture, Economic Research Service \cite{poverty_data} \\
\addlinespace[0.1em]
\hspace{1.5em}\hangindent=1.5em Median household income by county (2018) & \hangindent=1.5em United States Department of Agriculture, Economic Research Service \cite{poverty_data} \\
\addlinespace[0.1em]
\hspace{1.5em}\hangindent=1.5em Social vulnerability index for each county & \hangindent=1.5em \cite{svi_data} (Social Vulnerability Index) \\
\addlinespace[0.1em]
\midrule
\addlinespace[.1em]
\multicolumn{2}{l}{\textbf{Health Resources Availability}}\\
\addlinespace[0.1em]
\hspace{1.5em}\hangindent=1.5em \# of hospitals in each county & \hangindent=1.5em Kaiser Health News \cite{khn_data} \\
\addlinespace[0.1em]
\hspace{1.5em}\hangindent=1.5em \# of ICU beds in each county & \hangindent=1.5em Kaiser Health News \cite{khn_data} \\
\addlinespace[0.1em]
\hspace{1.5em}\hangindent=1.5em \# of full-time hospital employees in each county (2017) & \hangindent=1.5em Health Resources and Services Administration \cite{ahrf_data} (Area Health Resources Files) \\
\addlinespace[0.1em]
\hspace{1.5em}\hangindent=1.5em \# of MDs in each county (2017) & \hangindent=1.5em Health Resources and Services Administration \cite{ahrf_data} (Area Health Resources Files) \\
\addlinespace[0.1em]
\midrule
\addlinespace[.1em]
\multicolumn{2}{l}{\textbf{Health Risk Factors}}\\
\addlinespace[0.1em]
\hspace{1.5em}\hangindent=1.5em Heart disease mortality rate by county (2014-16) & \hangindent=1.5em Centers for Disease Control and Prevention \cite{heart_data} (Interactive Atlas of Heart Disease and Stroke)\\
\addlinespace[0.1em]
\hspace{1.5em}\hangindent=1.5em Stroke mortality rate by county (2014-16) & \hangindent=1.5em Centers for Disease Control and Prevention \cite{heart_data} (Interactive Atlas of Heart Disease and Stroke)\\
\addlinespace[0.1em]
\hspace{1.5em}\hangindent=1.5em Diabetes prevalence by county (2016) & \hangindent=1.5em Centers for Disease Control and Prevention \cite{diabetes_data} (Diagnosed Diabetes Atlas) \\
\addlinespace[0.1em]
\hspace{1.5em}\hangindent=1.5em Chronic respiratory disease mortality rate by county (2014) & \hangindent=1.5em Institute for Health Metrics and Evaluation \cite{resp_data} \\
\addlinespace[0.1em]
\hspace{1.5em}\hangindent=1.5em \% of smokers by county (2017) & \hangindent=1.5em County Health Rankings \& Roadmaps \cite{smoking_data}\\
\addlinespace[0.1em]
\hspace{1.5em}\hangindent=1.5em \% of adults with obesity by county (2016) & \hangindent=1.5em County Health Rankings \& Roadmaps \cite{smoking_data}\\
\addlinespace[0.1em]
\hspace{1.5em}\hangindent=1.5em Crude mortality rate by county (2012-16) & \hangindent=1.5em United States Department of Health and Human Services \cite{mortality_data} \\
\addlinespace[0.1em]
\midrule
\addlinespace[.1em]
\multicolumn{2}{l}{\textbf{Social Mobility}}\\
\addlinespace[0.1em]
\hspace{1.5em}\hangindent=1.5em Start date of stay at home order by county & \hangindent=1.5em Killeen et al. \cite{killeen2020county} \\
\addlinespace[0.1em]
\hspace{1.5em}\hangindent=1.5em \% change in mobility at parks, workplaces, transits, groceries/pharmacies, residential, and retail/recreational areas & \hangindent=1.5em Google LLC \cite{google_mobility_data} \\
\bottomrule
\end{tabular}
\end{adjustbox}
\caption[Relevant county-level features present in our data repository]{A list of select relevant features from across all county-level datasets contained in our COVID-19 repository grouped by feature topic. See Table~\ref{tab:data_county} for an overview of each of the individual county-level datasets.}
\label{tab:data_features_county}
\end{table}

\begin{table}
\centering
\rowcolors{2}{gray!20}{white}
\begin{adjustbox}{width=1.01\textwidth,center=\textwidth} 
\begin{tabular}[t]{p{3.5in}p{3.3in}}
\toprule
{\normalsize \textsc{\textbf{Description of Hospital-level Features}}} & {\normalsize \textsc{\textbf{Data Source(s)}}}\\
\midrule
\addlinespace[0.1em]
\hspace{.5em}\hangindent=1.5em CMS certification number & \hangindent=1.5em Centers for Medicares \& Medicaid Services \cite{cmi_data} (Case Mix Index File)  \\
\addlinespace[0.1em]
\hspace{.5em}\hangindent=1.5em Case Mix Index & \hangindent=1.5em Centers for Medicares \& Medicaid Services \cite{cmi_data} (Case Mix Index File); \cite{cms_teaching_data} (Teaching Hospitals)\\
\addlinespace[0.1em]
\hspace{.5em}\hangindent=1.5em Hospital location (latitude and longitude) & \hangindent=1.5em Homeland Infrastructure Foundation-Level Data \cite{hifld_hospital_data}; Definitive Healthcare \cite{dh_data} \\
\addlinespace[0.1em]
\hspace{.5em}\hangindent=1.5em \# of ICU/staffed/licensed beds and beds utilization rate & \hangindent=1.5em Definitive Healthcare \cite{dh_data} \\
\addlinespace[0.1em]
\hspace{.5em}\hangindent=1.5em Hospital type & \hangindent=1.5em Homeland Infrastructure Foundation-Level Data \cite{hifld_hospital_data}; Definitive Healthcare \cite{dh_data} \\
\addlinespace[0.1em]
\hspace{.5em}\hangindent=1.5em Trauma Center Level & \hangindent=1.5em Homeland Infrastructure Foundation-Level Data \cite{hifld_hospital_data} \\
\addlinespace[0.1em]
\hspace{.5em}\hangindent=1.5em Hospital website and telephone number & \hangindent=1.5em Homeland Infrastructure Foundation-Level Data \cite{hifld_hospital_data} \\
\bottomrule
\end{tabular}
\end{adjustbox}
\caption[Relevant hospital-level features present in our data repository]{A list of select relevant features from across all hospital-level datasets contained in our COVID-19 repository. See Table~\ref{tab:data_hospital} for an overview of each hospital-level dataset.}
\label{tab:data_features_hospital}
\end{table}

\begin{table}
\centering
\rowcolors{2}{gray!20}{white}
\begin{adjustbox}{width=1.0\textwidth,center=\textwidth} 
\begin{tabular}[t]{p{3in}p{3.9in}}
\toprule
{\normalsize \textsc{\textbf{County-level Dataset}}} & {\normalsize \textsc{\textbf{Description}}}\\
\midrule
\multicolumn{2}{l}{\textbf{COVID-19 Cases/Deaths Data}}\\
% \addlinespace[0.3em]
\hspace{1.5em}\hangindent=2em USAFacts \cite{usa_facts_data}*$^\dagger$ & \hangindent=1.5em Daily cumulative number of reported COVID-19-related death and case counts by US county, dating back to Jan. 22, 2020\\
% \addlinespace[0.3em]
\hspace{1.5em}\hangindent=2em The New York Times \cite{nytimes_data}$^\dagger$ & \hangindent=1.5em Similar to the USAFacts dataset, but includes aggregated death counts in New York City without county breakdowns\\
% \addlinespace[.3em]
\midrule
% \addlinespace[.3em]
\multicolumn{2}{l}{\textbf{Demographics and Socioeconomic Factors}}\\
% \addlinespace[0.3em]
\hspace{1.5em}\hangindent=2.5em Health Resources and Services Administration \cite{ahrf_data} (Area Health Resources Files)* & \hangindent=1.5em Includes data on health facilities, professions, resource scarcity, economic activity, and socioeconomic factors (2018-2019)\\
% \addlinespace[0.3em]
\hspace{1.5em}\hangindent=2.5em County Health Rankings \& Roadmaps \cite{smoking_data}* & \hangindent=1.5em Estimates of various health behaviors and socioeconomic factors (e.g., unemployment, education)\\
% \addlinespace[0.3em]
\hspace{1.5em}\hangindent=2.5em Centers for Disease Control and Prevention \cite{svi_data} (Social Vulnerability Index) & \hangindent=1.5em Reports the CDC's measure of social vulnerability from 2018\\
% \addlinespace[0.3em]
\hspace{1.5em}\hangindent=2.5em United States Department of Agriculture, Economic Research Service \cite{poverty_data} & \hangindent=1.5em Poverty estimates and median household income for each county\\
% \addlinespace[.3em]
\midrule
% \addlinespace[.3em]
\multicolumn{2}{l}{\textbf{Health Resources Availability}}\\
% \addlinespace[0.3em]
\hspace{1.5em}\hangindent=2.5em Health Resources and Services Administration \cite{ahrf_data} (Area Health Resources Files)* & \hangindent=1.5em Includes data on health facilities, professions, resource scarcity, economic activity, and socioeconomic factors (2018-2019)\\
% \addlinespace[0.3em]
\hspace{1.5em}\hangindent=2.5em Health Resources and Services Administration \cite{hpsa_data} (Health Professional Shortage Areas) & \hangindent=1.5em Provides data on areas having shortages of primary care, as designated by the Health Resources \& Services Administration\\
% \addlinespace[0.3em]
\hspace{1.5em} \hangindent=2.5em Kaiser Health News \cite{khn_data}* & \hangindent=1.5em \# of hospitals, hospital employees, and ICU beds in each county\\
% \addlinespace[.3em]
\midrule
% \addlinespace[.3em]
\multicolumn{2}{l}{\textbf{Health Risk Factors}}\\
% \addlinespace[0.3em]
\hspace{1.5em}\hangindent=2.5em County Health Rankings \& Roadmaps \cite{smoking_data}* & \hangindent=1.5em Estimates of various socioeconomic factors and health behaviors (e.g., \% of adult smokers, \% of adults with obesity)\\
% \addlinespace[0.3em]
\hspace{1.5em}\hangindent=2.5em Centers for Disease Control and Prevention \cite{heart_data} (Interactive Atlas of Heart Disease and Stroke)* & \hangindent=1.5em Estimated heart disease and stroke death rate per 100,000 (all ages, all races/ethnicities, both genders, 2014-2016)\\
% \addlinespace[0.3em]
\hspace{1.5em}\hangindent=2.5em Centers for Disease Control and Prevention \cite{diabetes_data} (Diagnosed Diabetes Atlas)* & \hangindent=1.5em Estimated percentage of people who have been diagnosed with diabetes per county (2016)\\
% \addlinespace[0.3em]
\hspace{1.5em}\hangindent=2.5em  Institute for Health Metrics and Evaluation \cite{resp_data}* & \hangindent=1.5em Estimated mortality rates of chronic respiratory diseases (1980-2014) \\
% \addlinespace[0.3em]
\hspace{1.5em}\hangindent=2.5em Institute for Health Metrics and Evaluation \cite{chronic_data} (Chronic Conditions) & \hangindent=1.5em Prevalence of 21 chronic conditions based upon CMS administrative enrollment and claims data for Medicare beneficiaries\\
% \addlinespace[0.3em]
\hspace{1.5em}\hangindent=2.5em  United States Department of Health and Hu- Overall mortality rates (2012-2016) for each county from the Na- man Services \cite{mortality_data} & \hangindent=1.5em Overall mortality rates (2012-2016) for each county from the National Center for Health Statistics\\
% \addlinespace[.3em]
\midrule
% \addlinespace[.3em]
\multicolumn{2}{l}{\textbf{Social Mobility}}\\
% \addlinespace[0.3em]
\hspace{1.5em}\hangindent=2.5em  Killeen et al. \cite{killeen2020county} (JHU Date of Interventions) & \hangindent=1.5em Dates that counties (or states governing them) took measures to mitigate the spread by restricting gatherings\\
% \addlinespace[0.3em]
\hspace{1.5em}\hangindent=2.5em Google LLC, \cite{google_mobility_data} (Google Community Mobility Reports)$^\dagger$ & \hangindent=1.5em Reports relative movement trends over time by geography and across different categories of places (e.g., retail/recreation, groceries/pharmacies)\\
% \addlinespace[0.3em]
\hspace{1.5em}\hangindent=2.5em Apple Inc. \cite{apple_mobility_data} (Apple Mobility Trends)$^\dagger$ & \hangindent=1.5em Uses Apple maps data to report relative (to Jan. 13, 2020)  volume of directions requests per country/region, sub-region or city\\
% \addlinespace[.3em]
\midrule
% \addlinespace[.3em]
\multicolumn{2}{l}{\textbf{Miscellaneous}}\\
% \addlinespace[0.3em]
\hspace{1.5em}\hangindent=2.5em  United States Census Bureau \cite{adjacency_data} (County Adjacency File)* & \hangindent=1.5em Lists each US county and its neighboring counties; from the US Census\\
% \addlinespace[0.3em]
\hspace{1.5em}\hangindent=2.5em  Bureau of Transportation Statistics \cite{plane_data} (Airline Origin and Destination Survey) & \hangindent=1.5em Survey data with origin, destination, and itinerary details from a 10\% sample of airline tickets in 2019 \\
% \addlinespace[0.3em]
\hspace{1.5em}\hangindent=2.5em  MIT Election Data and Science Lab \cite{voting_data} (County Presidential Data) & \hangindent=1.5em County-level returns for presidential elections from 2000 to 2016 according to official state election data records\\
% \addlinespace[.3em]
\bottomrule
\end{tabular}
\end{adjustbox}
\caption[List of county-level datasets (grouped by data category) present in our data repository]{A list of county-level datasets contained within in our COVID-19 repository grouped by data category. Datasets marked with $^\dagger$ are updated daily while all other sources are static. Datasets marked with an asterisk (*) were used in the predictors discussed in this work. Several datasets are relevant to multiple categories and are thus listed multiple times. See Table~\ref{tab:data_features_county} for an overview of select features from these county-level datasets.}
\label{tab:data_county}
\end{table}

\begin{table}
\centering
\rowcolors{2}{gray!20}{white}
\begin{adjustbox}{width=1.01\textwidth,center=\textwidth} 
\begin{tabular}[t]{p{3in}p{3.9in}}
\toprule
{\normalsize \textsc{\textbf{Hospital-level Dataset}}} & {\normalsize \textsc{\textbf{Description}}}\\
\midrule
% \addlinespace[0.3em]
\hspace{.5em}\hangindent=1.5em Homeland Infrastructure Foundation-Level Data \cite{hifld_hospital_data} & \hangindent=1.5em Includes number of ICU beds, and location for US hospitals\\
% \addlinespace[0.3em]
\hspace{.5em}\hangindent=1.5em Definitive Healthcare \cite{dh_data} & \hangindent=1.5em Provides data on number of licensed beds, staffed beds, ICU beds, and the bed utilization rate for hospitals in the US\\
% \addlinespace[0.3em]
\hspace{.5em}\hangindent=1.5em Centers for Medicares \& Medicaid Services \cite{cmi_data} (Case Mix Index File) & \hangindent=1.5em Reports the Case Mix Index (CMI) for each hospital\\
% \addlinespace[0.3em]
\hspace{.5em}\hangindent=1.5em Centers for Medicares \& Medicaid Services \cite{cms_teaching_data} (Teaching Hospitals) & \hangindent=1.5em Lists teaching hospitals along with address (2020) \\
\bottomrule
\end{tabular}
\end{adjustbox}
\caption[List of hospital-level datasets (grouped by data category) present in our data repository]{A list of hospital-level datasets contained within in our COVID-19 repository. Currently, all hospital-level sources are static. See Table~\ref{tab:data_features_hospital} for an overview of select features from these hospital-level datasets.}
\label{tab:data_hospital}
\end{table}

\paragraph{\emph{Comparison with the repository collated by Killeen et al.~\cite{killeen2020county} at Johns Hopkins University:}} Note that similar
but complementary county-level data was recently aggregated and released
in another study~\cite{killeen2020county}. Both our county-level repository
and the repository in \cite{killeen2020county} include data on COVID-19
cases and deaths, demographics, socioeconomic information, education, and
social mobility, albeit some are from different sources. For example, the
repository~\cite{killeen2020county} uses COVID-19 cases and deaths data
from the John Hopkins University CSSE COVID-19 dashboard by Dong et al.~\cite{dong2020interactive} whereas our data is pulled from~USAFacts \cite{usa_facts_data} and the New York Times \cite{nytimes_data}. The main difference, however, between the two repositories is that our data repository also includes data on COVID-19 health risk factors. Furthermore, while the repository in~\cite{killeen2020county} provides additional datasets at the state-level, we provide additional datasets at the hospital-level (given our initial goal of helping the allocation of medical supplies to hospitals, in partnership with the non-profit Response4Life). While their data repository contains both overlapping and complementary information to our repository, a thorough dataset-by-dataset comparison is beyond the scope of this work for two reasons: (i) We learned about this repository towards the completion of our work, and (ii) we were unable to find detailed documentation of how the datasets in their repository were cleaned.

\subsection{Data quality and bias}
\label{sec:data_biases}
Before introducing our prediction algorithms, it is vital to discuss the quality and limitations of the available COVID-19 data. Many downstream analyses, including ours, rely on accurate COVID-19 infection data, including accurate case and death counts. In this subsection, we focus our discussion and evaluation on the data quality of the county-level COVID-19 case and death count data. We also conduct some preliminary exploratory data analysis to shed light on the scale of bias and the possible directions of the biases in the data.

Though discussions on data quality issues and their possible consequences
are relatively sparse in the existing literature, Angelopoulos et al.~\cite{Angelopoulos2020On}
discuss a variety of possible data biases in the context of estimating the case fatality ratio. 
They proposed a method that can theoretically account for two biases: time lag and imperfect reporting of deaths and recoveries. 
Unfortunately, it is hard to evaluate their method's performance since the actual death counts due to COVID-19 remain unknown. Moreover, some data biases (e.g., under-ascertainment of mild cases) for estimating the case fatality ratio do not affect estimation of future death counts. 
Nonetheless, many of the ideas we explore here in uncovering possible biases
in the data are inspired by the work~\cite{Angelopoulos2020On}.\\

\paragraph{\emph{Imperfect reporting and attribution of deaths due to COVID-19:}} 
Numerous news articles have suggested that the official US COVID-19 death count is an underestimate~\cite{nytimes_missing_deaths}. According to The New York Times\footnote{\url{https://www.nytimes.com/interactive/2020/06/19/us/us-coronavirus-covid-death-toll.html}}, on April 5, the Council of State and Territorial Epidemiologists advised states to include both the confirmed cases based on laboratory testing, and probable cases---using specific criteria for symptoms and exposure. 
The Centers for Disease Control adopted these definitions, and national CDC data began including confirmed and probable cases on April 14. The infection data included in our data repository (USAFacts and NY Times) contains both the probable death and the confirmed deaths beginning April~14.
Although the probable death counts address imperfect reporting and attribution, it is unclear to what extent the problem is mitigated.
Going forward, we use the term recorded death counts and recorded case counts to reflect that the recorded counts are based on both confirmed and probable deaths and cases.\\

\begin{figure}
    \centering
      \includegraphics[width=\textwidth]{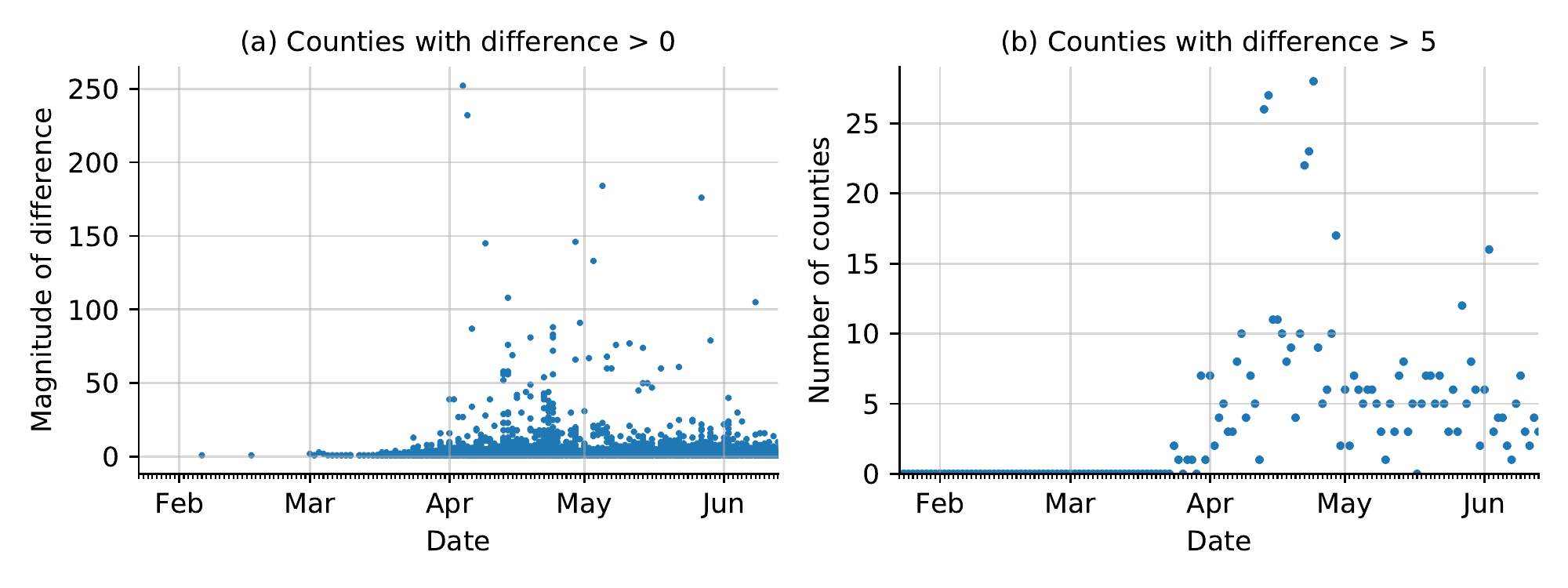}
      \vspace{-5mm}
    \caption[EDA: USAFacts vs NY Times datasets]{Plots illustrating the differences in county-wise recorded death counts between the USAFacts and NY Times datasets. In panel \textbf{(a)}, we plot the magnitude of difference in death counts for each county as a function of time, where one dot represents a particular county on a particular day. We notice that
    while there are counties where the discrepancy can be larger than $100$ deaths recorded in a day, the majority of the discrepancies are not large. The large discrepancies are possibly due to different data curation protocols used by USAFacts and NY Times.
    In panel \textbf{(b)}, we plot the number of counties that have a discrepancy of more than $5$ on any given day between the two datasets. We notice that on any given day, no more than 30 counties, i.e., $<1\%$  of the more than 3,100 counties, have a difference larger than $5$ between the two datasets.}
    \label{fig:usa_nytimes}
\end{figure}

\paragraph{\emph{Inconsistency across different data sources:}} There exist multiple sources of COVID-19 death counts in the US. In our data repository, we include data from~USAFacts \cite{usa_facts_data} and data from the New York Times \cite{nytimes_data}. According to USAFacts and the NY Times websites, they both collect data from state and local agencies or health departments and manually curate the data. 
However, these websites do not scrape data from those sources at the same time. While USAFacts states that ``they mostly collect data in the evening (Pacific Time)", NY Times mentions they update data throughout the day. 
Furthermore, while there are some discussions on how they collect and process the data on their websites, the specific data curation rules are not shared publicly. 
Possibly due to different scrapping times and curation rules, there are a few discrepancies in their case and death counts. In Figure~\ref{fig:usa_nytimes}(a), we plot the absolute difference in death counts from the two datasets for each county. In Figure \ref{fig:usa_nytimes}(b), we plot the number of counties whose recorded COVID-19 deaths on a given day differ by more than 5. The proportion of counties with observably different death-counts (difference $>5$) is in general small ($<1\%$), although sometimes the differences are quite significant ($>$100). Since two datasets are curated under different rules (which are unknown to us), it is not obvious how to combine them or assess their validity. For our analysis, we choose to use the USAFacts COVID-19 deaths data as they provide county-level death counts for New York City while the NY Times data aggregates the death counts over the five boroughs in that region.\\

\begin{figure}
    \centering
    \resizebox{\textwidth}{!}{
    \begin{tabular}{ccc}
        \includegraphics[width=0.33\textwidth, trim={0.1cm 0 0.5cm 0},clip]{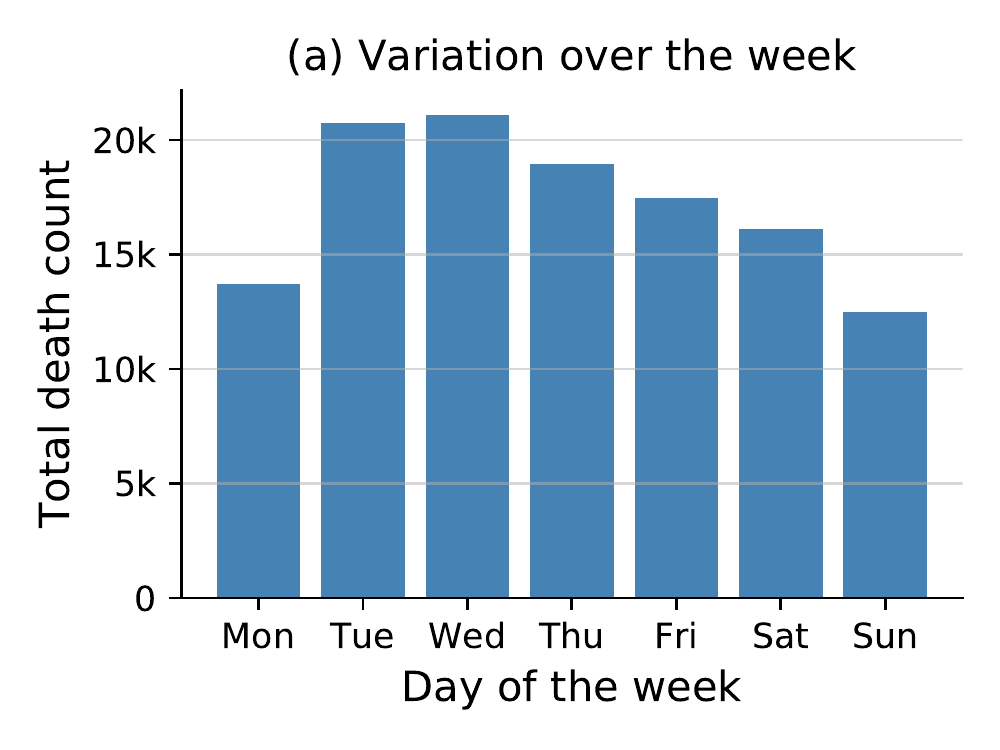} &
        %  trim={<left> <lower> <right> <upper>}
      \includegraphics[width=0.33\textwidth, trim={0.1cm 0 0.45cm 0},clip]{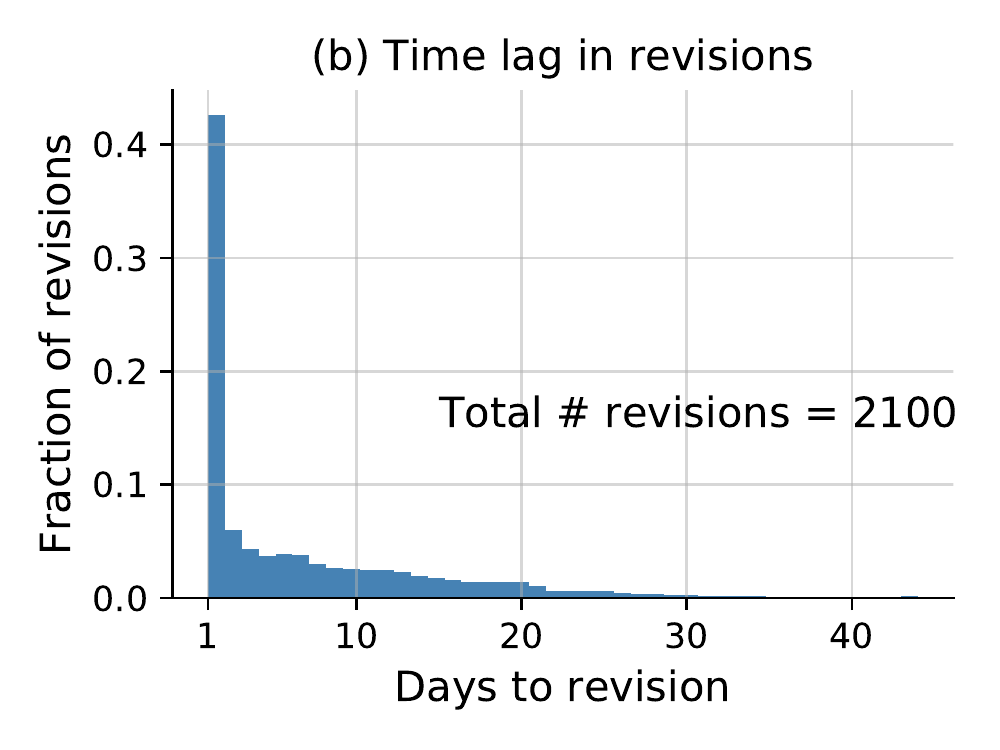}  & 
      \includegraphics[width=0.33\textwidth, trim={0.1cm 0 0.45cm 0},clip]{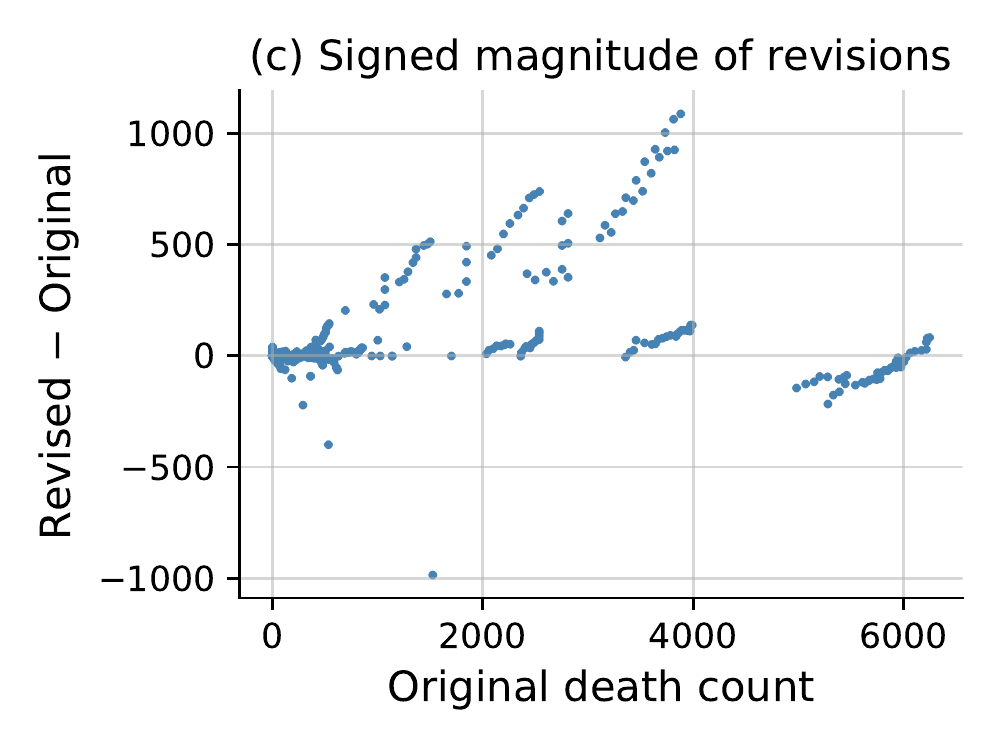} 
    %   \\
    % (a) & (b) & (c)
    \end{tabular}
    }
    \caption[EDA: Potential biases in USAFacts dataset]{EDA plots for identifying potential biases in USAFacts data. \textbf{(a)} Variation of recorded death counts over different days of the week. We observe that the total death counts (for March-June) shows a trend across different weekdays, and the total count is significantly smaller for Sunday and Monday. In panel \textbf{(b)}, we plot the histogram of the number of days after which the count for a given county on a given day is revised. In panel \textbf{(c)}, we present the signed magnitude of the change, i.e., the ``revised count minus the original count'' against the original count for each revision.}
    \label{fig:weekday_and_adjustment}
\end{figure}

\paragraph{\emph{Weekday patterns:}} The recorded case counts and death counts have a significant weekly pattern in both the USAFacts and NY Times data; such a pattern can possibly be attributed to the reporting delays as discussed in~\cite{Angelopoulos2020On}. We show the total number of deaths recorded on each day of the week in the USAFacts data in Figure \ref{fig:weekday_and_adjustment}(a). The total number of deaths on Monday and Sunday is significantly lower than that for any other day. We try to account for these weekly patterns in our prediction methods later in Appendix~\ref{sec:additional_models}.\\

\paragraph{\emph{Historical data revision:}} We observed that some of the historical infection data was revised after initially being recorded. According to USAFacts, these revisions are typically due to earlier mistakes from local agencies which revised their previously recorded death counts. Note that these data revisions are not related to the probable deaths as we discussed earlier and therefore we regard this phenomenon as a distinct source of bias. This kind of revision is not common: until June 21, we observe that only $2.1\%$ of counties across the U.S. had one or more historical revisions.   
Figure \ref{fig:weekday_and_adjustment}(b) shows a histogram of the amount of time from the initial record to the revision. It can be seen that almost half of the changes happen the day after the data was initially recorded. 
Figure \ref{fig:weekday_and_adjustment}(c) shows the signed magnitude of the change in death count that results from the revisions versus the initial recorded death counts. Note that there are a few stripes (consecutive upward revisions) in the plot. 
Each stripe corresponds to the revision of a particular county on different dates. Since the reported data is cumulative death counts, when the deaths from a few days ago get revised, all the data after that day until the day when the revision is made also get revised accordingly, thereby explaining the short stripy trends.

However, only $582$ out of 2100 revisions (around $27\%$) have absolute magnitude $>2$ deaths, and $354$ of these $582$ revisions (around $67\%$) are in the positive direction (i.e., more deaths than initially recorded). Furthermore, amongst the $61$ revisions with an absolute magnitude larger than $200$, almost all of them ($57/61$) lead to an increase in the number of recorded deaths. The four most significant downward revisions, i.e., the points with large negative ``revised-original count" in Figure~\ref{fig:weekday_and_adjustment}(c), correspond to counties in the Washington State. This finding can be corroborated by the media news that Washington State admitted errors in reporting the death counts, and subsequently lowered these counts in the revisions.\footnote{See \url{https://www.clarkcountytoday.com/news/washington-department-of-health-clarifies-covid-19-death-numbers/} and \url{https://www.king5.com/article/news/health/coronavirus/washington-coronavirus-testing-death-toll-mistake}, last accessed on July 24, 2020.}
It is natural for our predictions to vary if the training data (for a fixed period) varies with time, i.e., when the COVID-19 counts are adjusted for a backdate. Most of these revisions are minor, in which case the general performance of our predictors does not change significantly. However, when the revisions are a significant uptick, the predictions can become unstable for a few days (depending on the uptick, and the prediction-horizon). See Section~\ref{sub:results_clep} and \ref{sub:county_visual} for further discussion on these biases.
In this paper, we use the initial infection data available on June 21, 2020 to evaluate our algorithm performance, i.e., we do not use data that was revised after June 21. Nonetheless, we caution the reader to keep the following fact in mind while interpreting results from our work as well as other related COVID-19 studies: The recorded death counts themselves are an under-estimate and the consequent bias is hard to adjust for due to the lack of ground truth.

\section{Predictors for forecasting short-term death counts}
\label{sec:predictor}

\begin{figure}
    \centering
     \begin{tabular}{c}
    \includegraphics[width=0.9\textwidth]{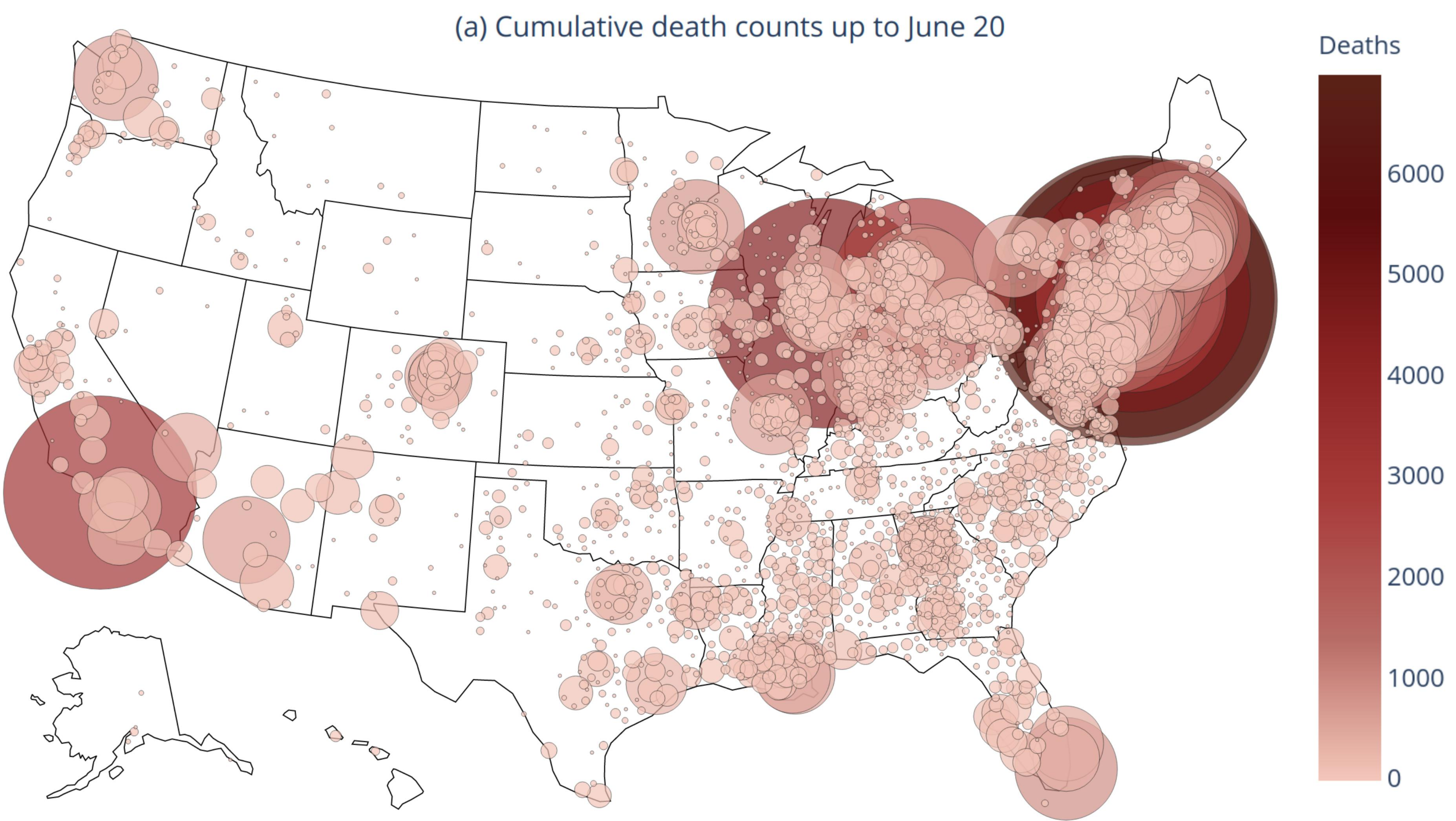}  \vspace{10mm}\\
    \includegraphics[width=0.9\textwidth]{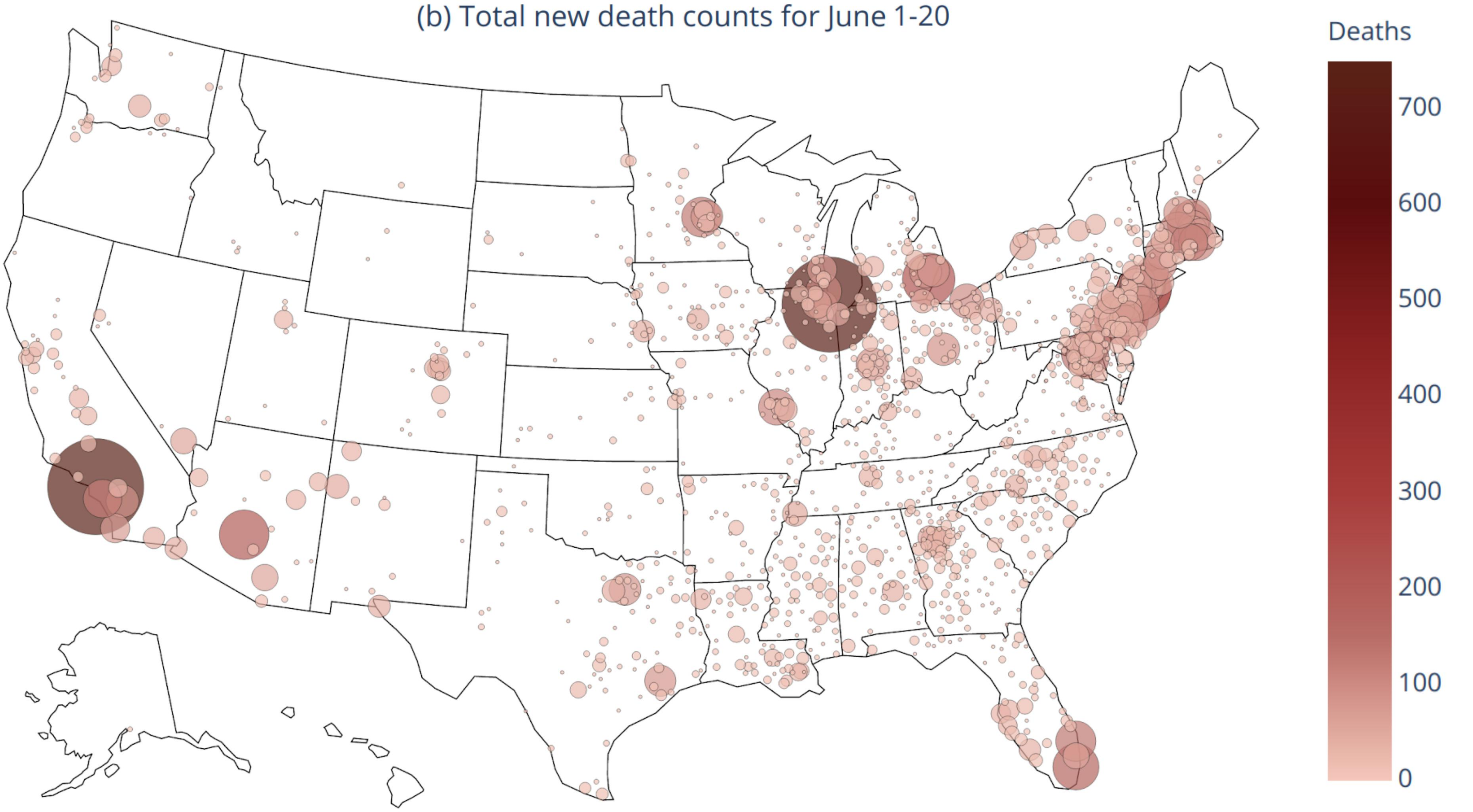}
    \end{tabular}
    \caption[Visualization of the COVID-19 outbreak in the US]{Visualization of the COVID-19 outbreak in the US. 
    We depict the cumulative recorded death counts up to June 20 in panel \textbf{(a)} and newly recorded death counts for the period June 1-20 in panel \textbf{(b)}. Each bubble denotes the death count for a county (the absence of a bubble denotes a zero count).
    The bubble size (area) is proportional to the death counts in the region. The two panels' bubble sizes are on the same scale, but the color scale is different as shown respectively on each plot.
    }
    \label{fig:total-deaths-map}
\end{figure}

Figure~\ref{fig:total-deaths-map} provides a visualization of the COVID-19 outbreak across the United States. We plot (a) the cumulative recorded death counts due to COVID-19 up to June 20, and (b) the new death counts from June 1 to June 20, 2020. Each bubble denotes a county-level count, a darker and larger bubble denotes a higher death count, and the absence of a bubble denotes that the count is zero. 
Panel (a) captures the extent of the outbreak in a region, while (b) captures the recent trends in the outbreak. The color scale differs between the two plots to better illustrate the respective counts in each plot, but the size scales are held constant between the two plots to help provide a comparison between the extent and recent trends of COVID-19. 
Overall, Figure~\ref{fig:total-deaths-map} clearly shows that the COVID-19 outbreak in the United States is incredibly dynamic both in time and across different regions. The worst-affected regions include the states of New York, New Jersey, Massachusetts, Michigan, Illinois, Florida, Louisiana, Georgia, Washington, and California.
Moreover, most of these areas continue to face a substantial COVID-19 burden in the first two-thirds of June.

We develop several different statistical or machine learning prediction algorithms to capture the dynamic behavior of COVID-19 death counts. Since each prediction algorithm captures slightly different trends in the data, we also develop various weighted combinations of these prediction algorithms. The five prediction algorithms or predictors for cumulative recorded death counts that we devise in this paper are as follows:

\begin{enumerate}
    \item \textbf{A separate-county exponential predictor (the ``separate'' predictors)}: a series of predictors built for  predicting cumulative death counts for each county using only past death counts from that county.
    \item \textbf{A separate-county linear predictor (the ``linear'' predictor)}: a predictor similar to the separate county exponential predictors, but uses a simple linear format, rather than the exponential format.
    \item \textbf{A shared-county exponential predictor (the ``shared'' predictor)}: a single predictor built using death counts from all counties, used to predict death counts for individual counties.
    \item \textbf{An expanded shared-county exponential predictor (the ``expanded shared'' predictor)}: a predictor similar to the shared-county exponential predictor, which also includes COVID-19 case numbers and neighboring county cases and deaths as predictive features. 
    \item \textbf{A demographics shared-county exponential predictor (the ``demographics shared'' predictor)}: a predictor also similar to the shared-county exponential predictor, but which also includes various county demographic and health-related predictive features.
\end{enumerate}

An overview of these predictors is presented in Table~\ref{tab:models}. 
We use the python package statsmodels~\cite{seabold2010statsmodels} to train all the five predictors with the same Poisson log-likelihood loss function (but the set of features for each predictor is different).\footnote{We use the default parameters in the Python statsmodels package (version 0.11.1) while training our predictors. For predictors (1) and (2), \emph{glm.fit} was used which always converged. For predictors (3)-(5), we first tried \emph{glm.fit\_regularized} since we experimented with the use of explicit $\ell_1$ and $\ell_2$ regularization in the beginning. Although, eventually the predictors (3)-(5) reported in this paper did not use any explicit $\ell_1$ or $\ell_2$ regularization, it turns out that default settings (algorithm and stopping criterion) in the functions \emph{glm.fit} and \emph{glm.fit\_regularized} are different leading to different implicit regularizations, and consequently different performance. We still chose to use \emph{glm.fit\_regularized} for fitting predictors (3)-(5) since it led to better performance for predictor (4) (which forms the basis of our best performing predictor CLEP). We also note that the function \emph{glm.fit\_regularized}, in fact, calls another function \emph{fit\_elasticnet} which uses block coordinate descent (BCD) by default to solve a generalized linear model. The default value for the maximum number of iterations of BCD ($\mathrm{max\_iter}$) is set to be 50, which, in some cases, resulted in early stopping (a form of implicit regularization) before the iterative algorithm converges.}
To combine the different trends captured by each of these predictors, we also fit various combinations of them, which we refer to as Combined Linear and Exponential Predictors (CLEP). CLEP produces a weighted average of the predictions from the individual predictors, where we borrow the weighting scheme from prior work~\cite{schuller_2002}. In this weighting scheme, a higher weight is given to those predictors with more accurate predictions, especially on recent time points. We find that the CLEP that combines only the \textit{linear predictor} and the \textit{expanded shared predictor} consistently has the best predictive performance when compared to the individual predictors and the CLEP that combines all five predictors. (We did not try all possible combinations to avoid over-fitting; also see Table~\ref{table:ensembleresults}). 
For the rest of this section, we expand upon the individual predictor models and the weighting procedure for the CLEP ensembles.

In addition,  Appendix~\ref{sec:additional_models} contains results on variants of the two best single predictors (linear and expanded shared), which include features for social-distancing and features that account for the under-reporting of deaths on Sunday and Monday (as observed in Figure~\ref{fig:weekday_and_adjustment}(a)). These additional features did not lead to better performance.

We note that although in this paper, we discuss our algorithms for predicting cumulative recorded death counts, the methods can be more generally applied to predict other quantities of interest, e.g., case counts, or new death counts for each day. Moreover, the combination scheme used for combining different predictors can be of independent interest in developing ensembling schemes with generic machine learning methods.

\begin{table}[ht]
    \centering
    \begin{footnotesize}
    \begin{tabularx}{1.01\textwidth}{X|X|X|X|X|X}\toprule
        \footnotesize
        Predictor name    & Type        & Fit separately to each county? & Fit jointly to all counties? & Use neighboring counties? & Use demographics? \\\hline
        %\midrule
        \thead[l]{Separate}            & Exponential &     \checkmark                           &                              &                           &                   \\
        Linear              & Linear      &   \checkmark                            &                              &                           &\\   
        \thead[l]{Shared}              & Exponential &                                &\checkmark                           &                           &                   \\
        \thead[l]{Expanded \\ shared}     & \thead[l]{Exponential} &                                &    \checkmark                          &          \checkmark                 &                   \\
        \thead[l]{Demographics\\ shared} & \thead[l]{Exponential} &                                &       \checkmark                       &                           &          \checkmark         \\   \bottomrule               
    \end{tabularx}
    \end{footnotesize}
    \caption[Overview of the five predictors]{Overview of the 5 predictors used here. The best model is a combination of the linear predictor and the expanded shared predictor (see Section~\ref{sec:combined-predictor}).}
    \label{tab:models}
\end{table}

\newcommand{\ehat}{\widehat{\operatorname{E}}}

\subsection{The separate-county exponential predictors (the ``separate'' predictors)}
The separate-county exponential predictor aims to capture the observed exponential growth of \mbox{COVID-19} deaths \cite{exponential-growth}. We approximate an exponential curve for death count separately for each county using the most recent 5 days of data from that county. These predictors have the following form:
\begin{align}
    \ehat [\textrm{deaths}_{t+1}^c \vert t]=\exp\big(\beta_0^c + \beta_1^c (t+1) \big),
    \label{eq:county-predictor}
\end{align}
where $\ehat [\textrm{deaths}_{t+1}^c \vert t]  $ denotes the (fitted) cumulative death count by the end of day $t+1$ for county~$c$, and it is trained on the data until day $t$, and computed
on the morning of day $t+1$.
Note that we use $t+1$ on the RHS of equation~\ref{eq:county-predictor} just for notational exposition, and in practice we just use $\beta_0^c + \beta_1^c t$ in the exponent in our code.

Here we fit a separate predictor for each county, and the coefficients $\beta_0^c$ and $\beta_1^c$ for each county $c$ are fit using maximum likelihood estimation under a Poisson generalized linear model (GLM) with $t$ as the independent variable and $\textrm{deaths}_t$ as the observed variable.
In simple words, on the morning of day $t+1$, the coefficients are estimated using the cumulative recorded death counts for day $\{t, t-1, t-2, t-3, t-4\}$. And to predict $k$-days-ahead cumulative death count on the morning of day $t+1$---denoted by  $\ehat [\textrm{deaths}_{t+k}^c \vert t]$---we simply replace $t+1$ with $t+k$ on the RHS of equation~\ref{eq:county-predictor}. Note that although the prediction $\ehat (\textrm{deaths}_{t+1}^c \vert t)$ is being made on day $t+1$, we call it 1-day-ahead prediction since it is made in the morning of day $t+1$ using the data for up to day $t$. Moreover, the recorded count $\textrm{deaths}_{t+1}^c$ is reported only late in the night of day $t+1$ or early morning of the next day~$t+2$.

If the first death in a county occurred less than 5 days prior to fitting the predictor, only the days from the first death were used for the fit. If there is less than three days' worth of data or the cumulative deaths remain constant in the past days, we simply use the most recent deaths as the predicted future value. We also fit exponential predictors to the full time-series (as opposed to just the most recent 5 days) of available data for each county. However, due to the rapidly shifting trends, these performed worse than our 5-day predictors. We also found that predictors fit using 6 days of data yielded similar results to predictors fit using 5 days of data, and using 4 days of data performed slightly worse.

To handle possible over-dispersion of data (when the variance is larger than the mean), we also explored estimating $\{\beta_0^c, \beta_1^c\}$ by fitting a negative binomial regression model (in place of Poisson GLM) with inverse-scale parameter taking values in $\{ 0.05, 0.15, 1\}$. However, we found that this approach yields a larger mean absolute error than the Poisson GLM for counties with more than 10 deaths.
% \footnote{To keep the paper readable, several exploratory, and hyper-parameter tuning results are not presented in this paper. 
% Most of our methods were tuned for data available till mid-April, and minimal tuning has been performed since then.
% }

\subsection{The separate-county linear predictor (the ``separate linear'' predictor)}
\label{sub:linear}
The separate linear predictor aims to capture linear growth,
based on the most recent 4 days of data in each county. In the early stages of tuning, we tried using 5 and 7 days of data, and obtained worse performance (see Appendix~\ref{fig:weekday_feat}.). The motivation for the linear model is that some counties are exhibiting sub-exponential growth. For these counties, the exponential predictors introduced in the previous section may not be a good fit to the data.
The separate linear predictors are given by
\begin{equation}
\label{eq:linear_model}
    \ehat[\textrm{deaths}_{t+1}^c \vert t] = \beta_0^c + \beta_1^c (t+1),
\end{equation}
where we fit the coefficients $\beta_0^c$ and $\beta_1^c$ via ordinary least squares using the cumulative death count for county $c$ for most recent 4 days.
Like equation~\ref{eq:county-predictor}, we use $t+1$ on the RHS simply for notational exposition.
Put simply, on the morning of day $t+1$, the coefficients $\{\beta_0^c, \beta_1^c \}$ are estimated using the death counts for day $t, t-1, t-2, t-3$. To predict $k$-days-ahead, i.e., predict cumulative death counts by the end of day $t+k$ on the morning of day $t+1$ (in our notation, $\ehat [\textrm{deaths}_{t+k}^c \vert t]$), we simply replace $t+1$ by $t+k$ on the RHS of equation~\ref{eq:linear_model}.

\subsection{The shared-county exponential predictor (the ``shared'' predictor)}

To incorporate additional data into our predictions, we fit a predictor that combines data across different counties. Rather than producing a separate predictor model for each county (as in the separate predictor approach above), we instead produce a single shared predictor that pools information from counties across the nation. The shared predictor is then used to predict future deaths in the individual counties. These changes allows us to leverage the early-stage trends from counties that are now much further along in the pandemic trajectory to inform the predictions for other current earlier-stage counties. 

The data underlying the shared predictor is slightly different from the separate county predictors. Instead of only including the most recent 5 days of data from each county, we include all days after the third death in each county. (In the earlier stages of tuning, we also tried including the counties after first and fifth death, and then selected the choice of third death due to better performance.) Thus the data from many of the counties extend substantially further back than 5 days, and for each county, $t=0$ is the day on which the third death occurred.
Instead of basing the exponential predictor prediction on time $t+1$ (as was the case for the separate predictors above), we base the prediction on the logarithm of the previous day's death count. This choice makes the counties comparable since the outbreaks began at different time points in each county. The shared predictor is given as follows:
\begin{align}
    \ehat [\textrm{deaths}_{t+1}^c|t]=\exp\bigg(\beta_0 + \beta_1 \log(\textrm{deaths}_{t}^c+1)\bigg),
    \label{eq:shared-predictor}
\end{align}
where $\ehat[\textrm{deaths}_{t+1}^c|t]$ denotes the (fitted) cumulative death count by the end of day $t+1$ for a county $c$, and $\textrm{deaths}_{t}^c$ denotes the recorded cumulative death count for that county by the end of day $t$. 
The coefficients $\beta_0$ and $\beta_1$ are shared across all counties and fitted by maximizing the log-likelihood corresponding to Poisson GLM (like that in the separate county predictor given by equation~\ref{eq:county-predictor}).
We normalize the feature matrix to have zero mean and unit variance before fitting the coefficients.
To predict $k$-days-ahead cumulative death count $\ehat [\textrm{deaths}_{t+k}^c \vert t]$, we first obtain the estimate $\ehat [\textrm{deaths}_{t+1}^c|t]$ using equation~\ref{eq:shared-predictor}. Next, we plug-in $\log(\ehat [\textrm{deaths}_{t+j}^c|t]+1)$ on the RHS of equation~\ref{eq:shared-predictor} to compute $\ehat [\textrm{deaths}_{t+j+1}^c|t]$ in a sequential manner for $j=1, \ldots, k-1$, and finally obtain $\ehat [\textrm{deaths}_{t+k}^c|t]$ ($k$-day-ahead prediction computed on the morning of day $t+1$).

\subsection{The expanded shared exponential predictor (the ``expanded shared'' predictor)}
\label{sub:expanded_shared}

Next, we expand the shared county exponential predictor to include other COVID-19 dynamic (time-series) features. In particular, we include the number of recorded \textit{cases} in the county, as this may give an additional indication to the severity of an outbreak. We also include the total sum of cumulative death (and case) counts in the \textit{neighboring} counties. 
Let $\textrm{cases}_t^c$, $\textrm{neigh\_deaths}^c_t$, $\textrm{neigh\_cases}^c_t$ respectively denote the (recorded) cumulative case count in the county $c$ at the end of day $t$, the total sum of cumulative death counts across all its neighboring counties at the end of day $t$, and the total sum of cumulative recorded case counts across all its neighboring counties at the end of day $t$. Then our (expanded) predictor to predict the number of recorded cumulative deaths $k$ days into the future is given by
\begin{align}
    \ehat[\textrm{deaths}_{t+1}^c|t] &= 
    \exp\bigg(\beta_0 + \beta_1 \log(\textrm{deaths}^c_{t}+1) + \beta_2 \log(\textrm{cases}^c_{t-k+1}+1) \notag\\ &\qquad\qquad+\beta_3\log(\textrm{neigh\_deaths}^c_{t-k+1}+1)+\beta_4\log(\textrm{neigh\_cases}^c_{t-k+1}+1) \bigg),
    \label{eq:shared_expanded_model}
\end{align}
where the coefficients $\{\beta_i\}_{i=0}^4$ are shared across all counties and are fitted using the Poisson GLM after normalization of each feature (in the exponent) to have zero mean and unit variance.
When \emph{fitting} the predictor on the morning of day $t+1$, we use the death counts for the county up to the end of day $\tidx$. However, we only use the new features (cases in the current county, cases in neighboring counties, and deaths in neighboring counties) up to the end of day $t-k+1$. Moreover, we normalize the feature matrix to have zero mean and unit variance before fitting the predictor.
While \emph{predicting} the death count for a given county $k$ days into the future (i.e, the cumulative death count by the end of day $\tidx+k$), we iteratively use the daily sequential predictions for the death counts for that county, and use the information for the other features only up to time $t$ (the time up to which we have data available).
More precisely, first we estimate $\ehat[\textrm{deaths}^c_{t+1}\vert t]$ by plugging in the normalized features $\log(\textrm{deaths}^c_{t})$, $\log(\textrm{cases}^c_{t-k+1})$,  $\log(\textrm{neigh\_deaths}^c_{t-k+1})$, and $\log(\textrm{neigh\_cases}^c_{t-k+1})$ in equation~\ref{eq:shared_expanded_model}, where the normalization is done across counties so that each feature has zero mean and unit variance. 
Then, for $j=1, 2, \ldots, k-1$, we recursively plug-in $\log(\ehat[\textrm{deaths}^c_{t+j} \vert t])$, $\log(\textrm{cases}^c_{t-k+j+1})$, $\log(\textrm{neigh\_deaths}^c_{t-k+j+1})$, $\log(\textrm{neigh\_cases}^c_{t-k+j+1}))$ in equation~\ref{eq:shared_expanded_model} (again after normalizing each of these features) to compute $\ehat[\textrm{deaths}^c_{t+j+1} \vert t]$, and finally obtain the compute
$\ehat[\textrm{deaths}^c_{t+k}\vert t]$ for $k$-day-ahead prediction made with data until day $t$.
It may be possible to jointly predict the new features along with the number of deaths, but we leave building such a predictor to future work.
As before, the predictor is fitted by including all days after the third death in each county.

\subsection{The demographics shared exponential predictor (the ``demographics shared'' predictor)}
\label{sec:demographics_predictor}
The demographics shared county exponential predictor is again very similar to the shared predictor. However, it includes several static county demographic and healthcare-related features to address the fact that some counties will be affected more severely than others, for instance, due to (a) their population makeup, e.g., older populations are likely to experience a higher death rate than younger populations, (b) their hospital preparedness, e.g., if a county has very few ICU beds relative to their population, they might experience a higher death rate since the number of ICU beds is correlated strongly (0.96) with the number of ventilators \cite{rubinson2010mechanical}, and (c) their population health, e.g., age, smoking history, diabetes, and cardiovascular disease are all considered to be likely risk factors for acute COVID-19 infection \cite{guan2020comorbidity, qi2020epidem, guan2020clinical, goh2020rapid, zhou2020clinical}. 

For a county $c$, given a set of demographic and healthcare-related features $d_1^c, \dots, d_m^c$ (such as median age, population density, or number of ICU beds), the demographics shared predictor is given by
\begin{align}
\label{eq:demo_shared}
    \ehat[\textrm{deaths}^c_{t+1}\vert t] = \exp\bigg( \beta_0 + \beta_1 \log(\textrm{deaths}^c_{t}+1) + \beta_{d_1} d^c_1 + \dots + \beta_{d_m} d^c_m\bigg).
\end{align}
Here the coefficients $\{\beta_0, \beta_1, \beta_{d_1}, \ldots, \beta_{d_m}\}$ are shared across all counties, and are fitted by maximizing the log-likelihood of the corresponding Poisson generalized linear model, where we include all the observations since the third death in each county. Moreover, we also normalize the feature matrix to have zero mean and unit variance before fitting the coefficients. The features we choose fall into three categories:
% \begin{small}
\begin{enumerate}
    \item County density and size: population density per square mile (2010), population estimate (2018)
    \item County healthcare resources: number of hospitals (2018-2019), number of ICU beds (2018-2019)
    \item County health demographics: median age (2010), percentage of the population who are smokers (2017), percentage of the population with diabetes (2016), deaths due to heart diseases per 100,000 (2014-2016).
\end{enumerate}
% \end{small}
The $k$-day-ahead predictions for this predictor are obtained in a very similar manner to the shared predictor~\eqref{eq:shared-predictor}: We first obtain the estimate $\ehat [\textrm{deaths}_{t+1}^c\vert t]$ using equation~\ref{eq:demo_shared} and then, sequentially plug-in  $\log(\ehat [\textrm{deaths}_{t+j}^c\vert t]+1)$ on the RHS of the equation~\ref{eq:demo_shared} (after normalization to obtain zero mean and unit variance) to compute $\ehat [\textrm{deaths}_{t+j+1}^c \vert t]$ in a sequential manner for $j=1, \ldots, k-1$.

\subsection{The combined predictors: CLEP}
\label{sec:combined-predictor}
Finally, we consider various combinations of the five predictors we have introduced above using an ensemble approach similar to that described in \cite{schuller_2002}. Specifically, we use the recent predictive performance (e.g., over the last week) of different predictors to guide an adaptive tuning of the corresponding weights in the ensemble. 
To simplify notation, let us denote the predictions for cumulative death count by the end of day $\tidx+k$---where the prediction is made on the morning of day $t+1$---by $\{\widehat y_{\tidx+k}^{m}\}$ with $m=1, \ldots, M$ denoting the index of various linear and exponential predictors.\footnote{Our predictions are released around 11:30 AM Pacific Time each day, both on GitHub and our website (\url{https://covidseverity.com}). The released predictions on day $t+1$ include the county-wise predictions for cumulative death counts by the end of day $t+1$ itself. To summarize, 1-day-ahead prediction for day $t+1$ is denoted by  $\ehat[\textrm{deaths}^c_{t+1}\vert t]$ earlier is now written simply as $\widehat{y}_{t+1}$. Similarly, the $k$-day-ahead prediction $\ehat[\textrm{deaths}^c_{t+k}\vert t]$ is denoted by $\widehat{y}_{t+k}$ in the simplified (and slightly abused) notation.} Then, their Combined Linear and Exponential Predictor (CLEP) is given by
\begin{align}
\label{eq:weightedaverage}
\widehat y_{\tidx+k}^{\textrm{CLEP}}  = 
\sum_{m=1}^M w_{\tidx+1}^m\widehat{ y}_{\tidx+k}^{\textrm{m}}.
\end{align}
Here the weight, $w_{t+1}^m$---used for combining the predictions made on the morning of day $t+1$---for predictor $m$, is computed according to the recent performance as follows:
\begin{align}
\label{eq:final_wt}
    w_{t+1}^{m} \propto \exp\left(-0.5\sum_{i = t-6}^{t} (0.5)^{t-i} \left|\sqrt{\widehat{y}_i^{m}} - \sqrt{ y_i}\right|\right),
\end{align}
where $\widehat{y}_i^{m}$ is the $3$-day-ahead prediction from the predictor $m$ trained on data up to time $i-3$ (and computed on the morning of day $i-2$).
In addition, the weights are normalized so that $\sum_{m = 1}^M w_{t+1}^m = 1$ for each $t+1$. The weights $\{w_{t+1}^m, m=1, \ldots, M\}$ are computed separately for each county. We now turn to our discussion on the general combination scheme that leads to the equation~\ref{eq:final_wt} with a certain choice of hyperparameters (and how those hyperparameters were chosen).

The weights in equation~\ref{eq:final_wt} are based on the general ensemble weighting format introduced in \cite{schuller_2002}. This general format is given by
\begin{align}
\label{eq:weights}
    w_{\tidx+1}^{m} \propto \exp\left(-c(1 - \mu) \sum_{i = t_0}^{\tidx} \mu^{t-i} \ell(\widehat{y}_i^{m} ,y_i)\right), 
\end{align}
where $\mu \in (0, 1)$ and $c>0$ are tuning parameters, $t_0$ represents some past time point, and
the weights are computed on the morning of day $t+1$. Since $\mu < 1$, the $\mu^{t - i}$ term represents the greater influence given to more recent predictive performance. For a given day $i$ and predictor $m$, we measure the predictive performance of the  predictor via the term $\ell(\widehat{y}_i^m,y_i)$, which denotes the loss incurred due to the discrepancy between its predicted number of deaths $\widehat{y}_i^{m}$ and the recorded death counts $y_i$. The hyperparameter $c$ controls the relative importance of predictors depending on their recent predictive performance. Given the same recent predictive performance and $\mu$, a larger $c$ gives a higher weight to the better predictors. The hyperparameter $t_0$ denotes the number of recent days used for evaluating the predictor performance to influence the weight~\eqref{eq:weights}.\\

\paragraph{\emph{Choice of hyper-parameters:}}
Equation~\ref{eq:final_wt} corresponds to equation~\ref{eq:weights} with appropriate hyper-parameters, $c$, $\mu$, $t_0$, and a specific loss format, $\ell$.
In \cite{schuller_2002}, the authors used the loss function $\ell(\widehat{y}_i^m,y_i) = |\widehat{y}_i^m-y_i|$, since their errors roughly had a Laplacian distribution. In our case, we found that this loss function led to vanishing weights due to our error distribution's heavy-tailed nature. To help address this, we apply a square root to the predictions and the true values, and define $\ell(\widehat{y}_i^m, y_i) = |\sqrt{\widehat{y}_i^m} - \sqrt{ y_i}|$. We found that this transformation improved performance in practice. We also considered a logarithmic transform instead of a square root (i.e.,
$\ell(\widehat{y}_i^m, y_i) = |\log({1+\widehat{y}_i^m}) - \log({1+ y_i})|$), but we found that using the logarithm yielded worse performance than using the square root transformation.\footnote{In our first submission on May 16, 2020 to arXiv, we had presented results for March 22 to May 10, 2020. During the preparation of manuscript, we had updated the transform to be the square-root transform in our code, but we did not update the CLEP equation in the paper, and erroneously reported that our CLEP weights used a logarithmic transform.}

To generate our predictions, we use the default value of $c$ in  \cite{schuller_2002} which is $1$. However, we change the value of $\mu$ from the default of $0.9$ to $0.5$ for two reasons: (i) we found $\mu = 0.5$ yielded better empirical performance, and (ii) it ensured that performance more than a week ago had little influence over the predictor. We chose $t_0 = t - 6$ (i.e., we aggregate the predictions of the past week into the weight term), since we found that performance did not improve by extending further back than 7 days. Moreover, the information from more than a week effectively has a vanishing effect due to our choice of $\mu$.

Finally, we found that for computing the weights in \eqref{eq:final_wt}, using $3$-day-ahead predictions in the loss terms $\ell(\widehat{y}_i^{m} ,y_i)$ led to best predictive performance; i.e., these weights are computed based on the 3-day-ahead predictions generated over the course of a week starting with the predictor built 11 days ago (for predicting counts 8 days ago) up to the predictor built 4 days ago (for predicting yesterday's counts). 
In principle, the five hyper-parameters---$c, \mu, t_0, \ell$, and the choice of the prediction horizon to use for evaluating the loss $\ell$---can be tuned jointly via a grid or randomized search. Nevertheless, to keep the computations tractable and our choices interpretable, we selected them sequentially. Moreover, a dynamic tuning of these hyper-parameters (over time) is left for future work (see last paragraph of Section~\ref{sec:comparison}).

\subsection{Ensuring monotonicity of predictions}
\label{sub:monotonicity}
In this work, we predict county-wise cumulative death count, which is a non-decreasing sequence. 
However, the predictors discussed in the previous sections need not provide monotonous estimates for different prediction horizon, i.e., $\ehat[\textrm{deaths}^c_{t+k}\vert t]$ may decrease as $k$ increases for a fixed $t$. 
Moreover, the predictors may estimate a future count that is smaller than the last observed cumulative death count, i.e., $\ehat[\textrm{deaths}^c_{t+k}\vert t] < \textrm{deaths}^c_{t}$.
In our setting, expanded shared predictor exhibited both these issues.
To avoid these pitfalls, we use post-hoc maxima adjustments for all the predictors as follows.
First, we replace the estimate $\ehat[\textrm{deaths}^c_{t+1}\vert t]$ by $\max\{\ehat[\textrm{deaths}^c_{t+1}\vert t], \textrm{deaths}_t^c\}$ to make sure that the predicted counts in the future are at least as large as the latest observed cumulative death counts.
Next, we iteratively replace the estimate $\ehat[\textrm{deaths}^c_{t+j}\vert t]$ by $\max\{\ehat[\textrm{deaths}^c_{t+j}\vert t], \ehat[\textrm{deaths}^c_{t+j-1}\vert t]\}$
for $j=2, 3, \ldots 21$. Imposing these constraints for the individual predictors also ensures the monotonicity of predictions by the CLEP. Note that we use these monotonous predictions (after the maxima calculations) to determine the weights in equation~\ref{eq:final_wt}.\footnote{We report partial results up to 21-day-ahead predictions, and detailed results up to 14-day-ahead predictions in Section~\ref{sec:results}. 
In the first arXiv submission of this work on May 16, 2020, we had not implemented monotonicity of predictions. The monotonicity implementation improved the overall results both for predictions and prediction intervals.
}

Note that even after imposing the previous monotonicity corrections, it is still possible that $\ehat[\textrm{deaths}^c_{t+k}\vert t] > \ehat[\textrm{deaths}^c_{t+k+1}\vert t+1]$ since the predictors are re-fitted over time. Hence, when plotted over time, $k$-day-ahead predictions, need not be monotonous with respect to $t$. For example, see the plots of 7-day-ahead predictions in Figure~\ref{fig:line} and 14-day-ahead predictions in Figure~\ref{fig:line_14_days}.

\section{Prediction intervals via conformal inference}
\label{sec:prediction_intervals}
Accurate assessment of the uncertainty of forecasts is necessary to help determine how much emphasis to put on them, for instance, when making policy decisions. As such, the next goal of the paper is to quantify the uncertainty of our predictions by creating prediction intervals. 
A common method to do so involves constructing (probabilistic) model-based confidence intervals, which rely heavily on the probabilistic assumptions made about the data.
However, due to the highly dynamic nature of \mbox{COVID-19}, assumptions on the distribution of death and case rate are challenging to check.
Moreover, such prediction intervals based on probability models are likely to be invalid when the underlying probability model does not hold to the desired extent.
For instance, a recent study~\cite{marchant2020learning} reported that the 95\% uncertainty credible intervals for state-level daily mortality predicted by the initial IHME model~\cite{uncertainty_IHME}, had a coverage of a mere 27\% to 51\% of recorded death counts over March 29 to April 2. The authors of the IHME model noted this behavior, and have since updated their uncertainty intervals so that they now provide more than $95\%$ coverage (where coverage is defined below in equation~\ref{eq:coverage}). However, while the previous releases of the intervals were based on asymptotic confidence intervals, the IHME authors have not precisely described the methodology for their more recent intervals.
In this section, we construct prediction intervals that attempt to avoid these pitfalls by taking into account the recent observed performance of our predictors; and later in Section~\ref{sub:mepi_results}, we show that these intervals obtain high empirical coverage while maintaining reasonable width.

\subsection{Maximum-absolute-Error Prediction Interval (MEPI)}
\label{sub:mepi}

We now introduce a generic method to construct prediction intervals for sequential or time-series data. 
In particular, we build on the ideas from conformal inference~\cite{vovk2005algorithmic} and make use of the past errors made by a predictor to estimate the uncertainty for its future predictions. 

To construct prediction intervals for county-level cumulative death counts caused by COVID-19, we calculate the largest (normalized absolute) error for the death count predictions generated over the past 5 days for the county of interest and use this value (the ``maximum absolute error'') to create an interval surrounding the future (e.g., tomorrow's) prediction. We call this interval the Maximum absolute Error Prediction Interval (MEPI).

Let $y_\tidx$ be the actual recorded cumulative deaths by the end of day $\tidx$, and $\ytk[\tidx]$ denote the estimate for $y_\tidx$ made $k$ days earlier (in our case on the morning of day $\tidx-k+1$) by a prediction algorithm. 
We call $\ytk[\tidx]$ the $k$-day-ahead prediction for day~$\tidx$. (Note that we suppress the dependence on county and prediction horizon $k$ for brevity and ease of exposition; and the notation here is slightly abused version of that used in Section~\ref{sec:combined-predictor}. In particular, we have $\ytk[\tidx] = \ehat[\textrm{deaths}_{t}^c\vert t-k]$ for some fixed county $c$.)
We define the normalized absolute error, $\err_{\tidx}$, of the prediction, $\ytk[\tidx]$, to be 
\begin{align}
\label{eq:norm_err}
    \err_{\tidx} :=  \Big|\frac{y_{\tidx}}{ \max\{\ytk[\tidx],1\}} - 1 \Big|.  
\end{align}
\noindent We use the normalization so that $y_\tidx$ (when non-zero) is equal to either $\widehat{y}_\tidx (1-\err_\tidx)$ or $ \widehat{y}_\tidx (1+\err_\tidx)$.
This normalization addresses the fact that the counts are increasing over time, and thus the un-normalized errors, $|y_t - \widehat{y}_t|$,  also tend to be increasing over time. The normalization ensures that the errors across time are comparable in magnitude, which is essential for the exchangeability of the errors (see Section \ref{sec:exchangeability}).

To compute the $k$-day-ahead prediction interval for day $\tidx+k$ on the morning of day $t+1$, we first compute the $k$-day-ahead prediction $\widehat{y}_{\tidx+k}$ ($=\ehat[\textrm{deaths}_{t+k}^c\vert t]$) using a CLEP. Next, we compute the normalized errors for the $k$-day-ahead predictions for the most recent 5 days $\err_{\tidx}, \err_{\tidx-1}, ..., \err_{\tidx-4}$ (5 days was chosen to balance the trade-off between coverage and length, see Appendix~\ref{sub:conformal-inference} for more details). The largest of these normalized errors is then used to define the \emph{maximum absolute error prediction intervals} (MEPI) for the $k$-day-ahead prediction as follows:
\begin{subequations}
\begin{align}
\label{eq:mepi_original}
    \widehat{\textrm{PI}}_{\tidx+k} &:= \left [\max\big\{\widehat y_{\tidx+k} (1-\err_{\text{max}}), y_\tidx\big\}, \ \widehat y_{\tidx+k} (1+\err_{\text{max}}) \right],\\ \quad\text{where}\quad 
    \err_{\text{max}}&:=\max_{0\leq j\leq 4} \err_{\tidx-j}; \label{eq:del_max}
\end{align}
\end{subequations}
where the lower bound for the interval includes a maxima calculation to account for the fact that $y_\tidx$ is a cumulative count, and thereby non-decreasing. This maxima calculation ensures that the lower bound for the interval is not smaller than the last observed value. 

For a general setting beyond increasing time-series, this maxima calculation can be dropped, and the MEPIs can be defined simply as
\begin{align}
\label{eq:mepi_general}
 [\widehat y_{\tidx+k}(1-\err_{\text{max}}),\ \widehat y_{\tidx+k}(1+\err_{\text{max}})].
\end{align}

In our case, we construct the MEPIs~\eqref{eq:mepi_original} separately for each county for the cumulative death counts.
We remind the reader that when constructing $k$-day-ahead MEPIs, the $\err_t$ defined
in equation~\ref{eq:norm_err} is computed using $k$-day-ahead predictions (our notation does not highlight this fact), so that the maximum error $\err_{\text{max}}$ would be typically different, say, for $7$-day-ahead and $14$-day-ahead predictions.

\subsection{Evaluation metrics}
\label{sub:evaluation_metric}
For any time-series setting, stationary or otherwise, the quality of a prediction interval can be assessed in terms of the percentage of time---over a not too short period---that the prediction interval covers the observed value of the target of interest (e.g., recorded cumulative death counts as in this paper). A good prediction interval should both contain the true value most of the time, i.e., have a good coverage, and have a reasonable width or length.\footnote{We use the terms width and length for an interval interchangeably in this paper.}
Indeed, one can trivially create very wide prediction intervals that would always contain the target of interest. 
We thus consider two metrics to measure the performance of prediction intervals: \emph{coverage} and \emph{normalized length}. 

Let $y_\tidx$ denote a positive real-valued time-series of interest, which in this case is the target variable: COVID-19 deaths ($\tidx$ denotes the time index). Let $\{\pint_{\tidx} = [a_\tidx, b_\tidx]\}$ denote the sequence of prediction intervals produced by an algorithm. The coverage of this prediction interval, $\textrm{Coverage}(\period)$, over a specified period, \period, corresponds to the fraction of days in this period for which the prediction interval contained the observed cumulative death counts. This notion of \emph{coverage} for streaming data has been used extensively in prior works on conformal inference~\cite{vovk2005algorithmic} and can be calculated for a given evaluation period $\period$ (which we set to be from April 11 to June 20) as follows:
\begin{subequations}
\begin{align}
\label{eq:coverage}
    \textrm{Coverage}(\period) &= \frac{1}{\vert \period\vert }\sum_{\tidx \in \period} \mathbb{I}(y_{\tidx}\in\pint_{\tidx}),
\end{align}
where $ \mathbb{I}(y_{\tidx}\in\pint_{\tidx})$ takes value $1$ if $y_t$ belongs to the interval $\pint_{t}$ and $0$ otherwise. The average \emph{normalized length} of the prediction intervals, $\textrm{NL}(\period)$, is calculated as follows:
\begin{align}
    \textrm{NL}(\period) &= \frac{1}{\vert \period\vert }\sum_{\tidx \in \period} \frac{b_\tidx-a_\tidx}{y_\tidx}.\label{eq:normalized_length}
\end{align}
\end{subequations}
In practice, we replace the denominator on the RHS of equation~\ref{eq:normalized_length} with $\max\{1, y_\tidx\}$ to avoid possible division by $0$. We use normalized length to address the fact that the death counts across different counties can differ by orders of magnitude.

Importantly, the definitions of coverage (equation~\ref{eq:coverage}) and the average length (equation~\ref{eq:normalized_length}) are entirely data-driven and do not rely on any probabilistic or generative modeling assumptions.

\subsection{Exchangeability of the normalized prediction errors} 
\label{sec:exchangeability}

While the ideas from MEPI are a special case of conformal prediction intervals~\cite{vovk2005algorithmic,shafer2008tutorial}, there are some key differences. While conformal inference uses the raw errors in predictions, MEPI uses the normalized errors, and while conformal inference uses a percentile (e.g., the 95th percentile) of the errors, MEPI uses the maximum. Furthermore, we only make use of the previous five days instead of the full sequence of errors. The reason behind these alternate choices is because the validity of prediction intervals constructed in this manner relies crucially on the assumption that the sequence of errors is exchangeable. Our choices are designed to make this assumption more reasonable. Due to the dynamic nature of \mbox{COVID-19}, considering a longer period (e.g., substantially longer than five days) would mean that it is less likely that the errors across the different days are exchangeable. Meanwhile, the normalization of the errors eliminates a potential source of non-exchangeability by removing the sequential growth of the errors resulting from the increasing nature of the counts themselves. Since we only use five time points to construct the interval, the 95th percentile can be approximated by the maximum.

We now provide some empirical evidence that the exchangeability of the past 5 \textit{normalized} errors for CLEP is indeed reasonable for both $7$-day-ahead and 14-day-ahead predictions, Figures~\ref{fig:average_rank_normalized_error} and \ref{fig:average_rank_normalized_error_14_days} in the Appendix, respectively.
For $k$-day-ahead prediction, we rank the errors $\{\err_{\tidx+k}, \err_{\tidx}, \err_{\tidx-1}, \ldots, \err_{\tidx-4}\}$ in increasing order so that the largest error has a rank of 6.
(Interested readers may first refer to Section~\ref{sub:mepi_coverage} for how exchangeability of these 6 errors is useful for establishing theoretical guarantees for MEPI.)
For a given $j \in \{0, ..., 4\}$, $\err_{t-j}$ denotes the error in $k$-day-ahead prediction for day $t-j$, where the prediction was made on the morning of day $t-j-k+1$, but the error can be computed only by the end of day~$t-j$ (or the morning of day $t-j+1$). 
If the errors were exchangeable, then for each of them, the rank has a uniform distribution on $\lbrace1,2,3,4,5,6\rbrace$, and in particular has a mean of 3.5. 
To approximate this numerically for 7-day-ahead predictions, we measure the rank of the errors $\Delta_{t+7}$, and $\Delta_{t-j}, j=0, \ldots, 4$, for each day~$t$ between March 26 to June 13, and take an average. 
Figure~\ref{fig:average_rank_normalized_error} plots the results for each of the $6$ worst hit counties as well as $6$ randomly-selected counties for these errors (see Section~\ref{sub:county_visual} for further discussion on these counties.). 
Corresponding results for 14-day-ahead predictions, where we rank the errors $\err_{t+14}$ 
and $\Delta_{t-j}, j=0, \ldots, 4$ is presented in Figure~\ref{fig:average_rank_normalized_error_14_days} in the Appendix. In both figures, we see that across most counties, the average rank for almost all the errors is around 3.5 as would be expected if the errors were exchangeable. Thus, the observations from Figures~\ref{fig:average_rank_normalized_error} and \ref{fig:average_rank_normalized_error_14_days} provide a heuristic justification for the construction of MEPI, albeit not a formal proof since average rank being close to 3.5 is not sufficient to claim exchangeability of the six errors.
Moreover, we refer the interested reader to Appendix~\ref{sub:conformal-inference} for further discussion on MEPIs, where we provide more evidence on why we chose only past 5 errors and normalization to define MEPI (see Figure~\ref{fig:error_metric_curve}).
Next, we derive a theoretical result for coverage with MEPI under the exchangeability of errors. 

\subsection{Theoretical guarantees for MEPI coverage}
\label{sub:mepi_coverage}
In order to obtain a rough baseline coverage for the MEPIs, we now reproduce some of the theoretical computations from the conformal literature. 
For a given county and a fixed time $t$, and a parameter $k$, if the six errors in the set  $\{\err_{\tidx+k}, \err_{\tidx}, \err_{\tidx-1}, \err_{\tidx-2}$, $\err_{\tidx-3}, \err_{\tidx-4}\}$ are exchangeable, then we have
\begin{align}
\label{eq:probability}
&\PP \left( y_{\tidx+k} \in \widehat{\textrm{PI}}_{\tidx+k} \right) 
=  \PP \left(\err_{\tidx+k} < \err_{\text{max}} \right) 
=  1-\PP \left(\err_{\tidx+k} = \err_{\text{max}} \right) 
= \frac56 \approx 0.83.
\end{align}
Recall the definition (equation~\ref{eq:coverage}) for Coverage($\period$)
for a given period of days $\period$.
Given equation~\ref{eq:probability}, we may believe that the Coverage$(\period)\approx 83\%$ holds for large $\vert\period\vert$, where the coverage was defined in equation~\ref{eq:coverage}. However, we now elaborate that a few challenges remain to take claim~\eqref{eq:probability} as a proof for the stronger claim that MEPI achieves 83\% coverage as defined by equation~\ref{eq:coverage}.

On the one hand, the probability in equation~\ref{eq:probability} is taken over the randomness in the errors, and the time-index $\tidx+k$ remains fixed. This observation, in conjunction with the law of large numbers, implies the following: Over multiple independent runs of the time-series, for a given county and a given time $\tidx+k$, the fraction of runs for which the MEPI $\widehat{\textrm{PI}}_{\tidx+k}$ contains the observed value $y_{\tidx+k}$ converges to $5/6$ as the number of runs goes to infinity. 
However, analyzing such a fraction over several different independent runs of the COVID-19 outbreak is not relevant for our work.

On the other hand, the evaluation metric we consider is the average coverage of the MEPI over a single run of the time-series, c.f., the definition (equation~\ref{eq:coverage}) for Coverage($\period$).
Thus, we require an online version of the law of large numbers in order to guarantee that Coverage$(\period) \to 83\%$ as $\vert \period\vert \to \infty$.
Such a law of large numbers, established in prior works~\cite{shafer2008tutorial}, has been crucial for establishing 
theoretical guarantees in conformal inference. 
In our case, this law---stated as Proposition 1 in Section 3.4 in their paper~\cite{shafer2008tutorial}---guarantees that, when the entire sequence of errors $\{\err_\tidx, \tidx \in \period \}$ for a given county is exchangeable, the corresponding Coverage$(\period)\approx 83\%$,
when the period $\period$ is large.
Unfortunately, such an assumption is both hard to check and unlikely to hold for the prediction errors obtained from 
CLEP for the COVID-19 cumulative death counts.

Despite the challenges listed above, later, we show in Section~\ref{sub:mepi_results} that  MEPIs with CLEP achieved good coverage with narrow widths for COVID-19 cumulative death count predictions.

\begin{figure}[ht]
    \centering
    \begin{tabular}{c}
        \includegraphics[width=0.85\textwidth]{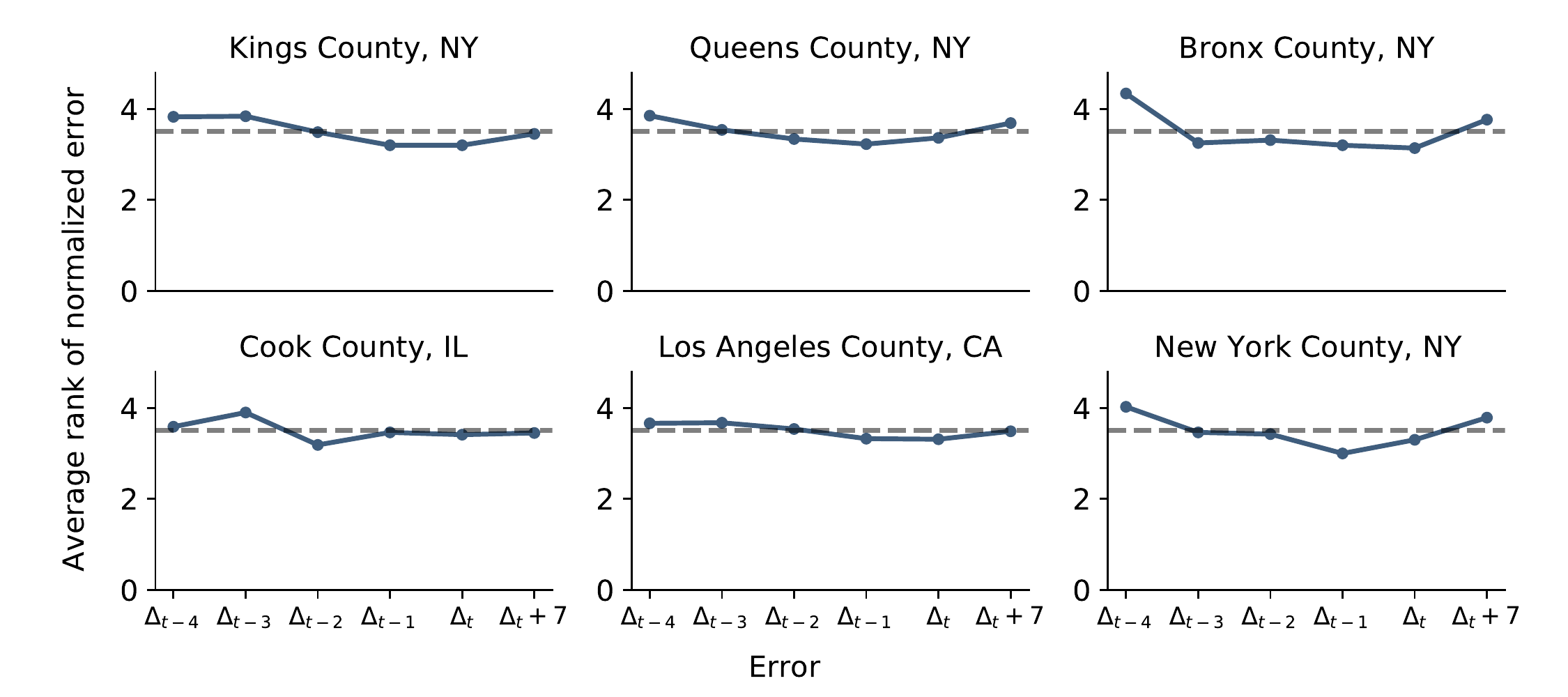} \vspace{-2mm} \\ \vspace{1mm}
        (a) Six worst-affected counties\\
        %\midrule \\
        \includegraphics[width=0.85\textwidth]{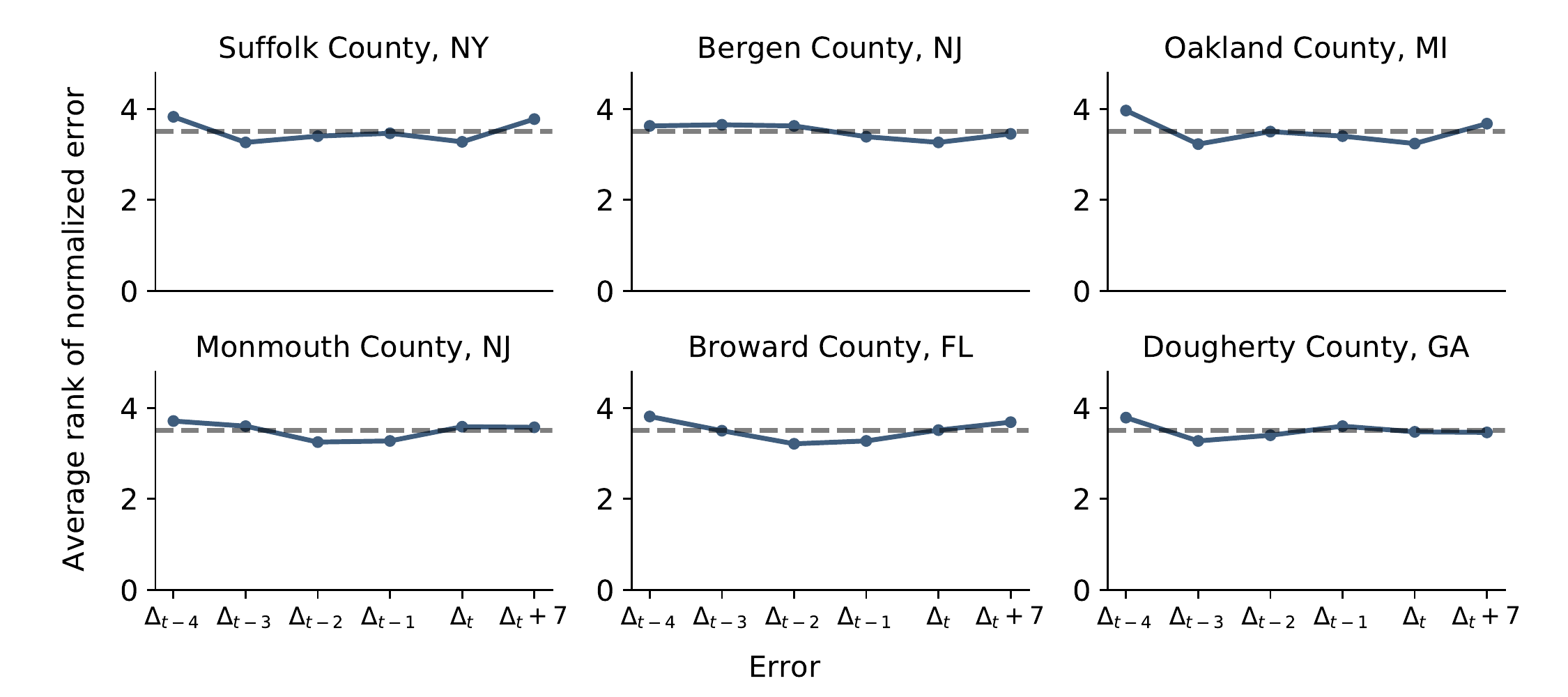} \vspace{-2mm}\\ 
        (b) Six randomly-selected counties
    \end{tabular}
    \caption[Investigating exchangeability of errors for 7-day-ahead CLEP predictions]{EDA plot for investigating exchangeability of normalized errors of \emph{7-day-ahead} CLEP predictions with its last $5$ errors made at time $t$, over the period $t =$ March 26, \ldots, Jun 13. We plot the average rank of the six errors $\{\err_{\tidx+7}, \err_{\tidx}, \err_{\tidx-1}, \ldots, \err_{\tidx-4}\}$ of our CLEP (with the expanded shared and linear predictors) for \textbf{(a)} the six worst affected counties, and \textbf{(b)} six random counties. We rank the errors $\{\err_{\tidx+7}, \err_{\tidx}, \err_{\tidx-1}, \ldots, \err_{\tidx-4}\}$ in increasing order so that the largest error has a rank of 6. If $\{\err_{\tidx+7}, \err_{\tidx}, \err_{\tidx-1}, \ldots, \err_{\tidx-4}\}$ are exchangeable for any day $t$, then the expected average rank for each of the six errors would be 3.5 (dashed black line).}
    \label{fig:average_rank_normalized_error}
\end{figure}

\section{Prediction results for March 22 to June 20}
\label{sec:results}
In this paper, we focus on predictive accuracy for up to 14 days. In this section, we first present and compare the results of our various predictors, and then give further examinations of the best performing predictor: the CLEP ensemble predictor that combines the expanded shared exponential predictor and the linear predictor (the best two among the individual predictors).
Finally, we report the performance of the coverage and length of the MEPIs for this CLEP.
(We note that CLEP that combined all five predictors performed worse since even bad performing predictors get non-zero weight~\eqref{eq:final_wt} in the ensemble and can adversely affect the prediction performance.)
A Python script to reproduce all the results in this section is made available on Github at
\url{https://github.com/Yu-Group/covid19-severity-prediction/tree/master/modeling}.
% \footnote{\url{https://github.com/Yu-Group/covid19-severity-prediction/blob/master/modeling/predict_all_deaths.py}}. don't link to specific file here, may get refactored -- Chandan

\subsection{Empirical performance of the single predictors and CLEP}
\label{sub:results_clep}
Table \ref{table:ensembleresults} summarizes the Mean Absolute Errors (MAEs) of our predictions for cumulative recorded deaths on raw, square-root and logarithm scale. We now explain how these errors are computed.

First, on the morning of day $\tidx+1$, we compute $\mathcal{C}_{\tidx}$---the collection of counties in the US that have at least 10 cumulative recorded deaths by the end of day $t$.  Let $\widehat{y}_{\tidx}^c$ and $y_{\tidx}^c$ respectively denote the predicted and recorded cumulative death count of county $c\in\mathcal{C}_{\tidx}$ by the end of day $t$.
We note that while the set of counties $\mathcal{C}_t$ varies with time, it is computable on the day the error is computed (i.e., $\mathcal C_t$ does not depend on future information). We define the set of counties in this manner, to ensure that only the counties with non-trivial cumulative death counts are included in our evaluation on a given day. Moreover this definition satisfies the condition $\mathcal C_{t} \subseteq \mathcal{C}_{t+1}$, that is, only new counties can be added in the set $\mathcal C_t$ as $t$ progresses (and a county once included is never removed).

Given the set $\mathcal C_t$, the mean absolute percentage error (MAPE), the raw-scale MAE, and the square-root-scale MAE for day $\tidx$ are given by
\begin{subequations}
\label{eq:maes}
\begin{align}
    \textrm{MAPE}_\tidx (\% \textrm{ error}) &= 100 \times \frac{1}{\vert\mathcal{C}_{\tidx}\vert}\sum_{c \in \mathcal{C}_{\tidx}} \frac{\left\vert\widehat{y}_{\tidx}^c - y_{\tidx}^c\right\vert}{ y_{\tidx}^c } ,\quad\text{and}\label{eq:mape} \\
    \textrm{Raw-scale MAE}_\tidx &= \frac{1}{\vert\mathcal{C}_{\tidx}\vert}\sum_{c \in \mathcal{C}_{\tidx}}\left\vert\widehat{y}_{\tidx}^c - y_{\tidx}^c\right\vert, \label{eq:raw_scale_mae}\\
     \textrm{Sqrt-scale MAE}_\tidx &= \frac{1}{\vert\mathcal{C}_{\tidx}\vert}\sum_{c \in \mathcal{C}_{\tidx}}\left\vert\sqrt{\widehat{y}_{\tidx}^c} - \sqrt{y_{\tidx}^c}\right\vert. \label{eq:sqrt_scale_mae}
\end{align}
\end{subequations}
The percentage error (MAPE) captures the relative errors of the predictors without attention to the scale of the counts while the raw-scale MAE would be heavily affected by the counties with large death counts. Both of these MAEs are commonly used to report the prediction performance for regression tasks with machine learning methods. We report the MAE at square-root-scale to be consistent with the square-root transform used in the CLEP weighting scheme~\eqref{eq:final_wt}.
Each row in Table~\ref{table:ensembleresults} corresponds to a single predictor, and we report different statistics of errors
made for $k$-day-ahead predictions over the period  $\tidx\in \{\text{March 22}, \dots, \text{June 20}\}$ for $k \in \{3, 5, 7, 14\}$.\footnote{As the expanded shared predictor is trained on counties with at least 3 deaths, there was not enough data to train 14-day-ahead CLEP that predicts recorded deaths before March 29. Hence for $\tidx\in \{\text{March 22}, \dots, \text{March 28}\}$, we use the 14-day-ahead predictions of the linear predictor to impute the 14-day-ahead predictions of the CLEP.} 
For any given predictor, we compute these errors for each day and report the 10th percentile (p10), 50th percentile (median), and 90th percentile (p90) values over the evaluation periods mentioned above.
From Table \ref{table:ensembleresults}, we find that the CLEP ensemble that combines the expanded shared exponential predictor and the separate county linear predictors has the best overall performance; with median MAPE of 8.18\%, 12.21\%, 15.14\% and 26.45\% for 3-, 5-, 7-, and 14-day-ahead predictions.\footnote{In the first version of this paper submitted on May 16, 2020 on arXiv, we reported results for the period March 22 to May 10, 2020 for 3-, 5-, 7-day ahead predictions. We made an error while computing aggregate statistics for 5-day and 7-day ahead predictions (reported in Table 3 of that version) and reported errors that were smaller than the actual errors made by CLEP. Correcting our aggregation code revealed that the actual performance of median of raw-scale MAE of CLEP for 5-day and 7-day predictions was worse by 16\% and 29\% respectively when compared to the reported errors.}
We note that the (p90) MAEs for the separate (exponential) and demographics shared predictors are too large, especially for larger horizons. 
For separate (exponential) predictors, we can attribute these large errors directly to the fact that exponential fit~\ref{eq:county-predictor} for large horizons is very likely to over-predict. 
On the other hand, the demographics shared predictor has large errors potentially due to over-fitting, and the recursive plug-in to obtain longer horizon estimates in the exponential fit~\ref{eq:demo_shared}. 
In the early stages of this project (and the COVID-19 outbreak in the US), these predictors had provided a reasonable fit for short-term (3- and 5-day-ahead) predictions in late-March to mid-April.

\begin{figure}
    \centering
    \resizebox{\textwidth}{!}{
    \begin{tabular}{ccc}
    \includegraphics[width=0.35\textwidth]{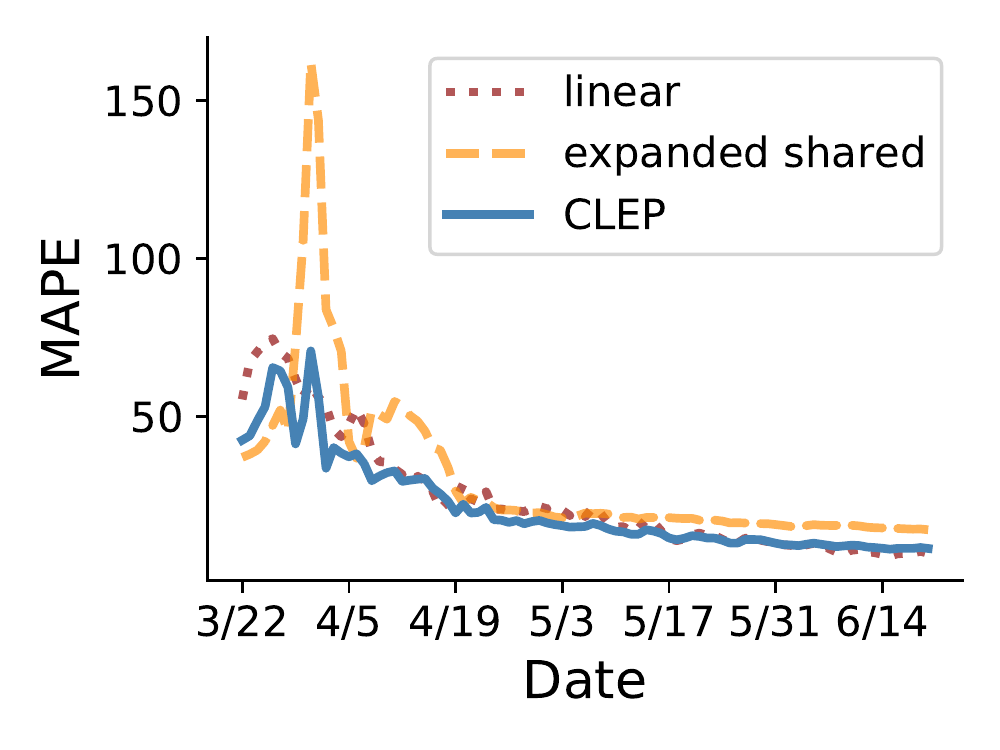} &
    \includegraphics[width=0.35\textwidth]{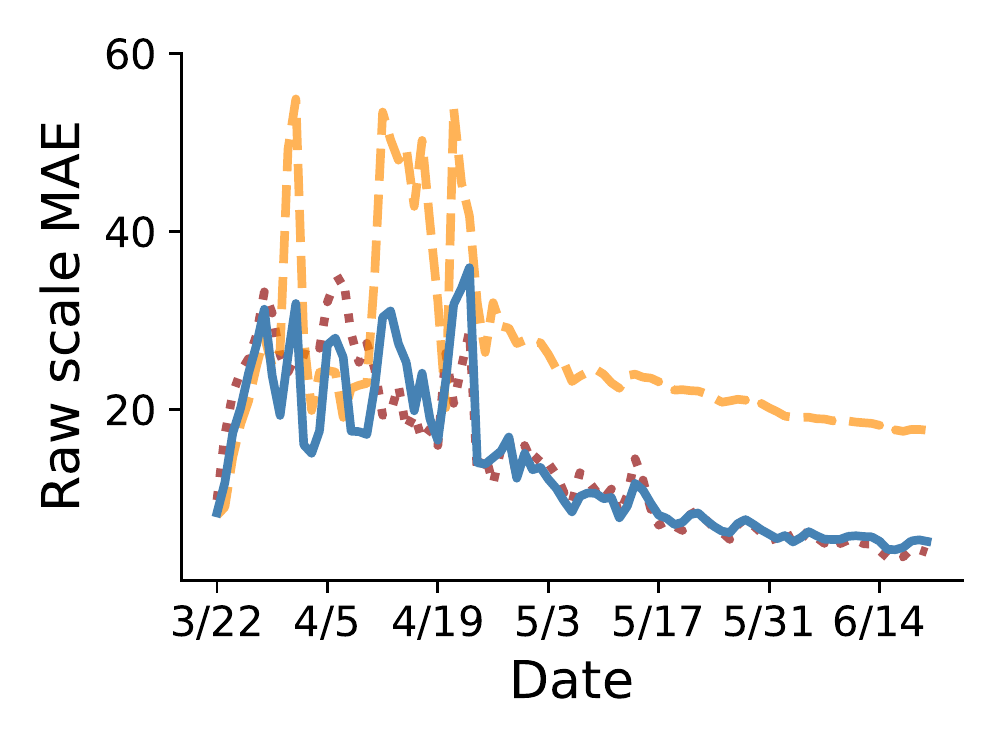} & 
    \includegraphics[width=0.35\textwidth]{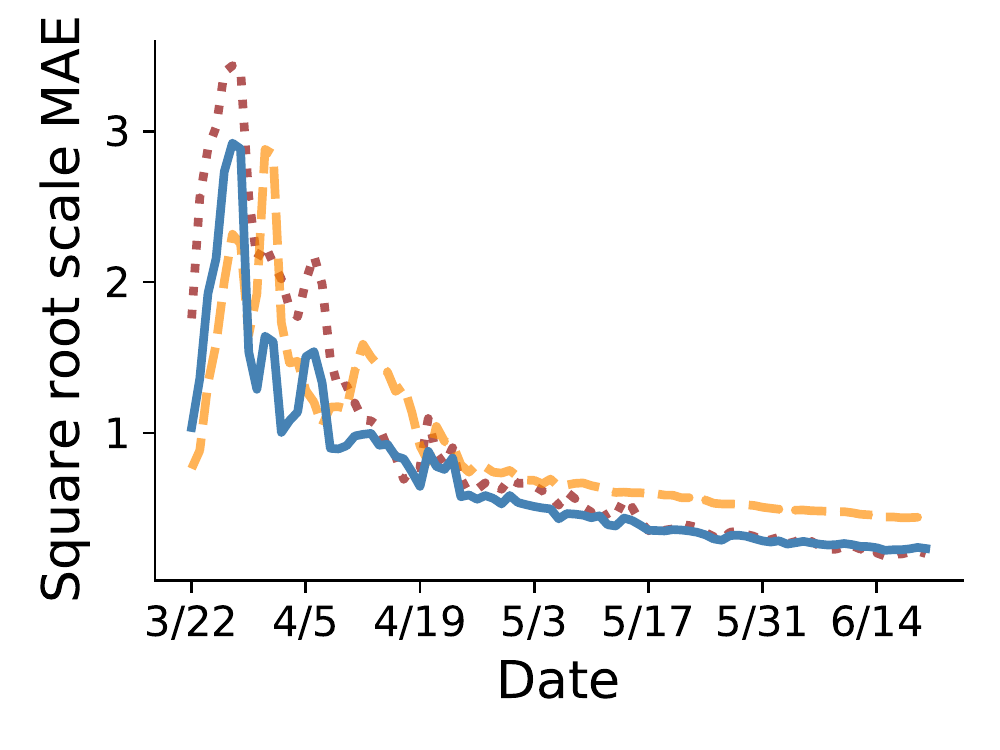} \\
         (a) MAPE &
         (b) Raw-scale MAE &
         (c) Square-root-scale MAE \vspace{2mm}\\       
    \end{tabular}
    }
    \caption[Mean absolute errors of different predictors for 7-day-ahead predictions]{Plots of mean absolute error (MAE) of different predictors for \emph{7-day-ahead} predictions from March 22 to June 20. We plot the \textbf{(a)} mean absolute percentage error (MAPE), \textbf{(b)} raw-scale MAE,  and \textbf{(c)} square-root-scale MAE versus time. Results are shown for expanded shared exponential predictor (orange dashed line), the separate county linear predictor (red dotted line), and the CLEP that combines the two predictors (solid blue line).}
    \label{fig:l1_over_time}
\end{figure}

In Figure~\ref{fig:l1_over_time}, all three errors from the display~\ref{eq:maes} as a function of time over the past 3 months for the expanded shared exponential predictor, the separate county linear predictor, and the CLEP that combines the two. 
We found that the MAE of the CLEP is often similar to, and usually slightly smaller than the smaller MAE of the two single predictors. 

\begin{figure}
    \centering
    \resizebox{\textwidth}{!}{
    \begin{tabular}{ccc}
    \includegraphics[width=0.33\textwidth]{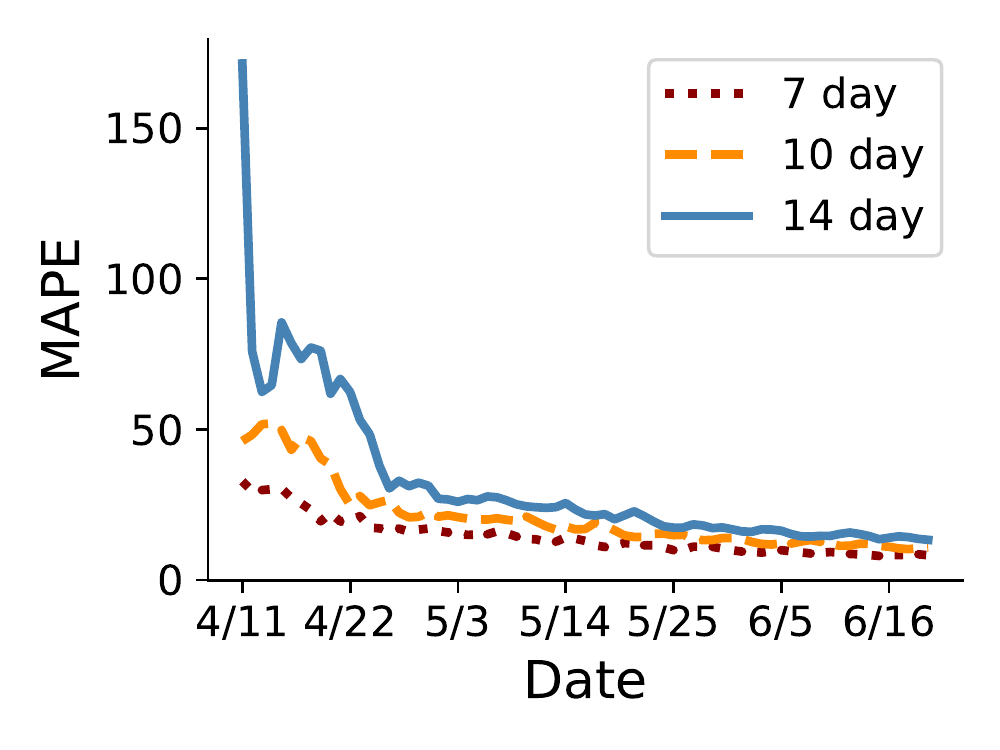} &
    \includegraphics[width=0.33\textwidth]{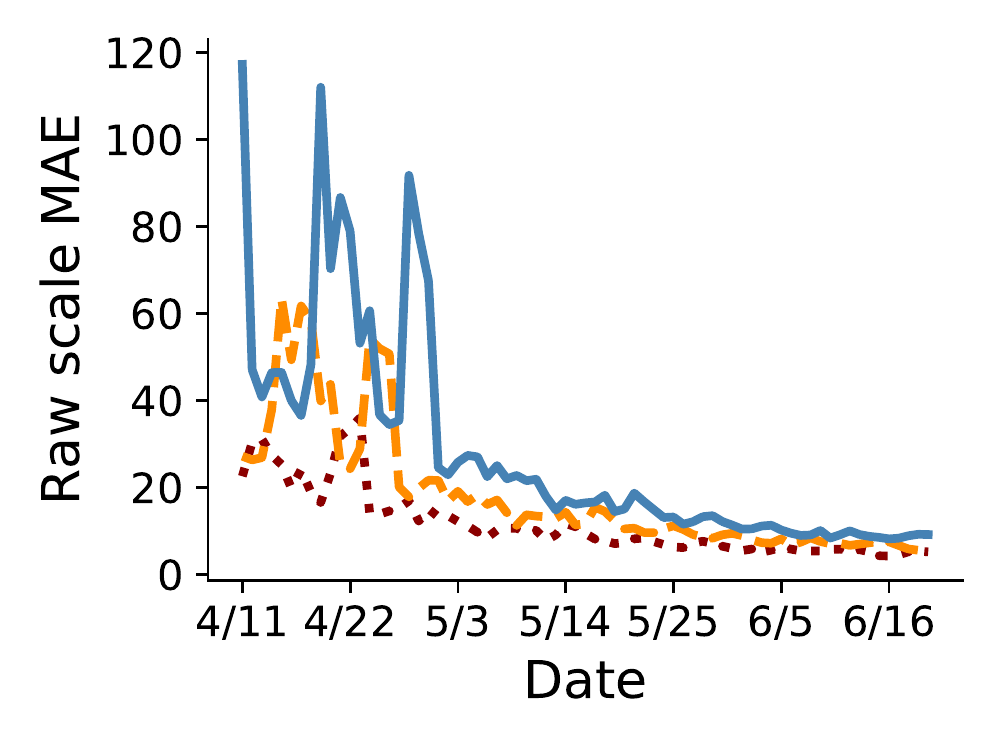}& 
    \includegraphics[width=0.33\textwidth]{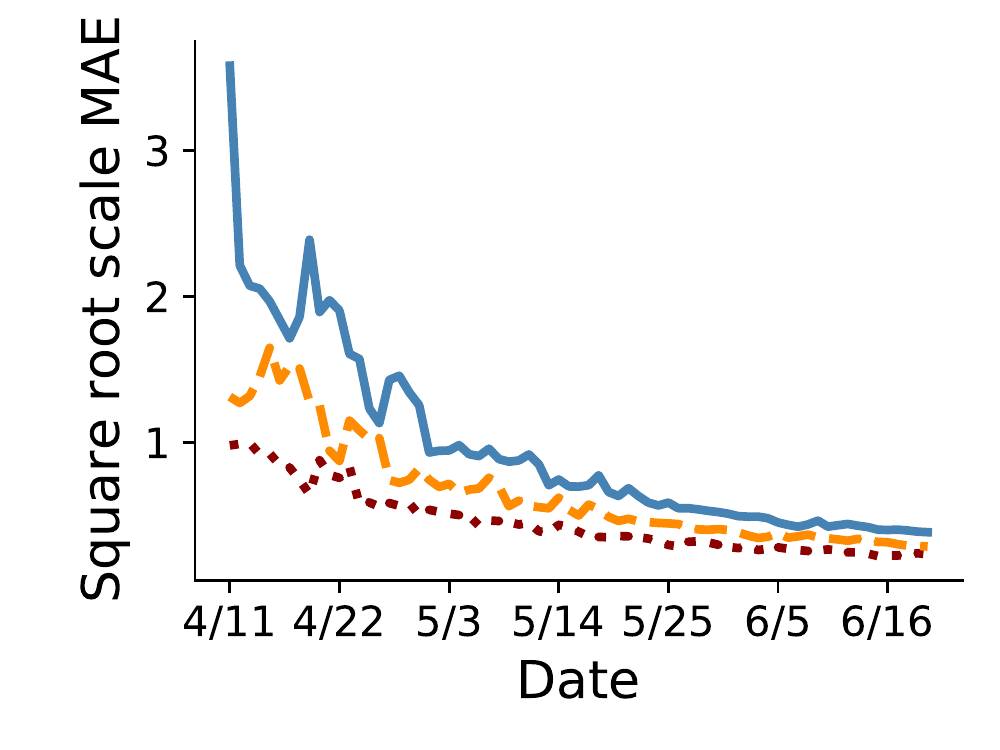} \\
         (a) MAPE &
         (b) Raw-scale MAE &
         (c) Square-root-scale MAE \vspace{2mm}\\   
    \end{tabular}
    }
    \caption[Mean absolute errors of CLEP for 7, 10 and 14-day-ahead predictions]{Plots of mean absolute error (MAE) of CLEP with different prediction horizons from April 11, 2020 to June 20, 2020. We plot the \textbf{(a)} mean absolute percentage error (MAPE), \textbf{(b)} raw-scale MAE,  and \textbf{(c)} square-root-scale MAE versus time, for  $k$-day-ahead predictions with $k\in\{7, 10, 14\}$. }
    \label{fig:longer_horizon_over_time}
\end{figure}

Next, in Figure~\ref{fig:longer_horizon_over_time}, we plot the performance of this CLEP for longer horizons. In particular, we plot the three MAEs, raw-scale, percentage-scale, and square-root-scale for $k$-day-ahead predictions for $k=7$, 10, and 14. 
Notice that the 7-day-ahead CLEP predictor has the lowest MAE and that the MAE increases as the prediction horizon increases. (Recall that precise statistics for 7-day-ahead and 14-day-ahead MAEs are listed in  Table~\ref{table:ensembleresults}.) The increases in MAE in mid-late April was caused due to the state of New York adding thousands of deaths (3,778) that were previously reported as ``probable'' to their counts on a single day, April 14. This change led the CLEP to greatly over-predict deaths in New York in mid-late April---(i) 7 days later on April 21 for the 7-day-ahead CLEP, (ii) 10 days later on April 24 for the 10-day-ahead CLEP, and (iii) 14 days later on April 28 for the 14-day-ahead CLEP. As further evidence that this is indeed the case, when we manually removed this uptick in the death counts in New York, the raw-scale MAE for the 14-day-ahead CLEP on April 28 was 29.5, which is much smaller than the original raw-scale MAE on April 28, which was 91.9 in Figure \ref{fig:longer_horizon_over_time}(b).

\begin{figure}
    \centering
    \resizebox{\textwidth}{!}{
    \begin{tabular}{ccc}
    \includegraphics[width=0.33\textwidth]{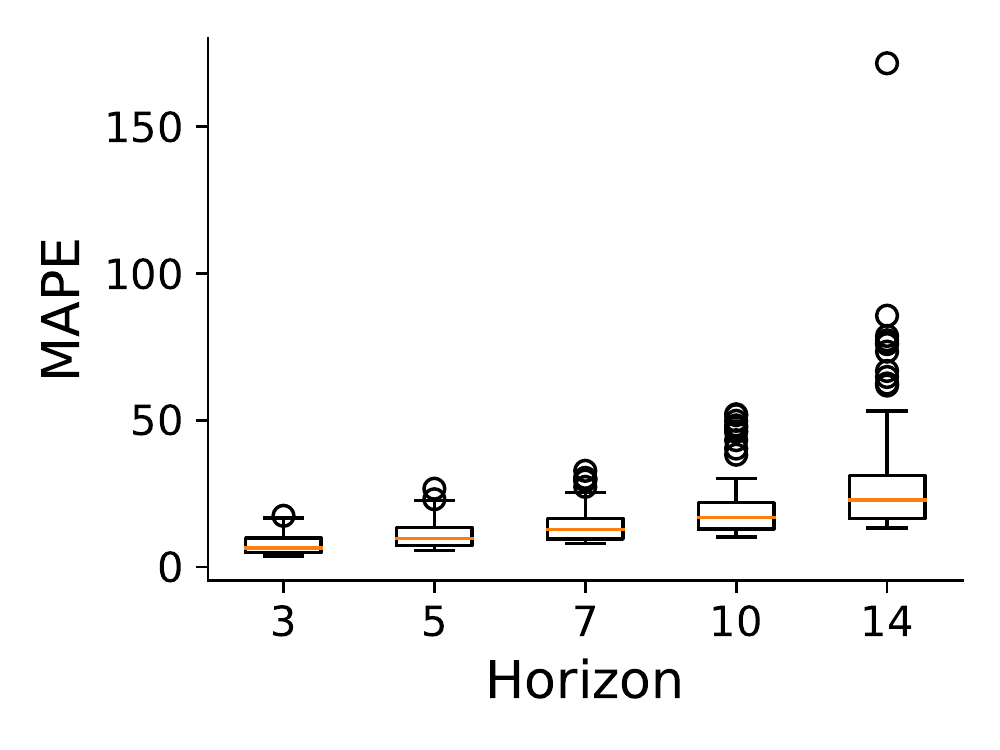} &
    \includegraphics[width=0.33\textwidth]{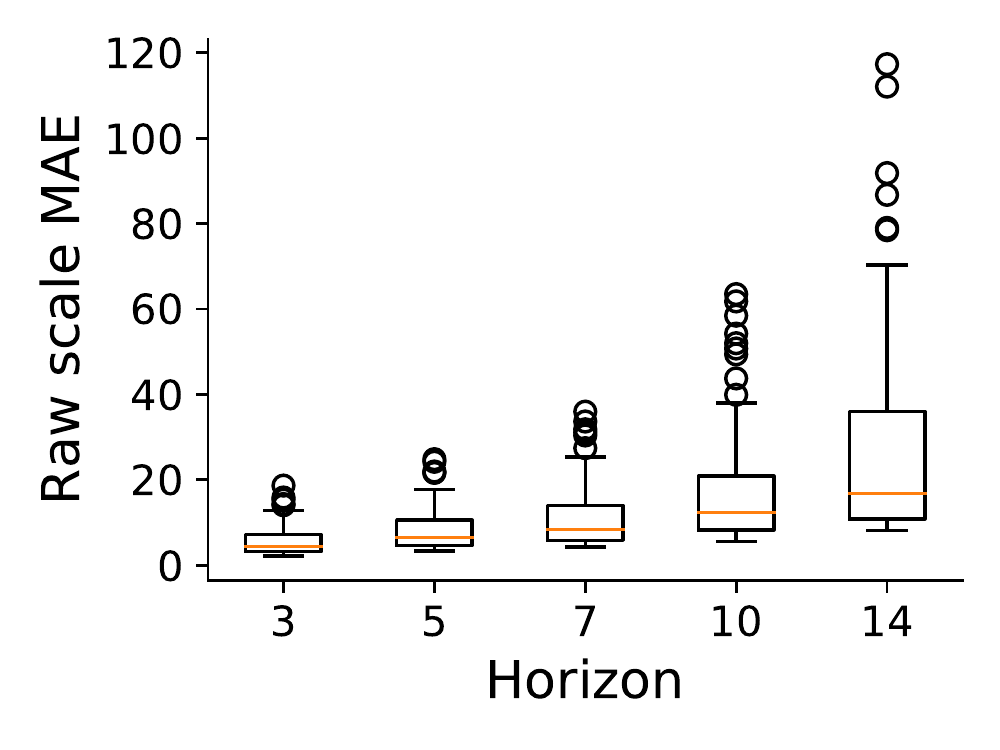} &
    \includegraphics[width=0.33\textwidth]{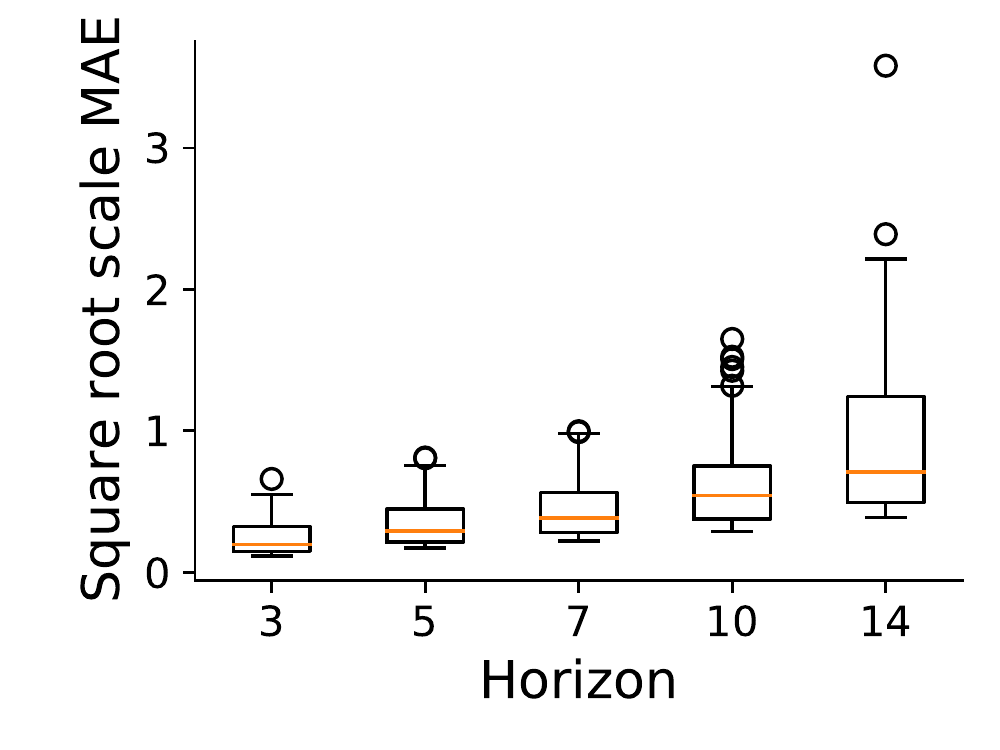} \\  
    (a) & (b) & (c) \vspace{5mm}\\ \midrule
    \includegraphics[width=0.33\textwidth]{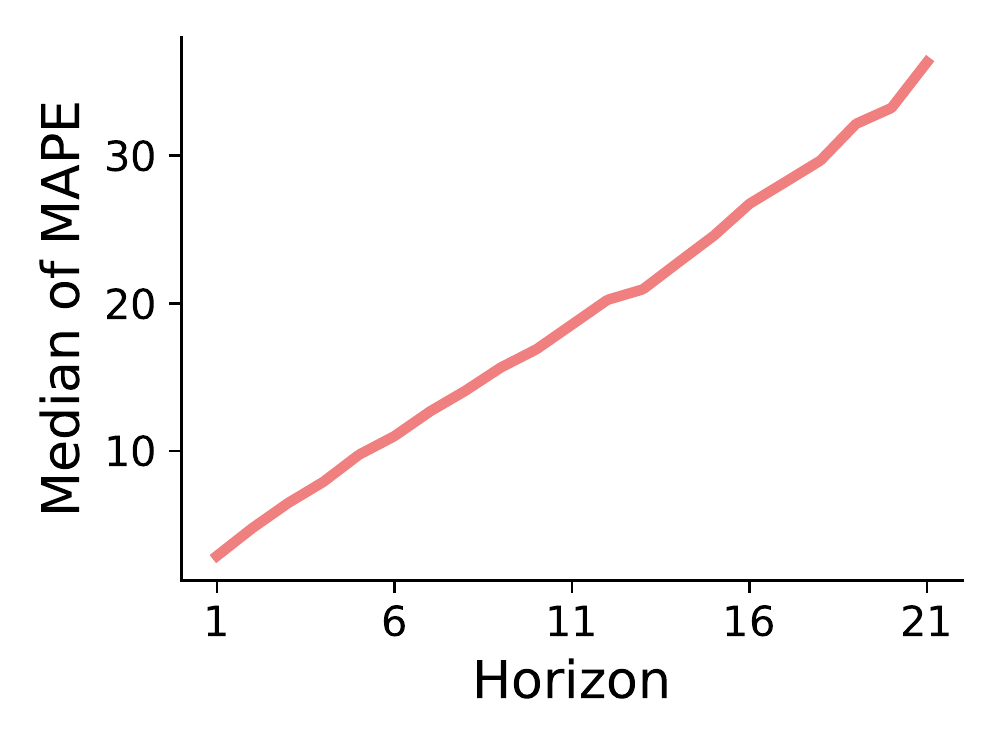} &
    \includegraphics[width=0.33\textwidth]{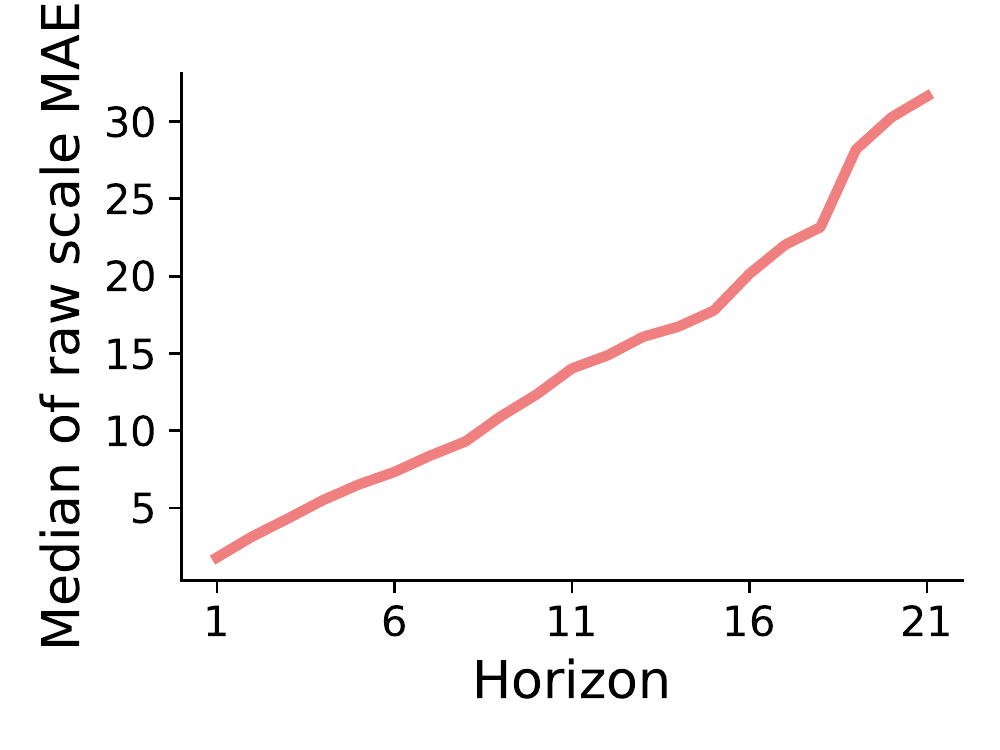} &
    \includegraphics[width=0.33\textwidth]{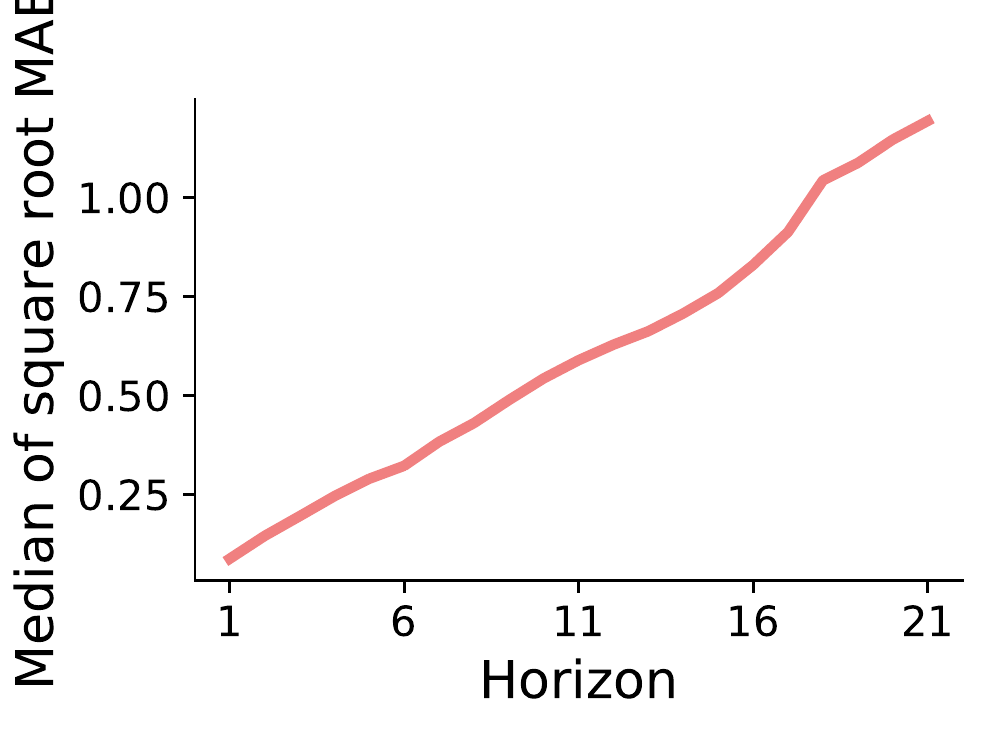} \\  
    (d) & (e) & (f) 
    \end{tabular}
    }
    \caption[Distribution of errors for 7, 10 and 14-day-ahead predictions]{Plots of MAE distribution of CLEP for different prediction horizon over the period April 11, 2020 to June 20, 2020. In panels \textbf{(a)}, \textbf{(b)} and \textbf{(c)}, we show the box-plots (over time) of the mean absolute percentage error (MAPE), raw-scale MAE, and square-root-scale MAE for $k$-day ahead predictions $k=\{3, 5, 7, 10, 14\}$. In panels \textbf{(d)}, \textbf{(e)} and \textbf{(f)}, we plot the median value of the different MAEs for even longer horizons, up to 21 days.}
    \label{fig:mae_21_days}
\end{figure}

We further evaluate the performance of CLEP for longer prediction horizons in Figure~\ref{fig:mae_21_days}, where all predictions were made over the period April 11-June 20. In panels (a-c) of Figure~\ref{fig:mae_21_days}, we show the box plots of the different MAEs for up to 14-day prediction horizon. From these plots, and the panels (d-f), we observe that the MAEs degrade roughly linearly with the horizon for up to 21 days.

%by doing the following modification to counties in the state of New York. For a given county, let $y_0$ denote its cumulative recorded deaths on April 14, and $y_{\text{CLEP}}$ denote the 1-day-ahead CLEP prediction for April 14 for that county. For dates on and after April 14, we subtract $y_0 - y_{\text{CLEP}}$ from the cumulative recorded deaths on that date. As such, the daily new deaths on April 14 was modified, and the daily new deaths after April 14 remains the same as the original data. 

Putting together the results from Table~\ref{table:ensembleresults}, Figures~\ref{fig:l1_over_time}, \ref{fig:longer_horizon_over_time} and \ref{fig:mae_21_days}, we find that
the adaptive combination used for building our ensemble predictor CLEP is able to leverage the advantages of linear and exponential predictors, and---by improving upon the MAE of single predictors---it is able to provide very good predictive
performance for up to 14 days in future.

\begin{table*}
\centering
\footnotesize
\begin{subtable}[t]{\textwidth}
\centering
\resizebox{\textwidth}{!}{
\begin{tabular}{@{}ccccccccccccc@{}} \toprule
&\multicolumn{3}{c}{3-day-ahead } & \multicolumn{3}{c}{5-day-ahead } &  
\multicolumn{3}{c}{7-day-ahead } &
\multicolumn{3}{c}{14-day-ahead }\\
\cline{2-4} \cline{5-7} \cline{8-10} \cline{11-13}
&  p10 & median  & p90 &  p10 & median  & p90  &  p10  & median  & p90 &  p10 & median  & p90\\ \midrule
separate        &      3.80 &     13.16 &     59.63 &      6.26 &     22.56 &    114.07 &      9.95 &     39.56 &    300.53 &      30.37 &     226.26 &    >1000%11321.46 
\\
shared          &      7.05 &     12.55 &     25.99 &     11.68 &     19.77 &     37.73 &     16.59 &     28.65 &     55.01 &      36.55 &      62.45 &      224.75 
\\
demographics    &     17.82 &     25.70 &     30.90 &     30.30 &     41.02 &     50.62 &     47.77 &     62.26 &    117.11 &     260.48 &     551.78 &  >1000 %4366271.82 
\\
expanded shared &      6.86 &      9.59 &     35.55 &     11.17 &     14.54 &     44.28 &     15.09 &     18.52 &     52.13 &      23.13 &      31.18 &     >1000%1235.76 
\\
linear          &    \bf  3.39 &      9.37 &     29.67 &    \bf  5.27 &     14.25 &     40.26 &    \bf  7.18 &     18.60 &     56.10 &      15.58 &      33.16 &      \bf 87.21 \\
CLEP        &      4.34 &     \bf 8.18 &   \bf  22.60 &      6.59 &   \bf  12.21 &    \bf 31.99 &      8.79 &    \bf 15.14 &   \bf  42.47 &    \bf  14.61 &     \bf 26.45 &       93.03 \\
\bottomrule
\end{tabular}
}
\caption{Summary statistics of mean absolute percentage error (MAPE)}
\end{subtable}\vspace{5mm}
\begin{subtable}[t]{\textwidth}
\centering
\resizebox{\textwidth}{!}{
\begin{tabular}{@{}ccccccccccccc@{}} \toprule
&\multicolumn{3}{c}{3-day-ahead} & \multicolumn{3}{c}{5-day-ahead} &  
\multicolumn{3}{c}{7-day-ahead} &
\multicolumn{3}{c}{14-day-ahead}\\
\cline{2-4} \cline{5-7} \cline{8-10} \cline{11-13}
&  p10 & median  & p90 &  p10 & median  & p90  &  p10  & median  & p90 &  p10 & median  & p90\\ \midrule

separate        &      2.35 &      8.10 &     25.13 &    \bf  3.67 &     13.94 &     57.03 &      5.33 &     24.30 &    124.61 &      14.58 &     105.64 &   >1000 %15421.75 
\\
shared          &      7.54 &     12.04 &     19.43 &     13.12 &     19.93 &     36.74 &     18.81 &     28.09 &     72.74 &      33.69 &      69.35 &     325.50 \\
demographics    &     17.47 &     48.35 &     54.54 &     35.41 &    108.47 &    119.71 &     59.29 &    217.64 &    243.56 &     675.96 &    >1000 %3184.97 
&  >1000 %572017.94 
\\
expanded shared &      8.07 &     10.69 &     14.32 &     13.10 &     16.68 &     23.02 &     18.24 &     22.95 &     42.84 &      29.90 &      36.56 &     329.21 \\
linear          &    \bf  2.15 &      \bf 5.93 &     13.81 &      3.67 &      9.49 &     20.02 &     \bf 4.91 &     12.05 &    \bf 26.89 &      10.24 &      25.47 &    \bf  56.73 \\
CLEP        &      2.76 &      5.98 &    \bf 11.93 &      4.09 &    \bf  8.64 &   \bf  18.67 &      5.42 &    \bf 10.64 &     27.29 &      \bf 9.18 &     \bf 22.50 &      81.77 \\
\bottomrule
\end{tabular}
}
\caption{Summary statistics of raw-scale MAE}
\end{subtable}\vspace{5mm}
\begin{subtable}[t]{\textwidth}
\centering
\resizebox{\textwidth}{!}{
\begin{tabular}{@{}ccccccccccccc@{}} \toprule
&\multicolumn{3}{c}{3-day-ahead } & \multicolumn{3}{c}{5-day-ahead } &  
\multicolumn{3}{c}{7-day-ahead } &
\multicolumn{3}{c}{14-day-ahead }\\
\cline{2-4} \cline{5-7} \cline{8-10} \cline{11-13}
&  p10 & median  & p90 &  p10 & median  & p90  &  p10  & median  & p90 &  p10 & median  & p90\\ \midrule
separate        &      0.11 &      0.39 &      1.66 &      0.19 &      0.63 &      2.91 &      0.25 &      1.03 &      3.83 &       0.67 &       3.40 &      18.26 \\
shared          &      0.22 &      0.40 &      0.76 &      0.36 &      0.63 &      1.22 &      0.50 &      0.90 &      1.82 &       1.09 &       2.06 &       5.70 \\
demographics    &      0.81 &      1.11 &      1.38 &      1.27 &      2.04 &      2.48 &      2.02 &      3.25 &      3.90 &       6.61 &      13.45 &      71.60 \\
expanded shared &      0.22 &      0.34 &      0.75 &      0.35 &      0.52 &      1.15 &      0.47 &      0.67 &      1.59 &       0.73 &       1.12 &      10.62 \\
linear          &    \bf  0.11 &      0.29 &      0.99 &    \bf  0.17 &      0.46 &      1.45 &    \bf  0.23 &      0.58 &      2.15 &       0.50 &       1.10 &       4.62 \\
ensemble        &      0.13 &    \bf  0.26 &     \bf 0.66 &      0.19 &    \bf  0.37 &   \bf   0.93 &      0.26 &    \bf  0.47 &    \bf  1.51 &     \bf  0.43 &    \bf   0.92 &    \bf   4.13 \\
\bottomrule
\end{tabular}
}
\caption{Summary statistics of sqrt-scale MAE}
\end{subtable}
\caption[Summary statistics of mean absolute errors for 3, 5, 7, 14-day-ahead predictions for March 22-June 20, 2020]{Summary statistics of Mean Absolute Errors (equation~\ref{eq:maes}) based on \textbf{(A)} the mean absolute percentage error (MAPE), \textbf{(B)}  the raw-scale MAE, and \textbf{(C)} the square-root-scale MAE. The results are presented for the 3, 5, 7, and 14-days-ahead forecasts for each of the predictors considered in this paper, and the CLEP that combines the expanded shared and separate linear predictors. The evaluation period is March 22, 2020 to June 20, 2020 (91 days). ``p10'', ``median'', and ``p90'' denote the 10th-percentile, median, and 90th-percentile of the 91 mean absolute errors computed daily in the evaluation period. The smallest error in each column is displayed in bold.}
\label{table:ensembleresults}
\end{table*}

\subsection{Performance of CLEP and MEPI at the county-level}
\label{sub:county_visual}
\begin{figure}
    \centering
    % \resizebox{\textwidth}{!}{
    \begin{tabular}{c}
         \includegraphics[width=\textwidth]{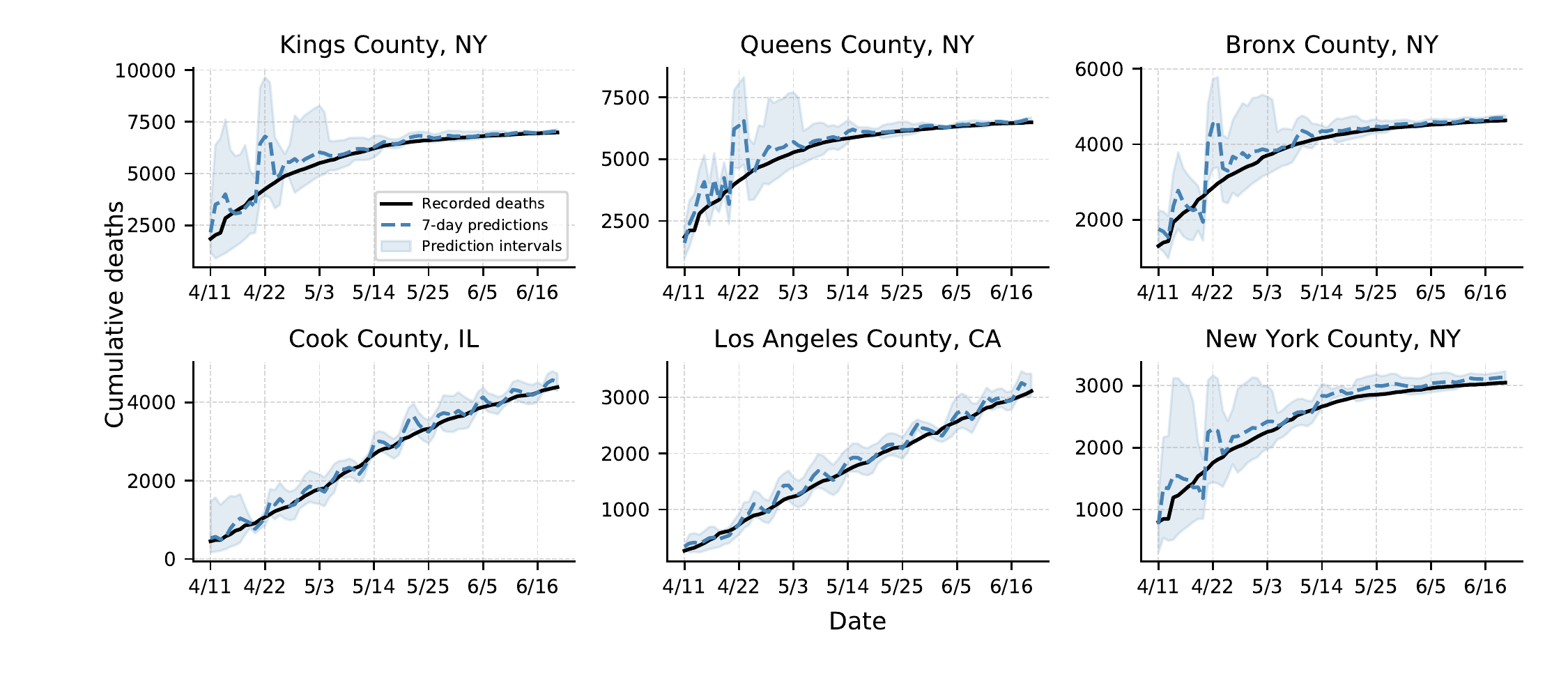} \vspace{-5mm} \\
         (a) Worst-affected counties on June 20\vspace{1mm} \\ \midrule
         \includegraphics[width=\textwidth]{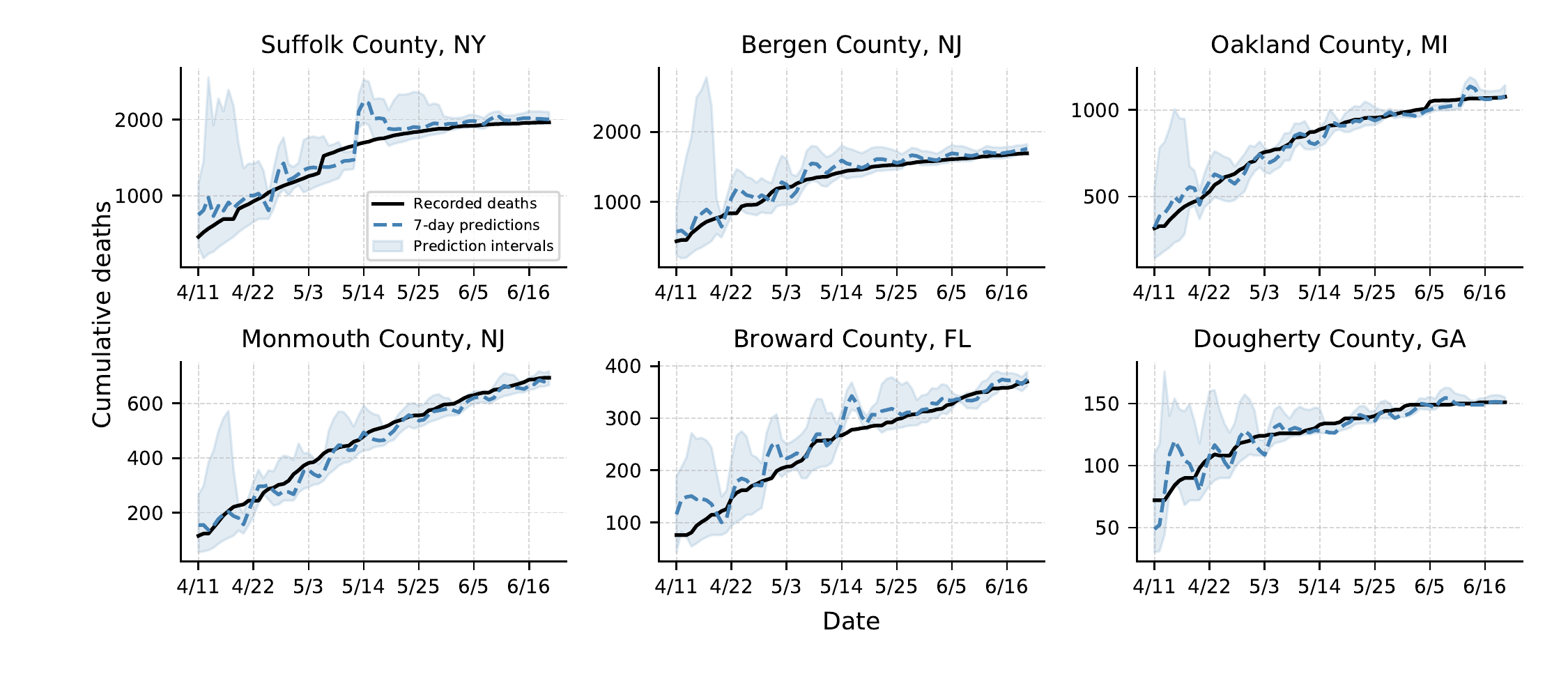} \vspace{-5mm}\\
         (b) Randomly-selected counties
    \end{tabular}
    % }
    \caption[County-wise visualization of 7-day-ahead CLEP and MEPI predictions]{
    A grid of line charts displaying the performance of \emph{7-day-ahead} CLEP and MEPI for the cumulative death counts due to COVID-19 between April 11, 2020 and June 20, 2020. The observed data is shown in black, CLEP predictions are shown in the dashed blue, and the corresponding 7-day-ahead MEPIs are shown as shaded blue regions. The prediction and prediction intervals for day $t$ ($t =$ April 11, \dots, June 20) are based on data up to day $t-7$. 
    In panel \textbf{(a)}, the MEPI coverage for the 6 counties are Kings (92\%), Queens (80\%), Bronx (90\%), Cook (90\%), Los Angeles (89\%) and New York (86\%). 
    In panel \textbf{(b)}, Suffolk (85\%), Bergen (96\%), Oakland (87\%), Monmouth (86\%), Broward (89\%) and Dougherty county (86\%). (It is worth noting that the theoretical coverage guarantees under certain assumptions is 83\%; see equation~\ref{eq:probability}.)}
    \label{fig:line}
\end{figure}

\begin{figure}
    \centering
    % \resizebox{\textwidth}{!}{
    \begin{tabular}{c}
         \includegraphics[width=\textwidth]{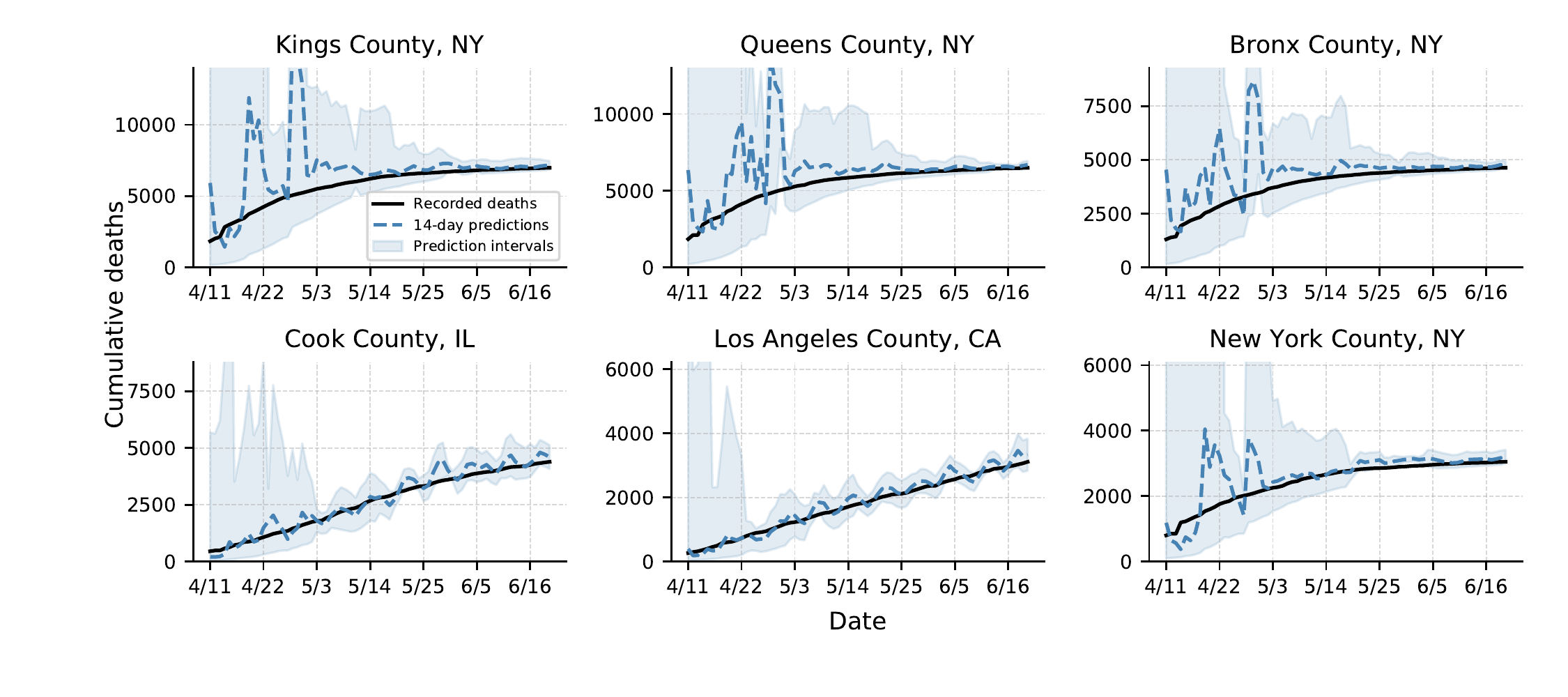} \vspace{-5mm} \\
         (a) Worst-affected counties on June 20\vspace{1mm} \\ \midrule
         \includegraphics[width=\textwidth]{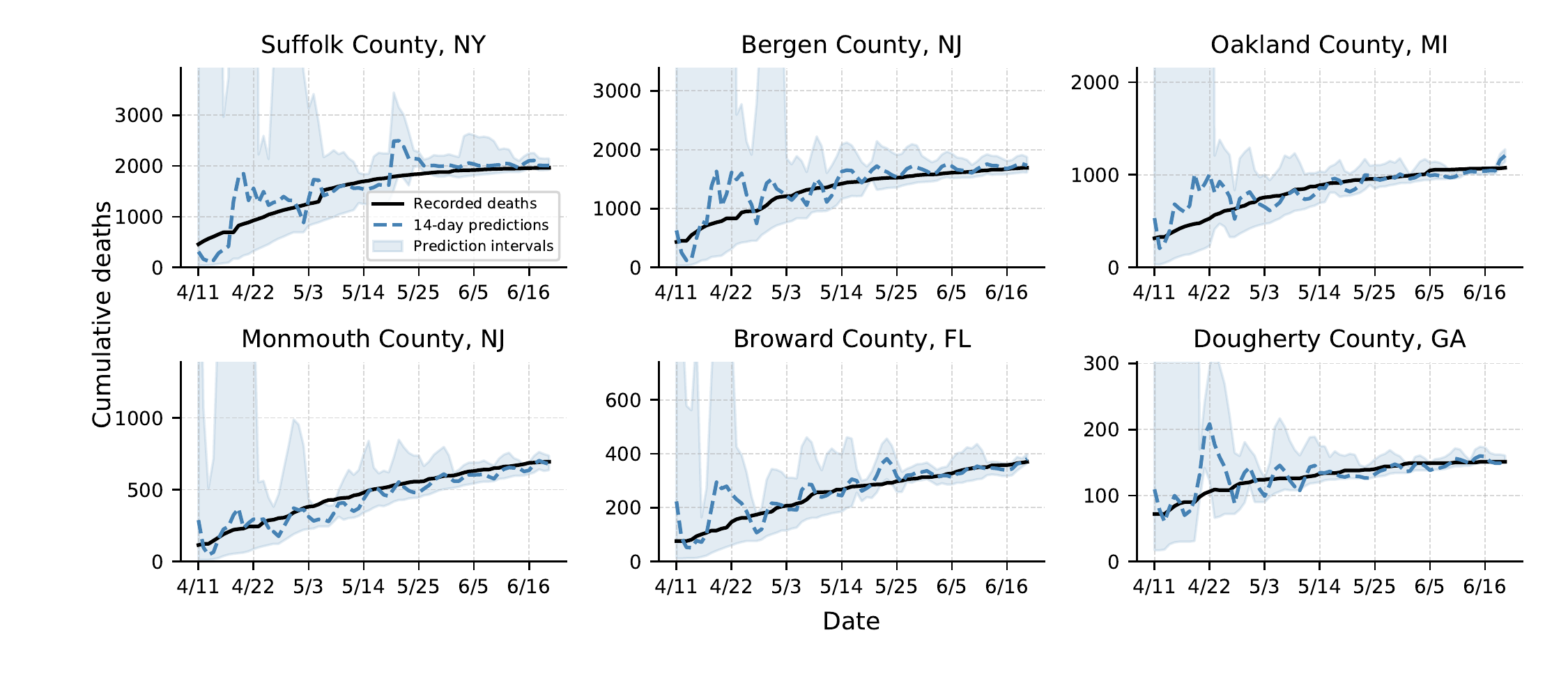} \vspace{-5mm}\\
         (b) Randomly-selected counties
    \end{tabular}
    % }
    \caption[County-wise visualization of 14-day-ahead CLEP and MEPI predictions]{
    A grid of line charts displaying the performance of \emph{14-day-ahead} CLEP and MEPI for the cumulative death counts due to COVID-19 between April 11, 2020 and June 20, 2020. The observed data is shown in black, CLEP predictions are shown in the dashed blue, and the corresponding 14-day-ahead MEPIs are shown as shaded blue regions. The prediction and prediction intervals for day $t$ ($t =$ April 11, \dots, June 20) are based on data up to day $t-14$. In panel \textbf{(a)}, the MEPI coverage for the 6 counties are Kings (99\%), Queens (99\%), Bronx (99\%), Cook (93\%), Los Angeles (96\%) and New York (89\%). 
    In panel \textbf{(b)}, Suffolk (92\%), Bergen (99\%), Oakland (89\%), Monmouth (94\%), Broward (97\%) and Dougherty county (94\%). Note that for these counties, the coverage of 14-day-ahead MEPIs is higher than that of 7-day-ahead MEPIs (shown in Figure~\ref{fig:line}) due to the wider intervals in the beginning of the period. (It is worth noting that the theoretical coverage guarantees under certain assumptions is 83\%; see equation~\ref{eq:probability}.)}
    \label{fig:line_14_days}
\end{figure}

Having examined the overall performance of our predictors, we now take a closer look at how our predictors are performing at the county level. In this section, we focus on the performance of our CLEP predictor (based on the best-performing CLEP of the expanded shared and separate linear predictor models) for the period April 11 to June 20 for 7-day and 14-day-ahead predictions (Figures~\ref{fig:line} and \ref{fig:line_14_days} respectively).
Since there are over 3,000 counties in the United States, we present results for two sets of counties:
\begin{itemize}
    \item The six worst-affected counties on June 20: Kings County, NY; Queens County, NY; Bronx County, NY; Cook County, IL; Los Angeles County, CA; and New York County, NY. 
    \item Six randomly-selected counties: Suffolk County, NY;  Bergen County, NY; Oakland County, MI; Monmouth County, NJ; Broward County, FL; Dougherty County, GA. 
\end{itemize}

In Figure~\ref{fig:line}, we present the 7-day-ahead prediction results for the worst-affected counties in the top panel (a), and for the randomly-selected counties in the bottom panel (b).
The solid black line denotes the recorded death counts, dashed blue line denotes the CLEP 7-day-ahead predictions, and
shaded blue region denotes the corresponding MEPI (prediction interval).  The predictions and prediction intervals for a given day $t$ ($t =$ April 11, \dots, June 20) are based on data up to day $t-7$.
Corresponding results for 14-day-ahead predictions are plotted in Figure~\ref{fig:line_14_days}.
Although the recorded cumulative death counts are monotonically increasing, our predictions for it (blue dashed lines in Figures~\ref{fig:line} and \ref{fig:line_14_days}) need not be since the predictions are updated daily. (Recall from Section~\ref{sub:monotonicity} that we did impose monotonicity constraints for predictions with respect to $k$ for $k$-day-ahead predictions made on a given day $t$.)

From Figure~\ref{fig:line}(a), we observe that, among the worst-affected counties, the CLEP appears to fit the data very well for Cook County, IL, and Los Angeles County, CA. After initially over-predicting the number of deaths in the four NY counties in mid-late April, our predictor also performs very well on these NY counties. Moreover, the MEPIs---plotted as blue shaded region---have reasonable width and appear to cover the recorded values for the majority of the days (detailed results on MEPIs are presented in Section~\ref{sub:mepi_results}).

The rapid increase in our predictions and MEPI interval widths observed in mid-late April in Figure~\ref{fig:line}(a) is caused due to a major upward revision of cumulative death counts by the New York state for Kings, Queens, Bronx, and New York counties on a single day, April 14. The effect of this sudden change in recorded counts on CLEP performance was also discussed in the previous section in the context of Figure~\ref{fig:longer_horizon_over_time}.
Our predictors quickly corrected themselves after the recorded counts stabilized. 

From Figure~\ref{fig:line}(b), we find that our predictors and MEPI perform well for each of our six randomly-selected counties (Broward County, FL, Dougherty County, GA, Monmouth County, NJ, Bergen County, NJ, and Oakland County, MI). However, notice again that for Suffolk County, NY, there is a sudden uptick in cumulative deaths on May 5, leading to a fluctuation in the predictions shortly thereafter, just like in the other NY counties in mid-April. 

In both panels, our predictions have higher uncertainty at the beginning of the examination period when recorded death numbers are low, which is reflected in the slightly wider MEPIs in the bottom left of each plot.

Next, in  Figure~\ref{fig:line_14_days}, we plot the visualizations for 14-day-ahead CLEP and MEPI predictions for the two sets of counties discussed above. On the one hand, overall, the CLEP predictions appear to track the cumulative death counts and MEPIs cover the observed death counts most of the times. On the other hand, we find that the MEPIs are pretty wide for all the counties in the beginning of the prediction period. This phenomenon is caused due to the fact that till mid-April, predictions made 14 days ago were trained on data with very few death counts, which led the expanded shared predictor to significantly underestimate the death counts in the beginning period (leading to larger normalized errors and thereby wider MEPIs). Furthermore, we also observe that for the counties in the NY State, CLEP greatly overestimates (sharp peak around April 22) the cumulative recorded death counts towards the end of April. Like in Figure~\ref{fig:line}, this overestimation is caused due to the major upward revision of the death counts in the NY State on April 14. It is worth noting that these sharp peaks for 14-day-ahead CLEP predictions in Figure~\ref{fig:line_14_days} occurs after 7 days of the sharp peaks observed for 7-day-ahead CLEP predictions in Figure~\ref{fig:line}.

\subsection{Empirical performance of MEPI}
\label{sub:mepi_results}
We now present the performance of our MEPI at the county level for cumulative death counts, with respect to both coverage (equation~\ref{eq:coverage}) and average normalized length~(equation~\ref{eq:normalized_length}) (see Section~\ref{sub:evaluation_metric}).
Since the average performance may change with time, we report the results for two time periods: 
$\{\text{April 11, \ldots, May 10} \}$, 
and $\{\text{May 11, \ldots, June 20} \}$.
We choose these time periods for a simple reason: The first draft of this paper
presented results until May 10, and thus the period May 11 to June 20 serves as additional out-of-sample validation. (Our methods were proposed and tuned prior to May 10, except for the change of transform from logarithm to square-root in equation~\ref{eq:final_wt}.)

\begin{figure}
    \centering
    \begin{tabular}{cc}
        \multicolumn{2}{c}{Evaluation period: April 11--May 10}\\
        \includegraphics[width=0.36\textwidth]{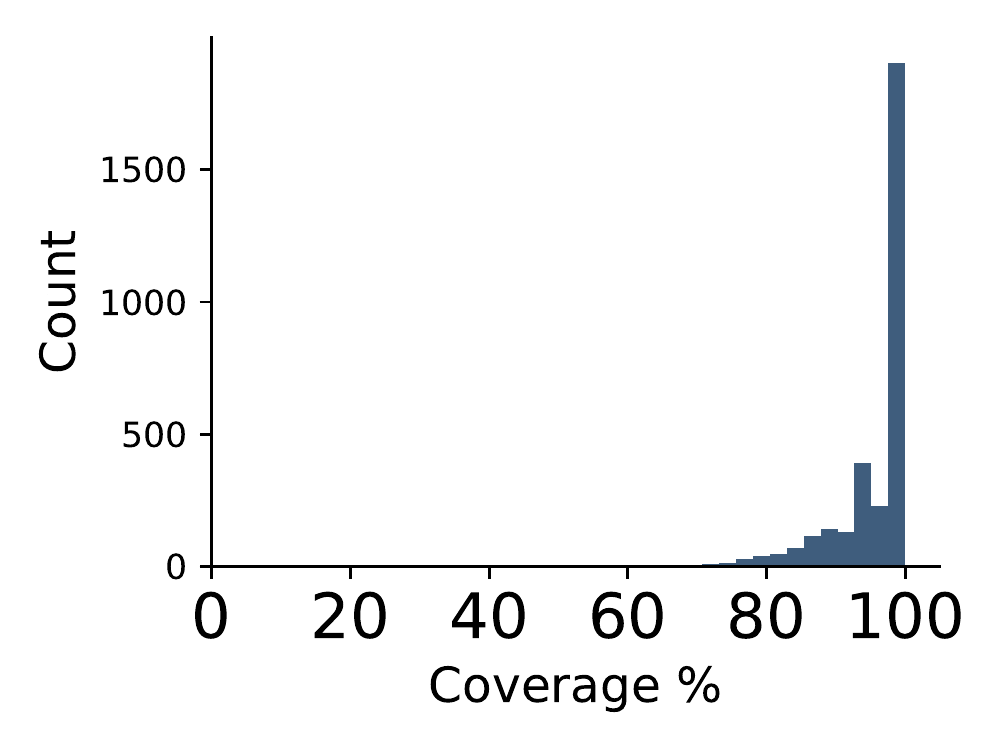} & 
        \includegraphics[width=0.36\textwidth]{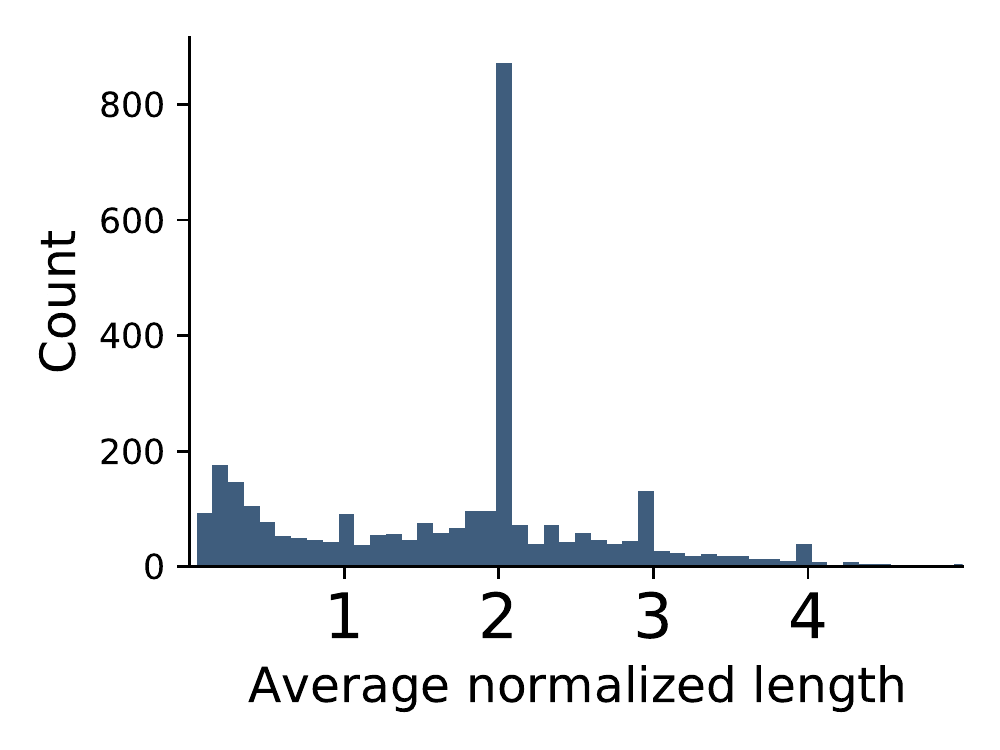}\\
        (a) Coverage for all counties &  
        (b) Average length for all counties\vspace{2mm}\\
         \midrule \\ 
         \multicolumn{2}{c}{Evaluation period: May 11--June 20}\\
         \includegraphics[width=0.36\textwidth]{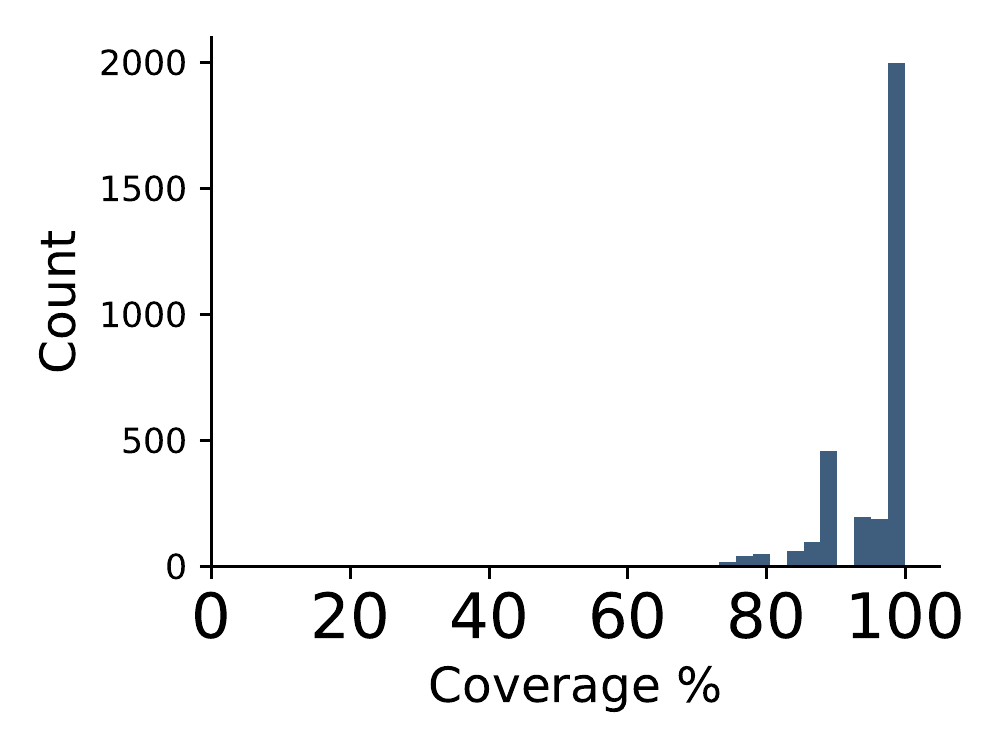} &
         \includegraphics[width=0.36\textwidth]{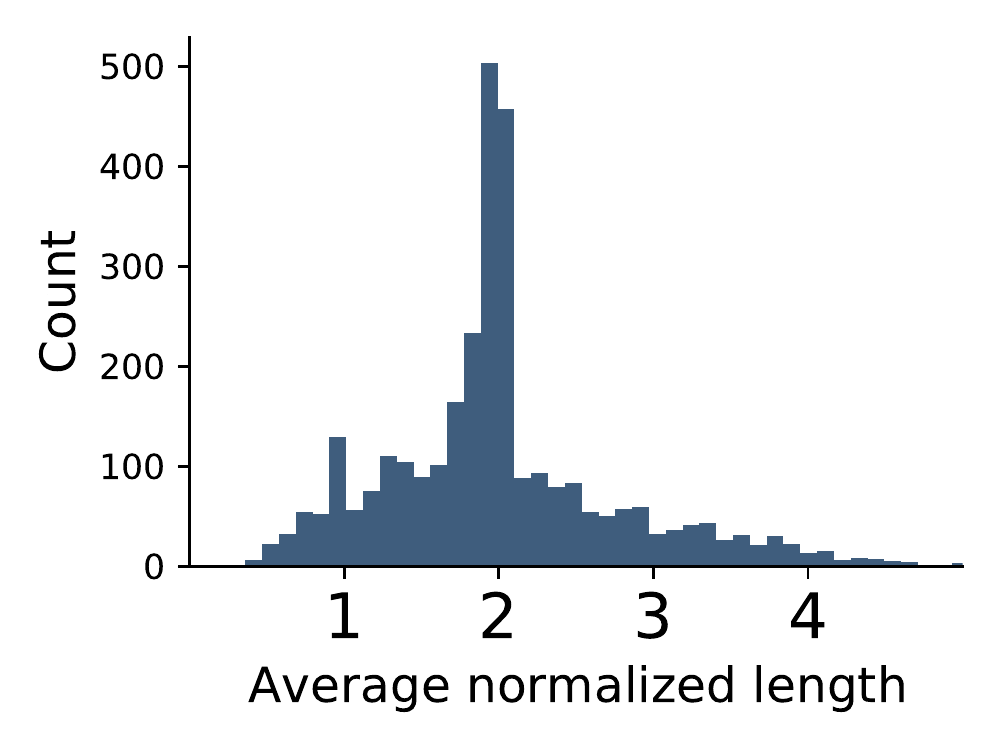} \\
        (c) Coverage for all counties &  
        (d) Average length for all counties\vspace{2mm}\\
         \midrule \\ 
        \includegraphics[width=0.36\textwidth]{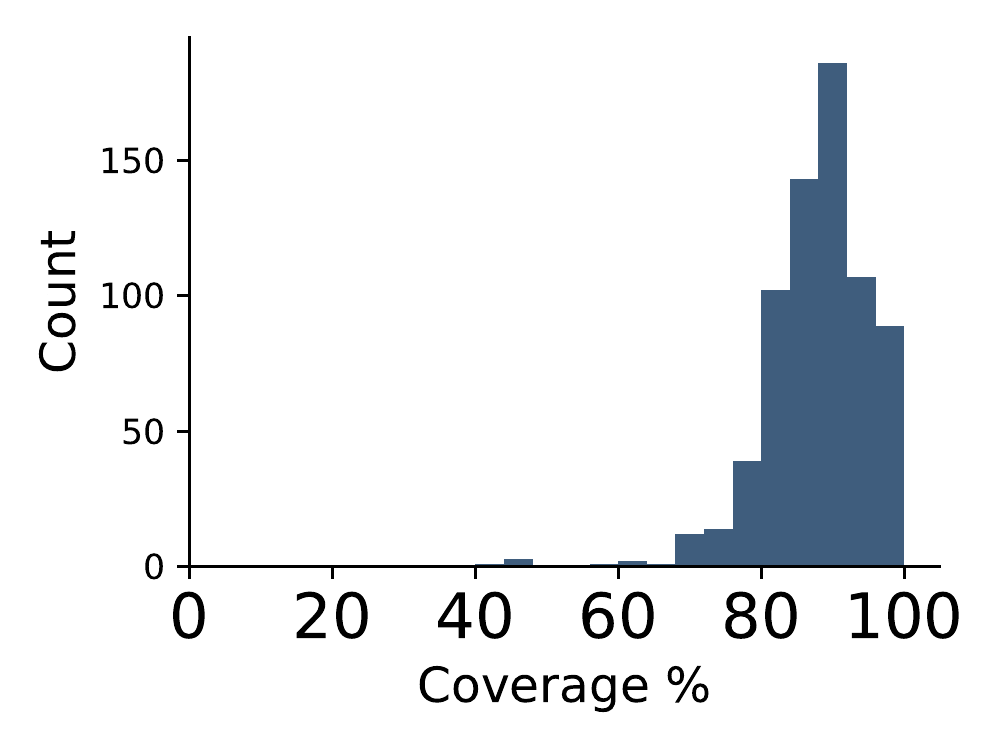} & 
        \includegraphics[width=0.36\textwidth]{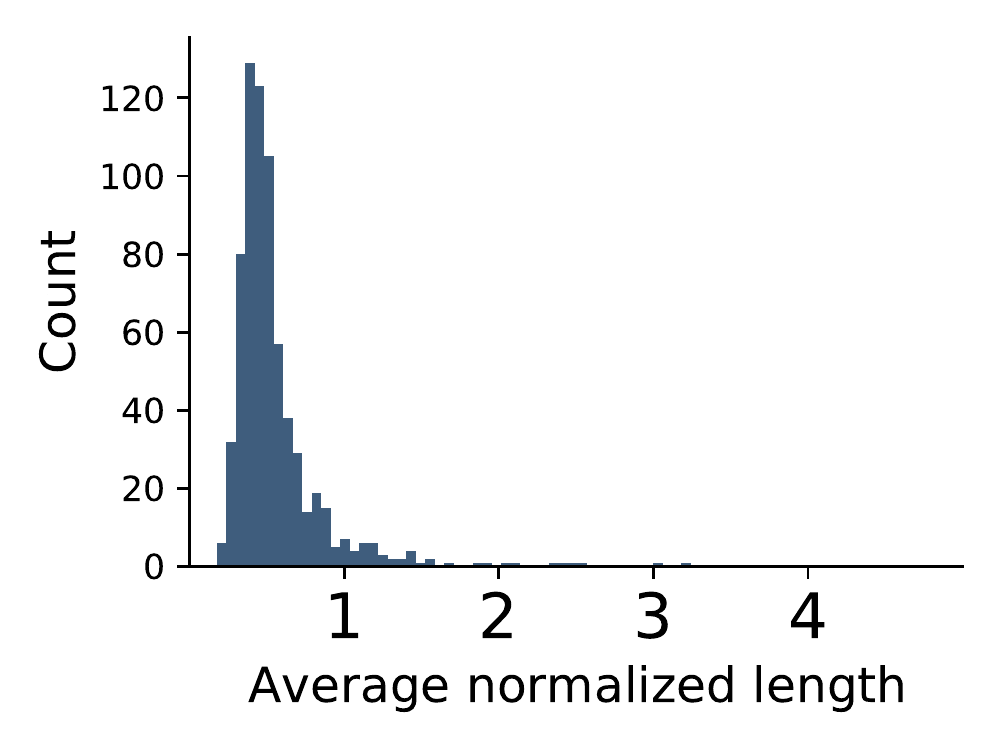} \\
        (e) Coverage for selected counties &  
        (f) Average length for selected counties\vspace{2mm}\\
    \end{tabular}
    \caption[MEPI coverage and lengths for 7-day-ahead predictions]{Histograms showing the performance of \emph{7-day-ahead  MEPI intervals} for county-level cumulative death counts. 
    For the top panels \textbf{(a)} and \textbf{(b)}, we compute the histogram for all counties spanning 
    April 11--May 10, 2020 and for the middle panels \textbf{(c)} and \textbf{(d)}, spanning
    May 11--June 20, 2020. 
    For the bottom two panels \textbf{(e)} and \textbf{(f)}, we only include counties that had at least 10 cumulative deaths by June 11, and the histogram is based on the county-specific periods, over the days April 11 to June 20, which only includes days for which the county's cumulative deaths is at least 10. See Figure~\ref{fig:mepi_coverage_and_length_14_days} for similar plots for
    14-day-ahead MEPIs.
    }
    \label{fig:mepi_coverage_and_length}\vspace{5mm}
\end{figure}

\begin{figure}
    \centering
    \begin{tabular}{cc}
        \multicolumn{2}{c}{Evaluation period: April 11--May 10}\\
        \includegraphics[width=0.36\textwidth]{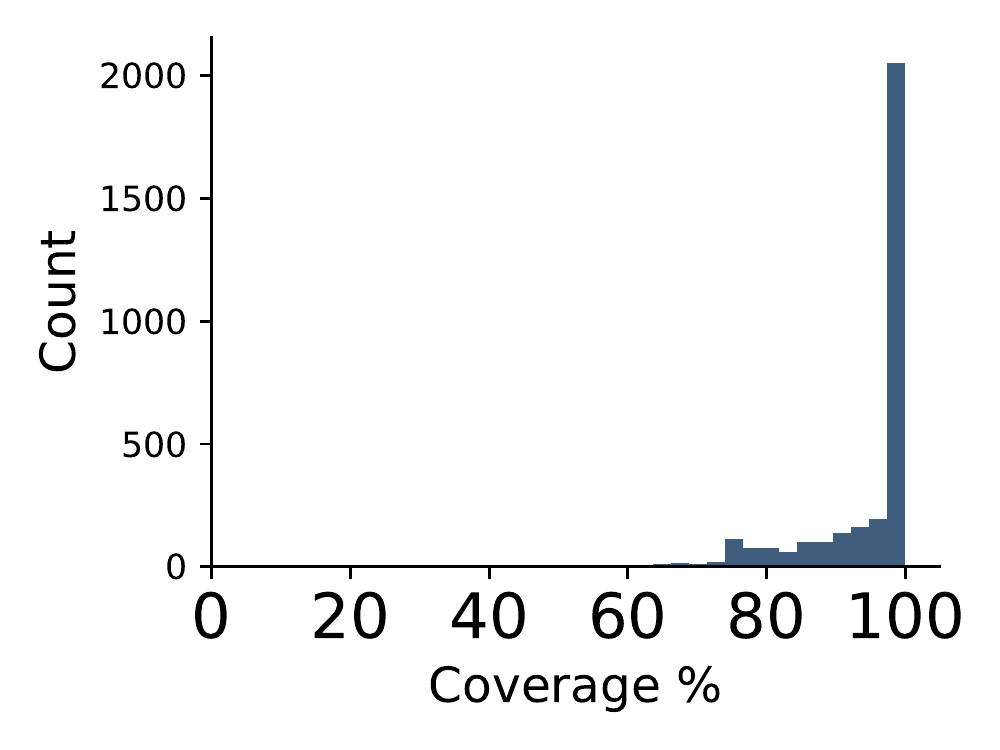} & 
        \includegraphics[width=0.36\textwidth]{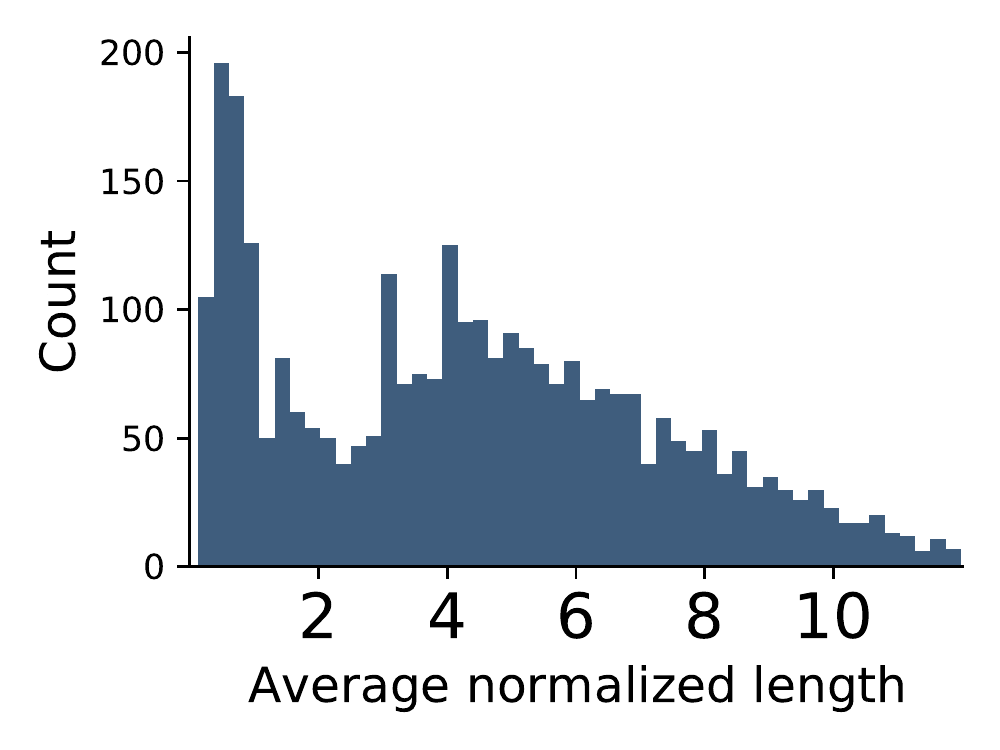}\\
         (a) Coverage for all counties &  
         (b) Average length for all counties\vspace{2mm}\\
         \midrule \\ 
         \multicolumn{2}{c}{Evaluation period: May 11--June 20}\\
         \includegraphics[width=0.36\textwidth]{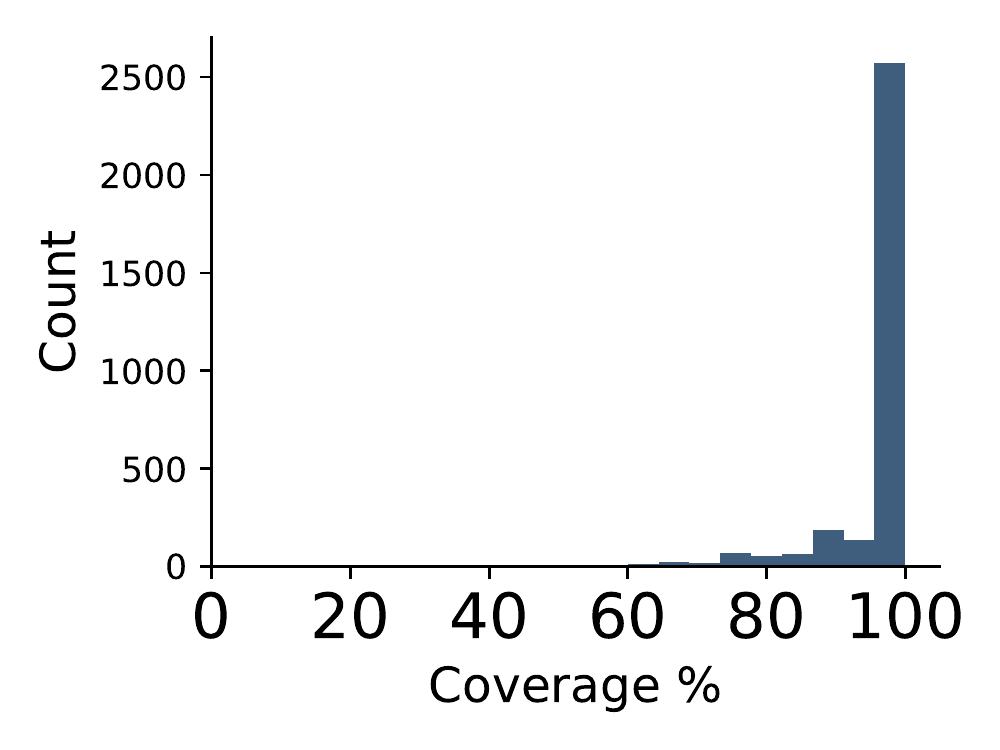} &
         \includegraphics[width=0.36\textwidth]{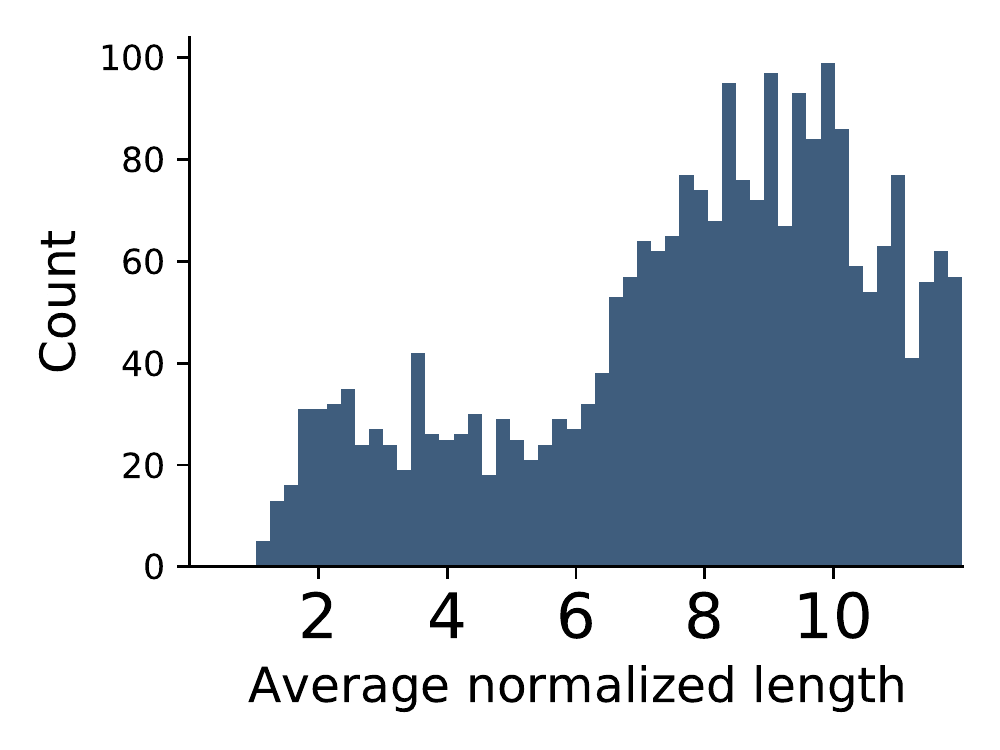} \\
         (c) Coverage for all counties &  
         (d) Average length for all counties \vspace{2mm}\\
         \midrule \\ 
        \includegraphics[width=0.36\textwidth]{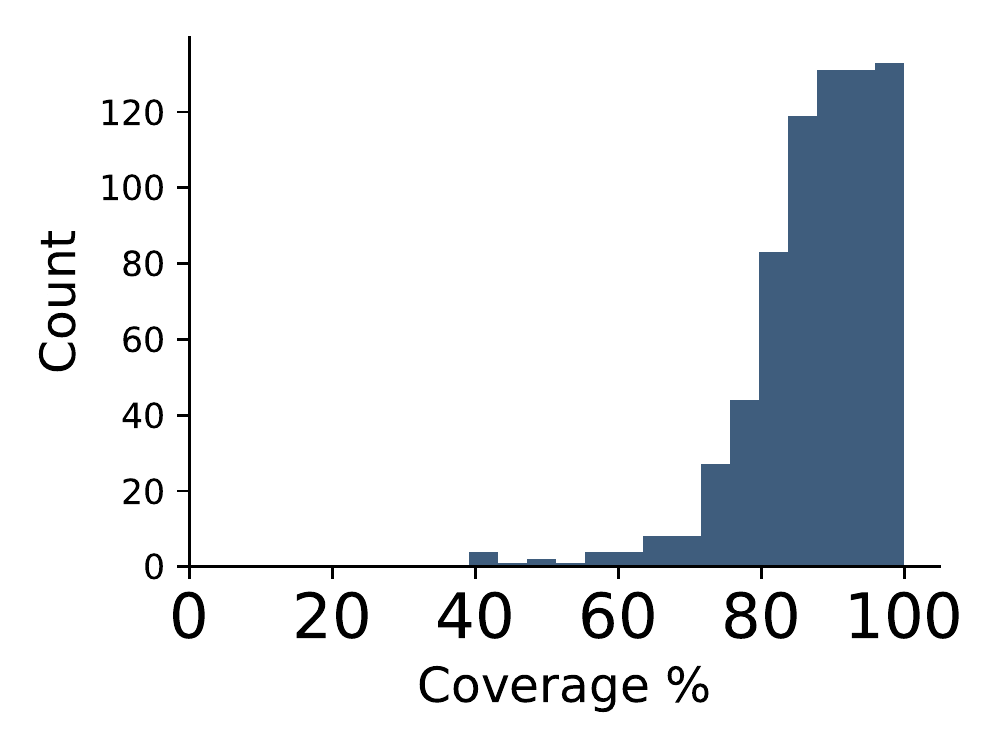} & 
        \includegraphics[width=0.36\textwidth]{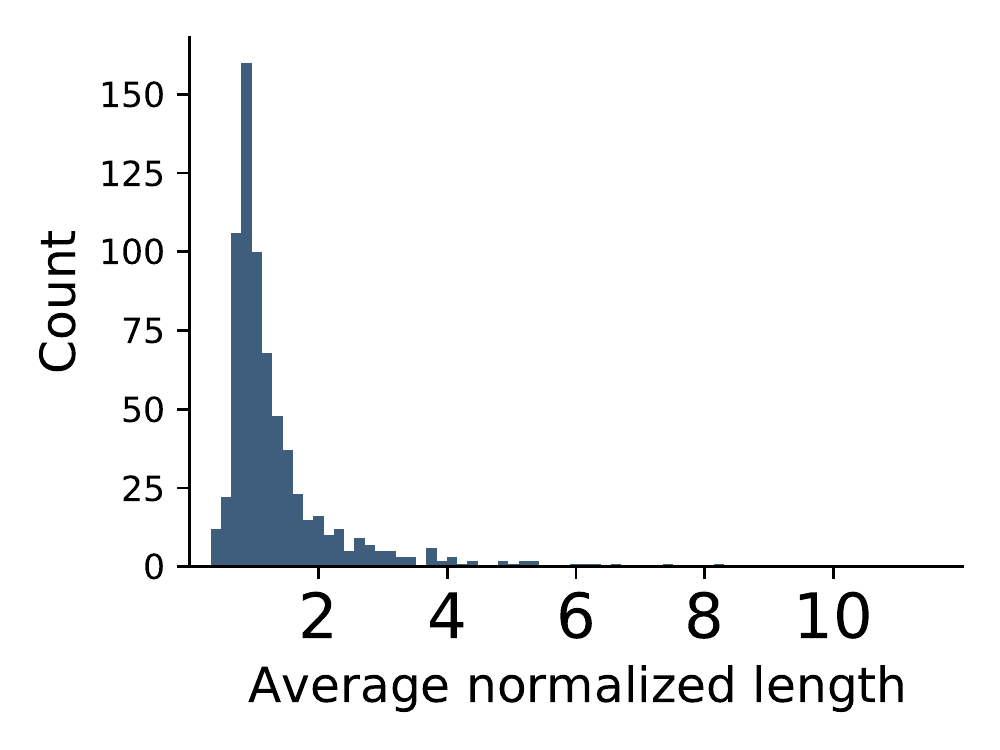} \\
        (e) Coverage for selected counties &  
        (f) Average length for selected counties\vspace{2mm}\\
    \end{tabular}
    \caption[MEPI coverage and lengths for 14-day-ahead predictions]{Histograms showing the performance of the \emph{14-day-ahead  MEPI intervals} for county-level cumulative death counts. 
    Panels \textbf{(a)} and \textbf{(b)} present histograms of coverage and average normalized length, respectively, for all counties over 
    April 11--May 10. Panels \textbf{(c)} and \textbf{(d)} present histograms of coverage and average normalized length, respectively, for all counties over May 11--June 20. 
    Panels \textbf{(e)} and \textbf{(f)} present histograms of coverage and average normalized length, respectively, only for counties that had at least 10 cumulative deaths by June 11, and the coverage and lengths are computed over the county-specific periods, over the days April 11 to June 20, for which the county's cumulative death count is at least 10.
    See Figure~\ref{fig:mepi_coverage_and_length} for similar plots for
    7-day-ahead MEPIs.
    }
    \label{fig:mepi_coverage_and_length_14_days}\vspace{5mm}
\end{figure}

We evaluate the 7-day-ahead and 14-day-ahead MEPIs, i.e., $k=7$ and 14 in equation~\ref{eq:mepi_original}, designed with the CLEP that combines the expanded shared and separate linear predictors, and 
summarize the results in Figure~\ref{fig:mepi_coverage_and_length} and Figure~\ref{fig:mepi_coverage_and_length_14_days} respectively.
We now discuss these results, first for 7-day-ahead and then for 14-day-ahead MEPIs.\\

\paragraph{\emph{Coverage:}} In panels (a) and (c) of Figure~\ref{fig:mepi_coverage_and_length}, we plot the histogram of observed coverage (equation~\ref{eq:coverage}) of 7-day-ahead MEPIs across all counties in the US for April 11--May 10, and May 11--June 20 respectively. Panel (e) of Figure~\ref{fig:mepi_coverage_and_length} shows the observed coverage of 7-day-ahead MEPIs for the 700 counties that had at least 10 deaths on June 11 (each such county has had significant counts for at least 10 days by the end of our evaluation period June 20).  
For each county, we include the days starting from either April 11, 2020 or the day of 10 deaths (if this day is after April 11) until June 20, 2020. 
On June 20, the median number of days since 10 deaths is 58. From these plots, we observe that for the majority of the counties, we achieve excellent coverage.
Finally, Figure~\ref{fig:mepi_coverage_and_length_14_days} shows the corresponding results for 14-day-ahead MEPIs, i.e., coverage for April 11--May 10 in panel (a), May 11--June 20 in panel (c), and over the county-specific period for 700 counties that had at least 10 deaths on June 11 in panel (e).

The coverage observed for the two periods in panels (a) and (c) of Figure~\ref{fig:mepi_coverage_and_length} are very similar.
For the earlier period---April 11 to May 10---in panel (a), the 7-day-ahead MEPIs have a median coverage of 100\% and mean coverage of 95.6\%. On the other hand, for the later period---May 11 to June 20---in panel (c), the 7-day-ahead MEPIs have a median coverage of 100\% and mean coverage of 96.2\%. However, the coverage is slightly decreased when restricting to counties with at least 10 deaths in panel~(e), for which we observe a median coverage of 88.7\%, and mean coverage of 87.9\%.
This observation is consistent with the fact that at the beginning of the pandemic, several counties had zero or very few deaths resulting in very good coverage with the prediction interval. On the other hand, note that smaller death counts would also imply a relatively larger normalized length for the MEPI intervals.

Figure~\ref{fig:mepi_coverage_and_length_14_days}(a), (c) and (e) show that, in general, our 14-day-ahead MEPIs achieve similar coverage as our 7-day-ahead MEPIs. For example, over the April 11--May 10 and May 11-June 20 periods, our prediction intervals have mean coverage of 95.0\% and 97.0\% (median is 100\% for both periods). For panel (e)---counties with at least 10 deaths on June 11---the coverage has a median of 89.7\% and a mean of 87.9\%. 

Overall, the statistics discussed above show that both 7-day-ahead and 14-day-ahead MEPIs achieve excellent coverage in practice.
In fact, for the counties with poor coverage, we show in Appendix~\ref{sub:poor_exchange} that there is usually a sharp uptick in the number of recorded deaths at some point in the evaluation period, possibly due to recording errors, or backlogs of counts. Modeling these upticks and obtaining coverage for such events is beyond the scope of this paper.\\

\paragraph{\emph{Normalized length:}} Next, we discuss the other evaluation metric of the MEPIs, their normalized length~\eqref{eq:normalized_length}. In panels (b) and (d) of Figure~\ref{fig:mepi_coverage_and_length}, we plot the histogram of the observed average normalized length of 7-day-ahead MEPIs for the periods April 11--May 10, and May 11--June 20 respectively. Panel (f) covers the same counties as did panel (e): those with at least 10 deaths for at least 10 days in the period April 11 to June 20, 2020. 

Recall that the normalized length is defined as the length of the MEPI over the recorded number of deaths~(equation~\ref{eq:normalized_length}).
And, more than 70\% of counties in the US recorded 2 or less COVID-19 deaths by May 1.  For these counties, having a normalized length of 2 means the actual length of the prediction interval is 4 (or less).  And thus, it is not surprising to see that the average normalized length of MEPI for a non-trivial fraction of counties is larger than 2 in panels (b) and (d).
When considering counties with at least 10 deaths in panel (f), the average normalized length over these (county-specific) periods is much smaller;
and the median is 0.470. 

Turning to 14-day-ahead MEPIs in Figure~\ref{fig:mepi_coverage_and_length_14_days}, panels (b) and (d) show that that the normalized length for 14-day-ahead MEPIs can be quite wide for counties with a small number of deaths. Nevertheless, panel~(f) shows that the 14-day-ahead MEPIs are reasonably narrow for counties with more than 10 deaths, with a median average normalized length of 1.027---which is roughly two times the median size of 0.470 for the 7-day-ahead MEPIs in Figure~\ref{fig:mepi_coverage_and_length}(f).

Overall, Figures~\ref{fig:mepi_coverage_and_length} 
and~\ref{fig:mepi_coverage_and_length_14_days} show that our MEPIs provide a reasonable balance between coverage and length, especially when the cumulative counts are not too small, for up to 14 days in future.

\section{Related work}
\label{sec:comparison}
Several recent works have tried to predict the number of cases and deaths related to COVID-19.
Even more recently, the Center for Disease Control and Prevention (CDC) has started aggregating results from several models.\footnote{Forecasts available at~\url{https://www.cdc.gov/coronavirus/2019-ncov/covid-data/forecasting-us.html}}
But to the best of our knowledge, ours was the first work focusing on predictions at the county-level. Besides, directly comparing other models' forecasting results to our own can be difficult for several other reasons: (1) the predictors mostly make strong assumptions and typically do not involve data-fitting, (2) we do not have access to a direct implementation of their predictors (or results), and (3) their predictors focus on substantially longer time horizons. Keeping these points in mind, we now provide a brief summary of recent work on predictive modeling for COVID-19.

Two recent works~\cite{IHME,ferguson2020impact} have modeled the death counts at the state level in the US.
The earlier versions of the model by Murray et al. (also referred to as the IHME model) was based on Farr's Law with feedback from Wuhan data.
On the other hand, the Imperial College model~\cite{ferguson2020impact}
uses an individual-based simulation models with parameters chosen based
on prior knowledge. On the topic of Farr's Law, we note that Bregman~\cite{Bregman}
used Farr's Law to predict that the total cases from the AIDS pandemic would diminish by the mid-1990s and the total number of cases would be around 200,000 in the entire lifetime of the AIDS pandemic. It is now estimated that 32 million people have died from the disease so far. While the AIDS pandemic is very different from the COVID-19 pandemic, it is still useful to keep this historical performance in mind.

Another approach uses exponential smoothing from time-series predictors to estimate day-level COVID-19 cases~\cite{elmousalami2020day}. 
In addition, several works use compartmental epidemiological predictors such as SIR, SEIR, and SIRD~\cite{fanelli2020analysis,pei2020initial,chime} to provide simulations at the national level.  
Other works~\cite{peak2020modeling,Hsiang2020} simulate the effect of social distancing policies either in the future for the US, or in a retrospective manner for China. Finally, several papers estimate epidemiological parameters retrospectively based on data from China~\cite{wang2020evolving,Kucharski2020}.

During the revision of our paper, another work was published by Chiang
et al.~\cite{chiang2020hawkes} that appeared in medRxiv on June 8, 2020
\footnote{Accessed at \url{https://www.medrxiv.org/content/10.1101/2020.06.06.20124149v1.full.pdf}} after the submission of our paper to arXiv in mid-May. Chiang et al. use models based on Hawkes' process to provide county-level predictions for new daily cases as well as new death counts. Of note is that the authors also explore a CLEP with adaptive tuning of $c$ and $\mu$ (whereas we used fixed values for these parameters). Such a tuning approach might present a promising improvement of CLEP performance in general and we plan to investigate adaptive tuning of various hyper-parameters in the CLEP in our own future work. Unfortunately, we were unable to properly reproduce the CLEP results provided in Chiang et al.'s work using their provided documentation. During a private email exchange the authors kindly provided further information regarding some of our questions about their methodology\footnote{In particular,
via this email exchange we learned that: (i) they had implemented the adaptive tuning of CLEP; and (ii) they had computed the \% error (in Table S1 of their paper) for total new counts over the entire $k$-day-block (for $k$-day-ahead predictions) summed over all the counties in a given quantile thereby explaining the (surprising at first) decrease in \% error as the prediction horizon $k$ increases.}, but several of their choices make it difficult to compare their work to ours. For instance, their work focuses on daily counts, rather than cumulative counts as ours does. 
More importantly, their prediction numbers are not available on their GitHub repository (both for the period till May 20 analyzed in their paper and the days since then).
The authors do not report the performance of their confidence intervals in the paper, and report the MAE performance metric only for the counties that fall in the top quantiles of cumulative counts at the end of the evaluation period. Such a quantile-based group of counties is not interpretable (since it is time-varying and not spatially meaningful) and does not allow for real-time use, since one must wait until the end of the evaluation period to calculate the performance.
In addition, the authors compute their predictions in blocks of days, e.g., once a week for the 7-day-ahead predictions (rather than daily as in this paper). 
Thus, from our point of view, these decisions unfortunately make their work ill-suited to real-time usage for making fast-paced policy decisions related to COVID-19.

\section{Impact: a hospital-level severity index for distributing medical supplies} 
\label{sec:impact}

Together with the non-profit Response4Life\footnote{\url{https://response4life.org/}}, our models have been used to determine which hospitals are most urgently in need of medical supplies, and have subsequently been directly involved in the distribution of medical supplies across the country. To do this, we translate our forecasts into the COVID pandemic severity index, which is a simple measure of the COVID-19 outbreak severity for each hospital. 

To generate this hospital-level severity index, we divided the total county-level deaths among all of the hospitals in the county proportional to their number of employees. Next, for each hospital, we computed its percentile among all US hospitals with respect to total deaths so far and also with respect to predicted new deaths in the next seven days. These two percentiles are then averaged to obtain a single score for each hospital. Finally, this score is quantized evenly into three categories: low, medium, and high severity. Evaluation and refinement of this index are ongoing, as more hospital-level data becomes available.
The interested reader can find a daily-updated map of the COVID pandemic severity index and additional hospital-level data at our website \url{https://covidseverity.com}.

\section{Conclusion}
\label{sec:conclusion}
In this paper, we made three key contributions. We (1) introduced a data repository containing COVID-19-related information from a variety of public sources, (2) used this data to develop CLEP predictors for short-term forecasting at the county level (up to 14 days), and (3) introduced a novel yet simple method MEPI for producing prediction intervals for these predictors. By focusing on county-level predictions, our forecasts are at a finer geographic resolution than those from a majority of other relevant studies. By comparing our predictions to real observed data, we found that our predictions are accurate and that our prediction intervals are reasonably narrow and yet provide good coverage. We hope that these results will be useful for individuals, businesses, and policymakers to plan and cope with the COVID-19 pandemic and that our data repository and forecasting and interval methodology will be useful for academic purposes. Indeed, our results are already being used to determine the hospital-level need for medical supplies and have been directly influential in determining the distribution of these supplies.

Our data repository will be useful both for educational purposes, as well as for other teams interested in analyzing the data underlying the COVID-19 pandemic. Our CLEP ensembling techniques and MEPI methodology can be applied to other models for COVID-19 forecasting, as well as to online methods and time-series analysis more broadly. Our data, codes and models can be found at \url{https://covidseverity.com}.

Last but not the least, inspired by the recent work~\cite{chiang2020hawkes},
we are beginning our investigation into adaptive tuning (over time) of $\mu$, $c$ and other hyperparameters for CLEP, in the hope to improve its performance.

\section*{Acknowledgements}
\label{sec:acknowledgements}

Bin Yu acknowledges the support of CITRIS Grant 48801, and a research award by Amazon.
The authors would like to thank many people for help with this effort. Our acknowledgement section is unusually long because it reflects the nature of an ER-like collaborative project. We were greatly energized by the tremendous support and the outpouring of help that we received not only from other research groups, but also high school students, medical staff, ER doctors, and several volunteers who signed up at Response4Life.

We would like to first thank the Response4Life team (Don Landwirth and Rick Brennan in particular), and volunteers for building the base for this project. We would also like to thank Max Shen's IEOR group at Berkeley: Junyu Cao, Shunan Jiang, Pelagie Elimbi Moudio for helpful inputs in the early stages of the project. 
We thank Aaron Kornblith and David Jaffe for advice from a medical perspective, especially Aaron for a great deal of useful feedback.
We want to mention special thanks to Sam Scarpino for sharing data with us.
 
We would like to thank Danqing Wang, Abhineet Agarwal, and Maya Shen for their help in improving our visualization website \url{https://covidseverity.com} over the summer, and support from Google, in particular Cat Allman and Peter Norvig. 
We would also like to thank the high school students Matthew Shen, Anthony Rio, Miles Bishop, Josh Davis, and Dylan Goetting for helping us to collect valuable hospital related information.

Finally, we acknowledge helpful input on various data and modeling aspects from many others including Ying Lu, Tina Eliassi-Rad, Jas Sekhon, Philip Stark, Jacob Steinhardt, Nick Jewell, Valerie Isham,  Sri Satish Ambati, Rob Crockett, Marty Elisco, Valerie Karplus, Marynia Kolak, Andreas Lange, Qinyun Lin,  Suzanne Tamang, Brian Yandell and Tarek Zohdi. We also acknowledge the helpful feedback from the anonymous referees which enhanced the readability of the paper, and led to inclusion of detailed discussion on data biases, performance of CLEP and MEPI for 14-day-ahead predictions results in the paper, and exploration of the variants of few predictors with additional features (discussed in Appendix~\ref{sec:additional_models}).

\subsection*{Disclosure Statement}
The authors have no conflicts of interest to declare.

%Begin appendix section(s)
\appendix

% Add appendices here:
\section{Predictors with additional features}
\label{sec:additional_models}

We now describe a few additional features that were considered to potentially improve our predictors (but did not lead to
any significant improvements).
We included these features after our first submission (on May 16, 2020), and hence tried the new features only in the context of the CLEP that combines the expanded shared and linear predictors.\footnote{We note that the expanded shared predictor in this appendix is implemented without the monotonicity adjustment (discussed in Section~\ref{sub:monotonicity}). The linear predictor does not need such adjustments in our setting. 
Since our attempts with new features considered in this appendix did not lead to any improvement, we did not re-do the investigation with the monotonicity adjustment. We leave any further investigation with these new features (or their variants) for future work.}

\subsection{Social-distancing feature}
Here we consider adding a social distancing feature to the expanded shared model (discussed in Section~\ref{sub:expanded_shared}). We included an indicator feature in equation~\ref{eq:shared_expanded_model} for every county that takes value $1$ on a day if at least two weeks have passed since social distancing was first instituted in a county, and 0 otherwise. We chose two weeks as the time lag to account for the two week progression time for the illness to the recovery of the COVID-19.
We found it necessary to regularize this predictor since, without regularization, our 7-day-ahead predictions became infinite in some cases. We regularized this model with the elastic net and an equal penalty of .01 for both $\ell_1$ and $\ell_2$ regularization.

\begin{figure}
    \centering
    \includegraphics[width=0.5\textwidth]{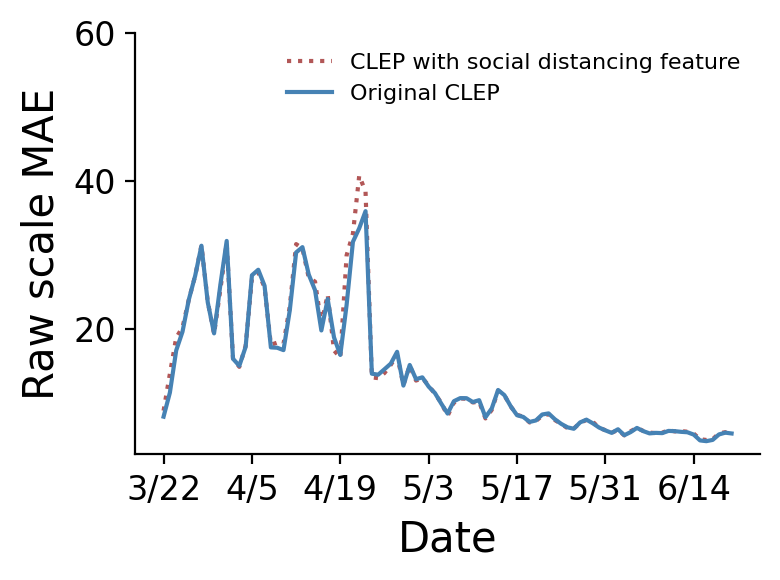}
    \caption[Mean absolute errors with social distancing feature]{Plots of raw-scale MAE for \emph{7-day-ahead} predictions of two variants of CLEP combining expanded shared and linear predictors: CLEP with a social distancing indicator feature for whether social distancing was in place in a county for more than two weeks or not, and the original CLEP considered in the main paper. The social distancing feature is included in the expanded shared model.}
    \label{fig:social_distancing}
\end{figure}

We now evaluate the performance of the two variants of the expanded shared model by ensembling each of them with the separate linear model as done for the original CLEP in the main paper.
In particular, we compare CLEP with the social-distancing feature included in the expanded shared predictor, with the original CLEP from the main paper, for $7$-day-ahead prediction of the recorded cumulative death counts.  
We found that the new variant (with the social distancing feature) performed slightly worse than our original CLEP. Over the period March 22 to June 20, the original CLEP has a mean (over time) raw-scale MAE~(equation~\ref{eq:maes}) of 13.95, while the social-distancing variant has an MAE of 14.2.
In Figure~\ref{fig:social_distancing}, we plot the behavior of raw-scale MAEs with time for the evaluation period from March 22 to June 20. We observe that the performance of the new CLEP variant is similar to that of the original CLEP, with the exception of a couple of the peaks, where the new CLEP variant performs slightly worse. 

\subsection{Weekday feature}
As illustrated in Section~\ref{sec:data_biases} and Figure~\ref{fig:weekday_and_adjustment}(a), the COVID-19 death counts are under-reported on Sunday and Monday, which could potentially lead to increased errors for our prediction algorithm. 

\begin{figure}
    \centering
    \resizebox{\textwidth}{!}{
    \begin{tabular}{cc}
    \includegraphics[width=0.49\textwidth]{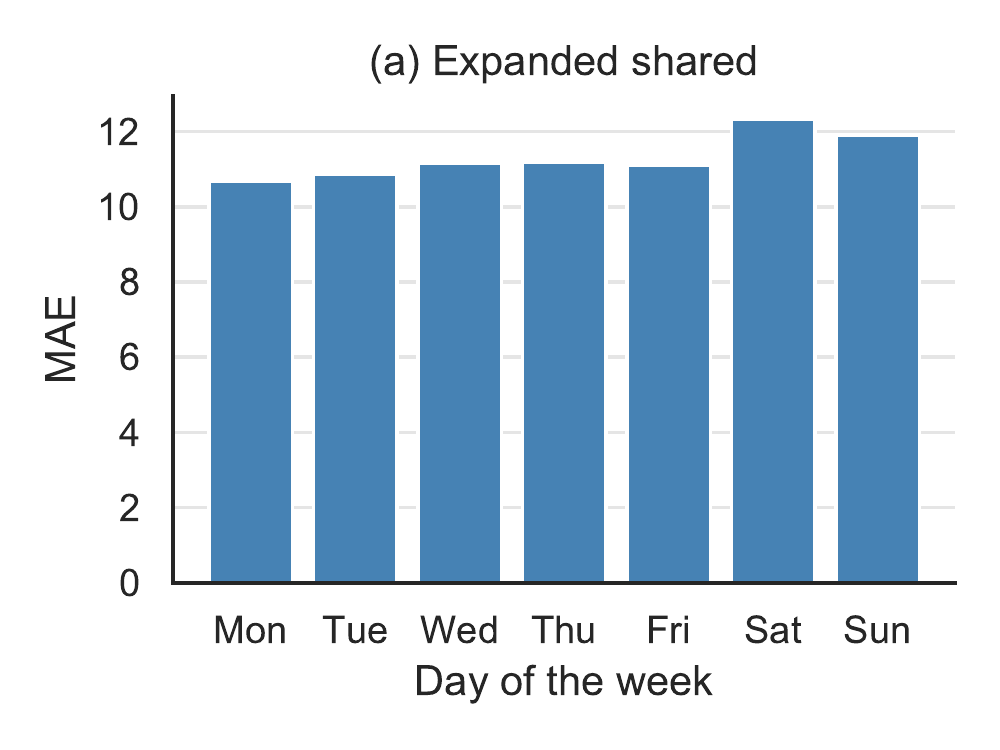}&  
    \includegraphics[width=0.49\textwidth]{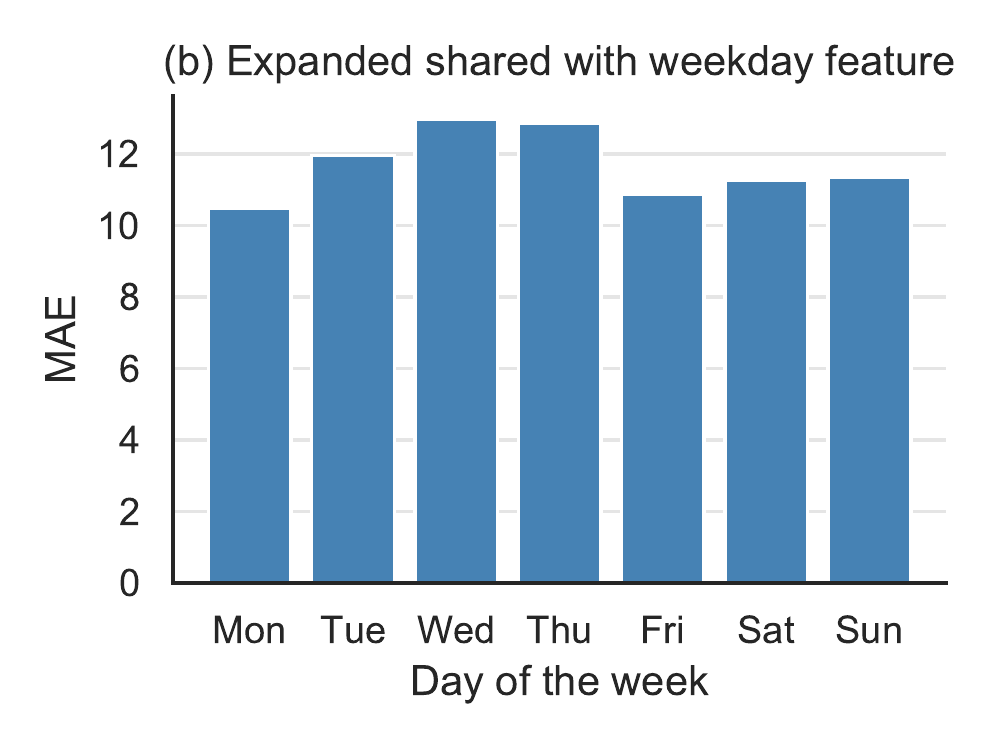}
    \end{tabular}
    }
    \caption[Mean absolute errors for expanded shared predictor with weekday feature]{Mean raw-scale MAEs by weekdays for the expanded shared predictor for \emph{3-day-ahead} predictions with and without the weekday feature. The MAE for a given day is the MAE for the 3-day-ahead prediction \textit{computed on that day}. So the MAE for Wednesday is the MAE for predicting the cumulative deaths on Friday.}
    \label{fig:weekday_feat}
\end{figure}

To address this, we first investigated our expanded shared predictor's (equation~\ref{eq:shared_expanded_model}) performance for 3-day-ahead prediction on a per weekday basis and plotted them in Figure~\ref{fig:weekday_feat}(a). We can observe the average raw-scale MAE is slightly higher for the days when the 3-day-ahead period included both Sunday and Monday. For example, 3-day-ahead predictions made on Saturday would require making predictions for Saturday, Sunday, and Monday, and that made on Sunday would require making predictions for Sunday, Monday, and Tuesday. 

To help account for this bias, we introduced an additional indicator feature in equation~\ref{eq:shared_expanded_model} that takes a value of 1 when the day---for which the prediction is made---is either Sunday or Monday, and 0 otherwise. 
For instance, when we make 3-day-ahead predictions on Saturday, this feature would take value $0$ while computing the prediction for Saturday, and $1$ when we compute the prediction for Sunday and Monday.
We plot the error distribution over days of this new variant in Figure~\ref{fig:weekday_feat}(b). For the new variant, we find that the raw-scale MAEs for the days, when the 3-day-ahead period does not include both Sunday and Monday, typically have higher MAE. Overall when averaging across all days for March 22 to June 20, we find that the new variant of expanded shared predictor performed slightly worse than the original version. The raw-scale MAE~\eqref{eq:raw_scale_mae} for the new variant is 11.7, while the original variant had a raw-scale MAE of 11.5.

Next, we experiment with adding a weekday feature to the separate linear predictors (discussed in Section~\ref{sub:linear}) for 3-day-ahead predictions, by adding a binary feature that takes value $1$ if the day---for which the prediction is made---is either Sunday or Monday, and $0$ otherwise.
Thus, the new variant of the separate linear predictors is given by
\begin{align}
\label{eq:linear_weekday_model}
    \ehat[\textrm{deaths}_{t+1}^c|t] = \beta_0^c + \beta_1^c (t+1) + \beta_2^c v_{t+1},
\end{align}
where $v_{t+1}$ indicates whether day $t+1$ is Sunday/Monday or not. For 3-day-ahead predictions ($\ehat[\textrm{deaths}_{t+3}^c|t]$), we simply replace $t+1$ by $t+3$, and $v_{t+1}$ by $v_{t+3}$ on the RHS of equation~\ref{eq:linear_weekday_model}.

\begin{figure}
    \centering
    \resizebox{\textwidth}{!}{
    \begin{tabular}{cc}
    \includegraphics[width=0.49\textwidth]{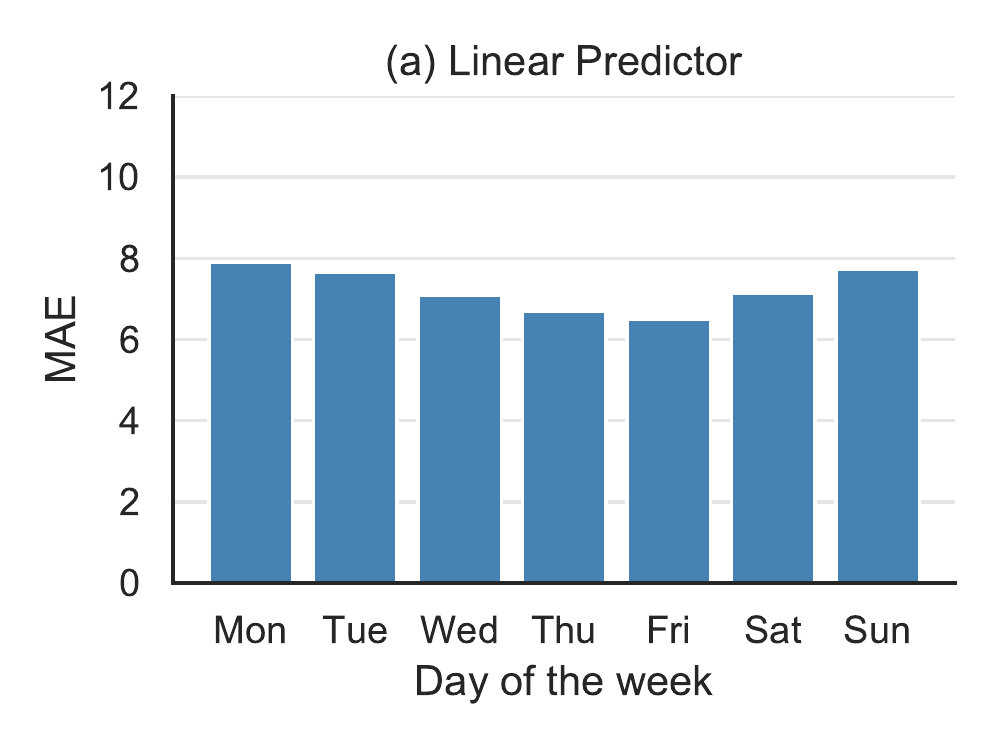}&  
    \includegraphics[width=0.49\textwidth]{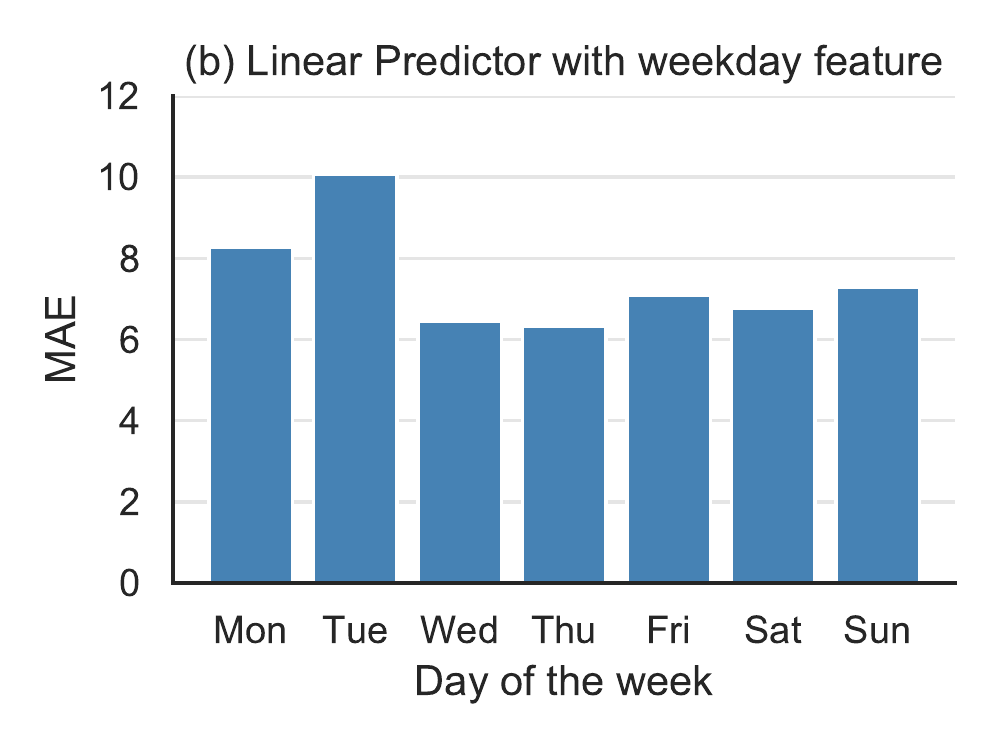}
    \end{tabular}
    }
    \caption[Mean absolute errors for separate linear predictor with weekday feature]{Mean raw-scale MAEs by weekdays for the separate linear predictors for \emph{3-day-ahead} predictions with and without the weekday feature. The MAE for a given day is the MAE for the 3-day-ahead prediction \textit{computed on that day}. So the MAE for Wednesday is the MAE for predicting the cumulative deaths on Friday.}
    \label{fig:weekday_feat_linear}
\end{figure}

Recall that the original separate linear predictors in Section~\ref{sub:linear} were fit only with the four most recent days data. For some choices of days, the new feature $v_{t}$ takes only a single value $0$ in the training data. For instance, when day $t+1$ is Saturday, we have $v_{t}=v_{t-1}=v_{t-1}=v_{t-2}=v_{t-3}=0$, i.e., the new feature is identically zero in the training data. For such cases, the parameter $\beta_2^c$ is not identifiable. To address this issue of non-identifiability, for these experiments, we use the 7 most recent days to fit the linear predictors. 
For comparison, using the 7 most recent days instead of the 4 most recent days for the original linear predictor increases the raw-scale MAE~\eqref{eq:raw_scale_mae} from $7.0$ to $7.2$. Adding the new indicator feature $v_t$ increases this error further to $7.4$. As with the expanded shared model, we plot the mean raw-scale MAE per day of the week for 3-day-ahead predictions in Figure~\ref{fig:weekday_feat_linear}. With the weekday feature (panel~(b)), we see that errors for most days are lower than that without the weekday feature (panel~(a)), but this gain is offset by a high error for predictions made on Tuesday.

\section{Further discussion on MEPI}
\label{sub:mepi_further}
We now first shed light on why MEPI had slightly worse coverage for some of the counties.
And then, we provide a further discussion on various choices made for designing MEPI.

\subsection{Counties with poor coverage}
\label{sub:poor_exchange}
\begin{figure}[ht]
    \centering
    \includegraphics[width=\textwidth]{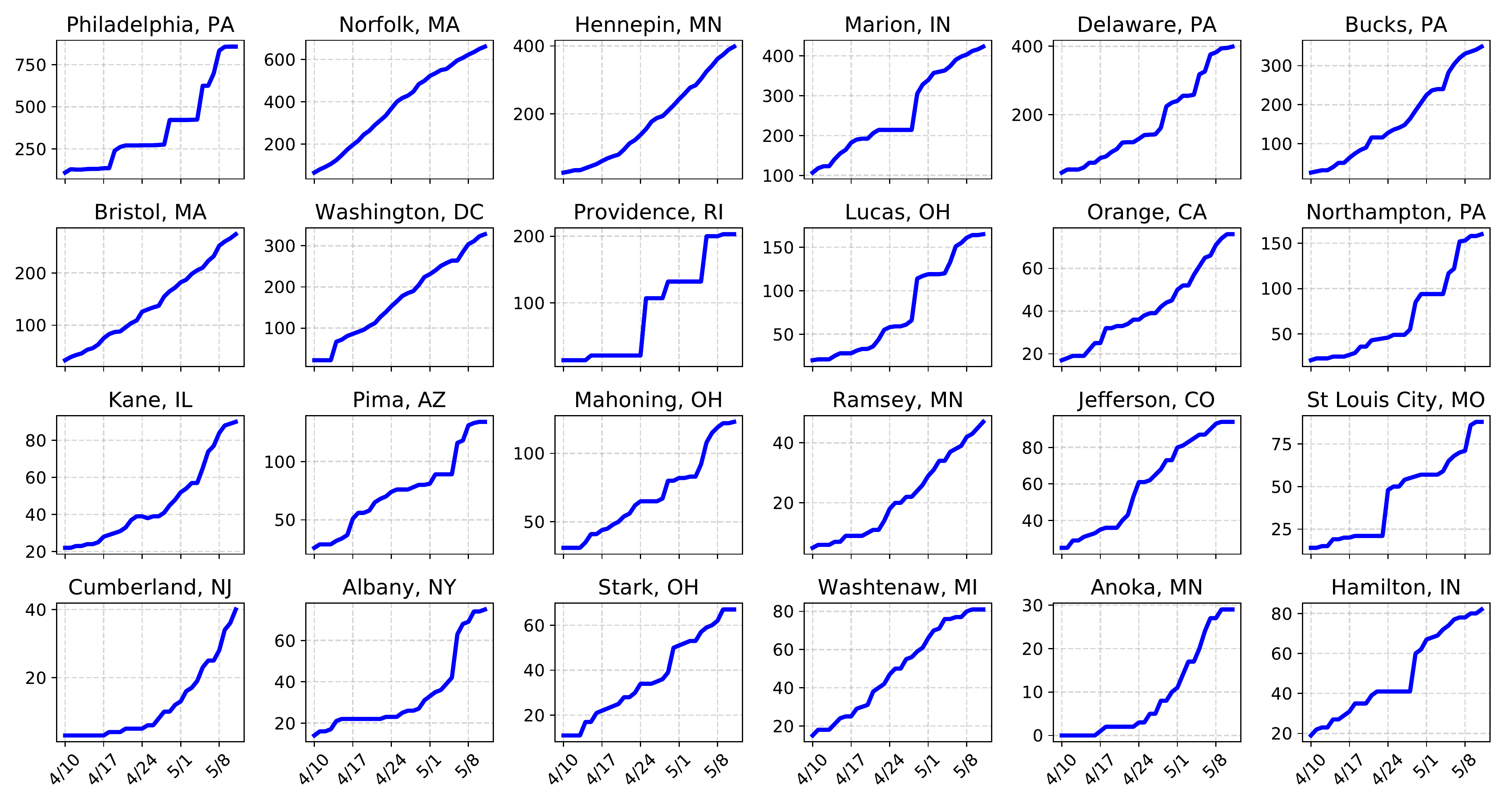}
    \caption[Visualization of death counts for counties with poor coverage by MEPI]{The cumulative death count data from the 24 worst affected counties where the coverage of the \emph{7-day-ahead} MEPIs is below 0.8 (in Figure~\ref{fig:mepi_coverage_and_length}(a)).}
    \label{fig:mepi_outliers}
\end{figure}

While Figure~\ref{fig:mepi_coverage_and_length}(a) shows that MEPI intervals achieve higher than 83\% coverage for the vast majority of counties over the April 11--May 10 period, there are also counties with coverage below the targeted level. We provide a brief investigation of counties where the coverage of MEPIs for cumulative death counts is below 0.8 in Figure~\ref{fig:mepi_coverage_and_length}(a). Among 198 such counties, Figure~\ref{fig:mepi_outliers} shows the cumulative deaths from April 11 to May 10 of the worst-affected 24 counties. Many of these counties exhibit a sharp uptick in the number of recorded deaths similar to that which we encountered in New York, possibly due to reporting lag. For instance, Philadelphia (top row, first column from left) only recorded 2 new deaths between April 28 and May 3, but recorded 201 new deaths on May 4, which brought the cumulative deaths on May 4 to 625.

\subsection{MEPI vs conformal interence}
\label{sub:conformal-inference}

Recall that the MEPI (equation \ref{eq:mepi_original}) can be viewed as a special case of conformal prediction interval~\cite{vovk2005algorithmic,shafer2008tutorial}. 
Here, we provide further discussion on this connection and discuss the assumptions under which the MEPI should achieve good coverage.
A general recipe in conformal inference with streaming data is to compute the past several errors of the prediction model and use an $s$-percentile value for some suitable $s$ (e.g., $s=95$) to construct the prediction interval for the future observations. 
At a high-level, theoretical guarantees for conformal prediction intervals rely on the assumption that the sequence of errors is exchangeable.
Roughly, the proof proceeds as follows: the exchangeability of the residuals ensures that the rankings of future residuals are uniformly distributed. Hence, the probability of the future residuals being in the top $s$-percentile is no larger than $s$, thereby obtaining the promised $s\%$-coverage. 
For more details, we refer the reader to the excellent tutorial~\cite{shafer2008tutorial} and the book~\cite{vovk2005algorithmic}.

Given the dynamic nature of COVID-19, it is unrealistic to assume that the prediction errors are exchangeable over a long period. As the cumulative death count grows, so too will the magnitude of the errors. Thus our MEPI scheme deviates from the general conformal recipe in two ways. We compute a \emph{maximum error over the past 5 days}, and we \emph{normalize} the errors.
Each of these choices---of normalized errors and looking at only past $5$ errors--- is designed to make the errors more exchangeable. 
Moreover, given that we take $5$ data points to bound the future error, computing a maximum over them is a more conservative choice (e.g., when compared to taking median or a percentile-based cut-off).Furthermore, in order to compute a $95$-percentile value, we need to consider errors for at least the past $20$ days. Exchangeability is not likely to hold for such a long horizon.\\

\begin{figure}[ht]
    \centering
    \begin{tabular}{c}
        \includegraphics[width=0.8\textwidth]{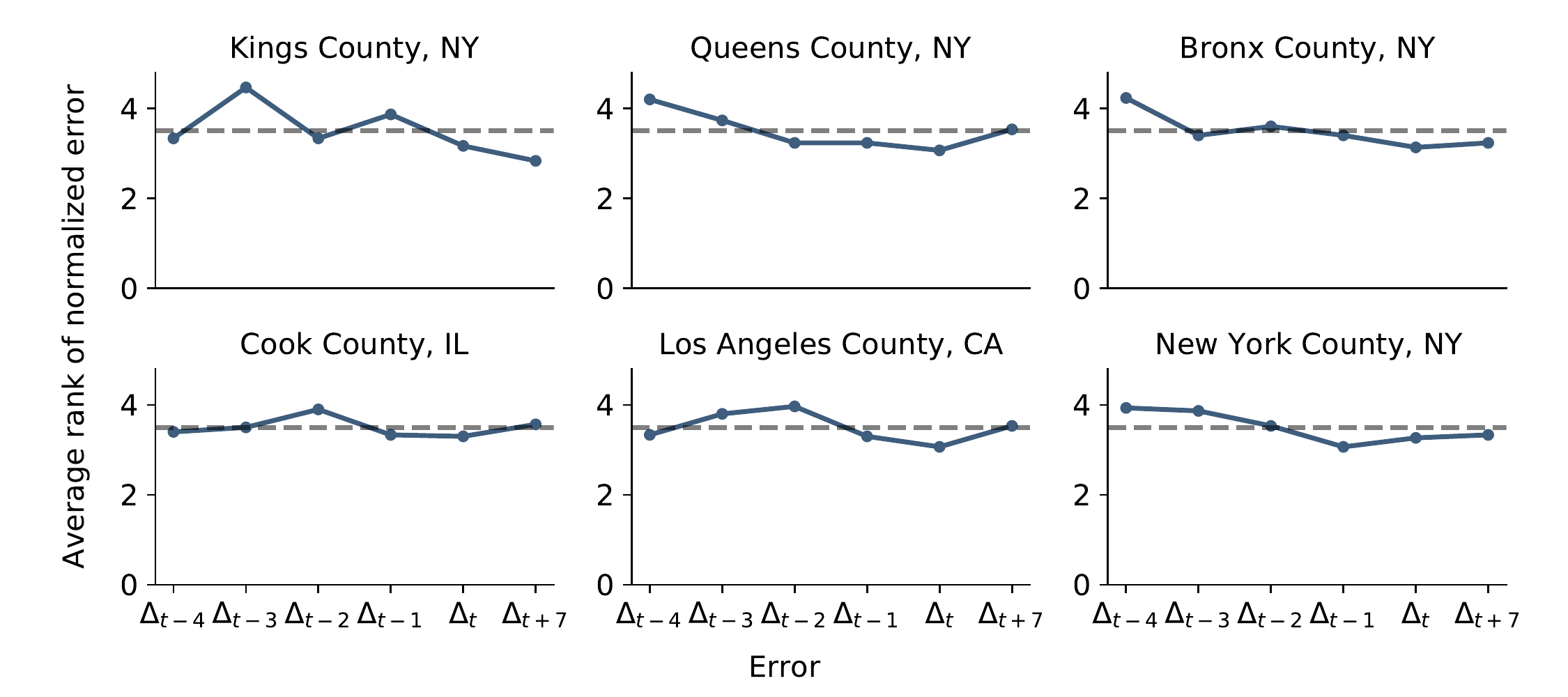} \vspace{-2mm}\\
        (a) Rank distribution of normalized errors over 5 days \vspace{2mm}\\
        \midrule
        \includegraphics[width=0.8\textwidth]{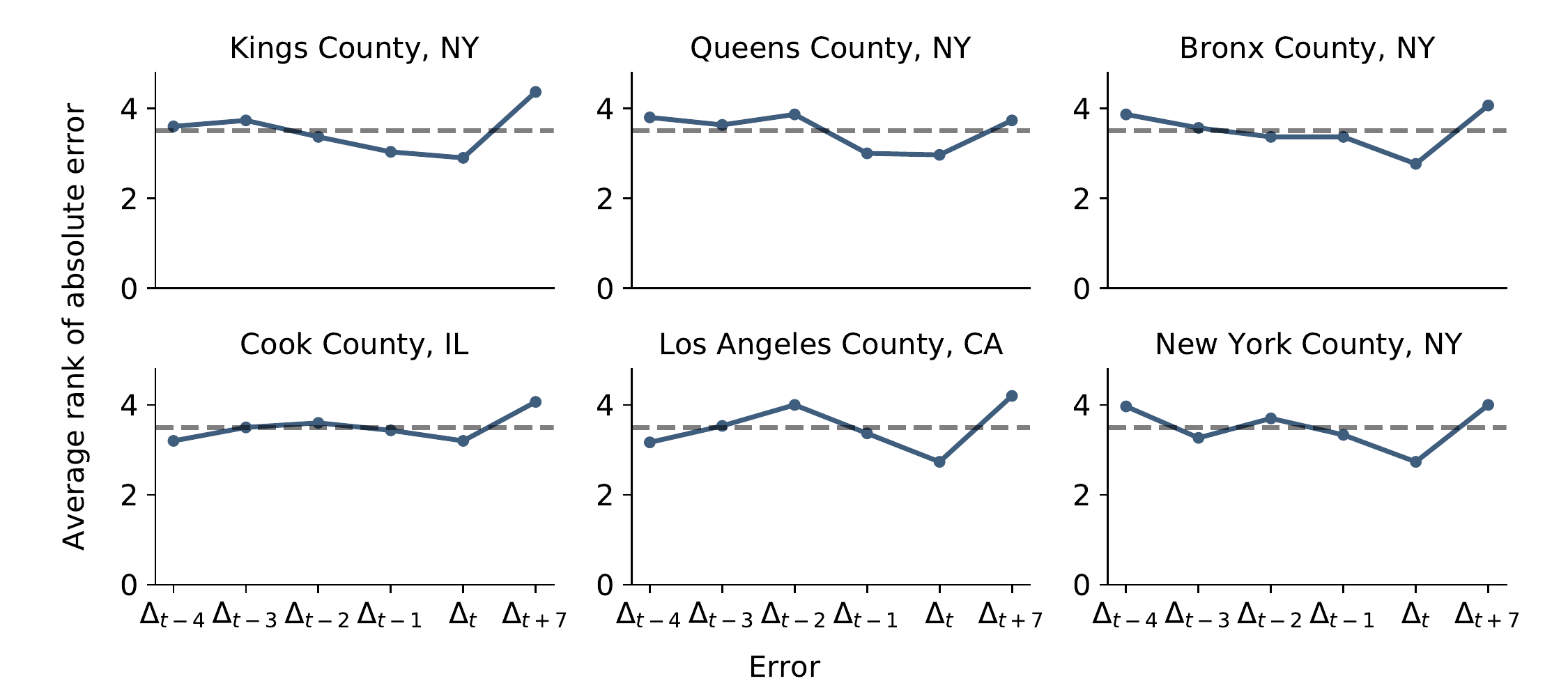} \vspace{-2mm}\\
        (b) Rank distribution of absolute unnormalized errors over 5 days \vspace{2mm}\\
        \midrule
        \includegraphics[width=0.8\textwidth]{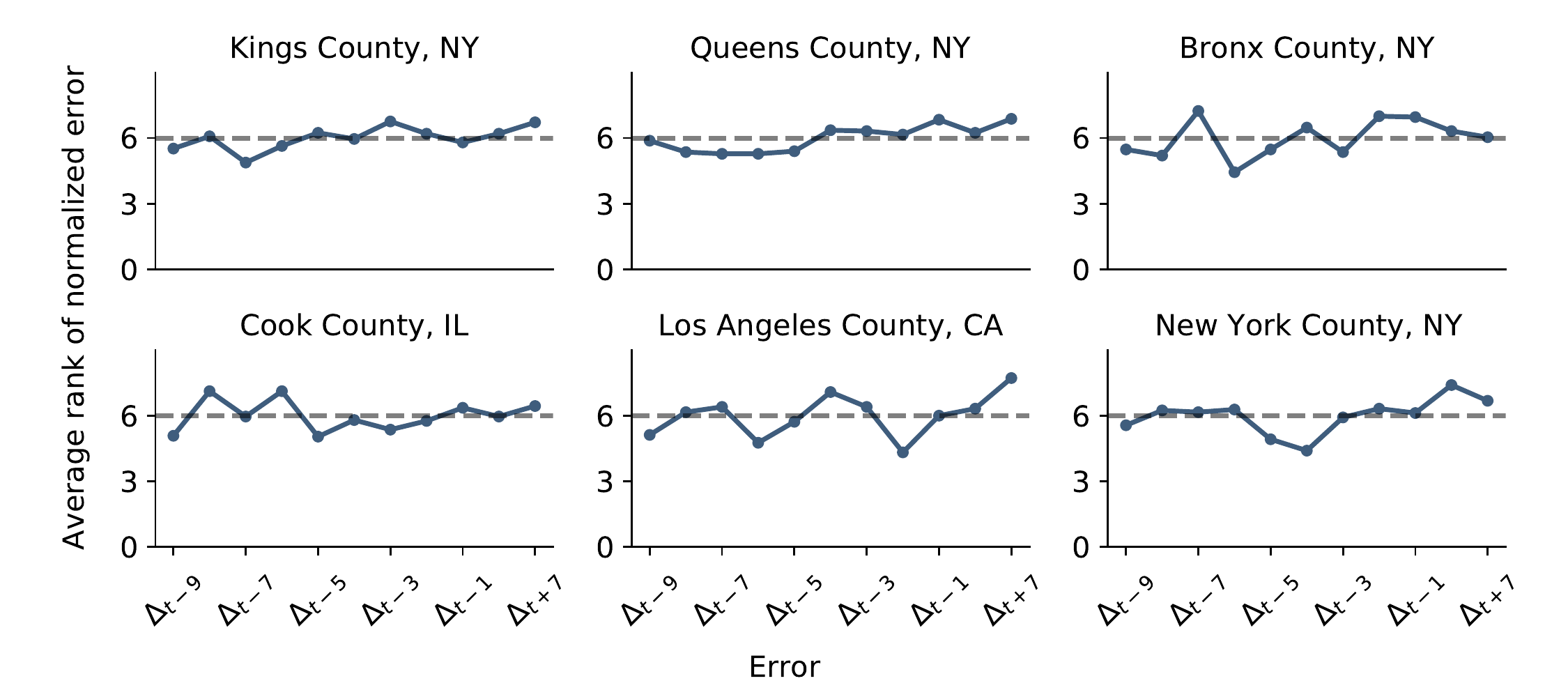} \\
        (c) Rank distribution of normalized errors over a longer window of 10 days
    \end{tabular}
    \caption[EDA: Evidence for our choices in 7-day-ahead MEPI construction]{EDA plot with unnormalized and normalized errors for \emph{7-day-ahead} predictions made by CLEP, computed over $t =$ March 26, \dots, April 25. \textbf{(a)} The rank distribution of normalized  errors of our CLEP (with the expanded shared  and linear predictors) for the six worst affected counties; \textbf{(b)}
    the absolute unnormalized errors of our CLEP for the six worst affected counties  and \textbf{(c)} the rank distribution of the normalized errors over a longer window.}
    \label{fig:error_metric_curve}
\end{figure}

\paragraph{\emph{Normalized vs. unnormalized errors:}}
We now provide some numerical evidence to support our choice of normalized errors to define the MEPI.
Figure~\ref{fig:error_metric_curve}(a) shows the rank distribution of normalized errors of our 7-day-ahead CLEP predictions for the six worst-affected counties over over an earlier period (March 26--April 25), and Figure~\ref{fig:error_metric_curve}(b) shows the (unnormalized) $\ell_1$ errors $|\widehat{y}_t-y_t|$ over the same period. We found that in Figure~\ref{fig:error_metric_curve}(b), the $\ell_1$ errors on days $t-4, t-3, t-2, t-1, t$ and $t+7$ do not appear to be exchangeable. Recall that under exchangeability conditions, the expected average rank of each of these six $\ell_1$ errors would be 3.5. However, for all six counties, the average rank of the absolute error on day 
$t+7$ is larger than 4. This indicates that the future absolute error tends to be higher than past errors, and using the $\ell_1$ error $|\widehat{y}_t-y_t|$ in place of the normalized error $\err_t$ can lead to substantial underestimation of future prediction uncertainty.\\

\paragraph{\emph{Longer time window:}}
In Figure~\ref{fig:error_metric_curve}(c), we show the rank distribution of normalized errors over a longer window of 10 days. We found that due to the highly dynamic nature of COVID-19, these errors appear to be even less exchangeable. Under exchangeability conditions, the expected average rank of each of these 11 errors would be 6. However, we found that the average rank substantially deviates from this expected value for many days in this longer window for all displayed counties. 

Overall, we believe that putting together the observations from Figures~\ref{fig:average_rank_normalized_error} and \ref{fig:error_metric_curve} yield reasonable justification for the two choices we made to define MEPI (equation~\ref{eq:mepi_original}), namely, the $5$-day window (versus the entire past), and the choice of normalized errors (versus the unnormalized absolute errors).\\

\begin{figure}[ht]
    \centering
    \begin{tabular}{c}
        \includegraphics[width=0.85\textwidth]{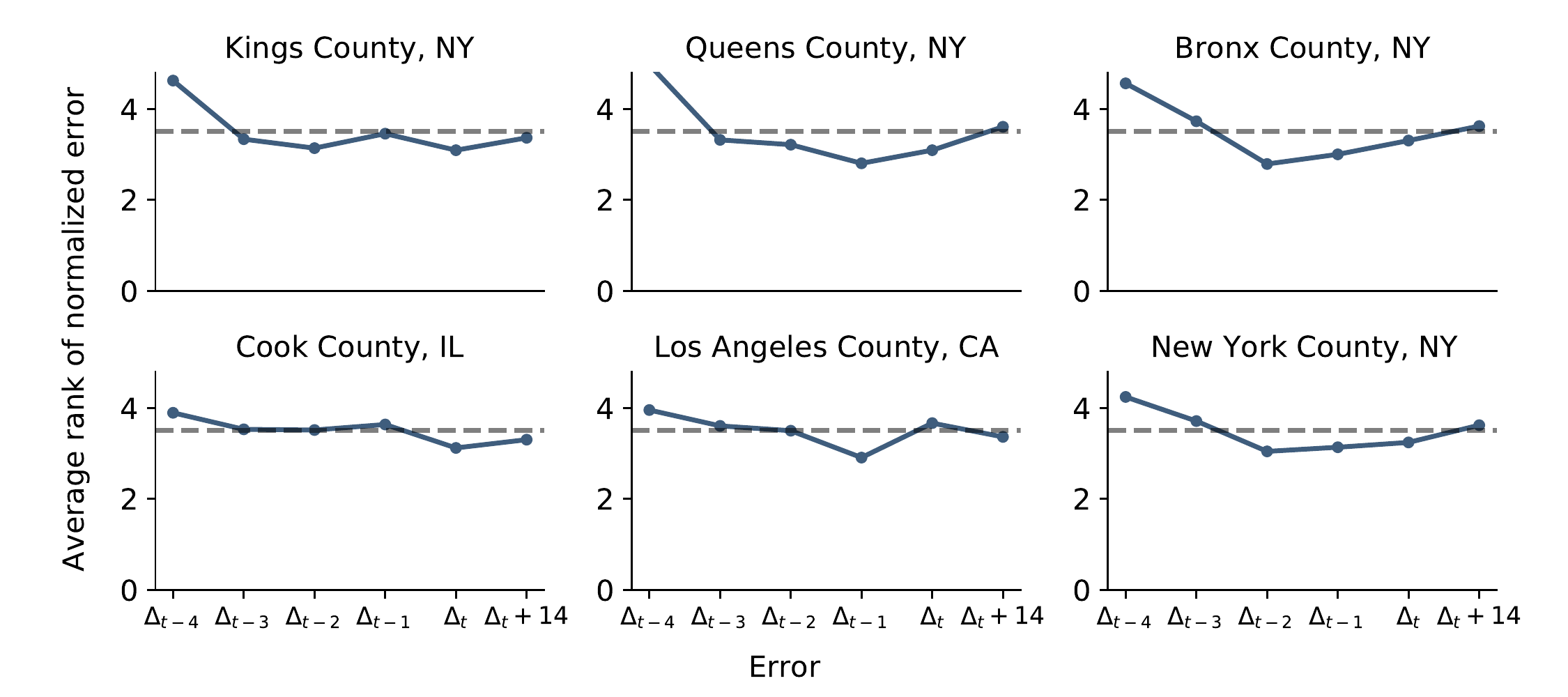} \vspace{-2mm}\\ \vspace{1.4em}
        (a) Six worst-affected counties\\
        %\midrule \\
        \includegraphics[width=0.85\textwidth]{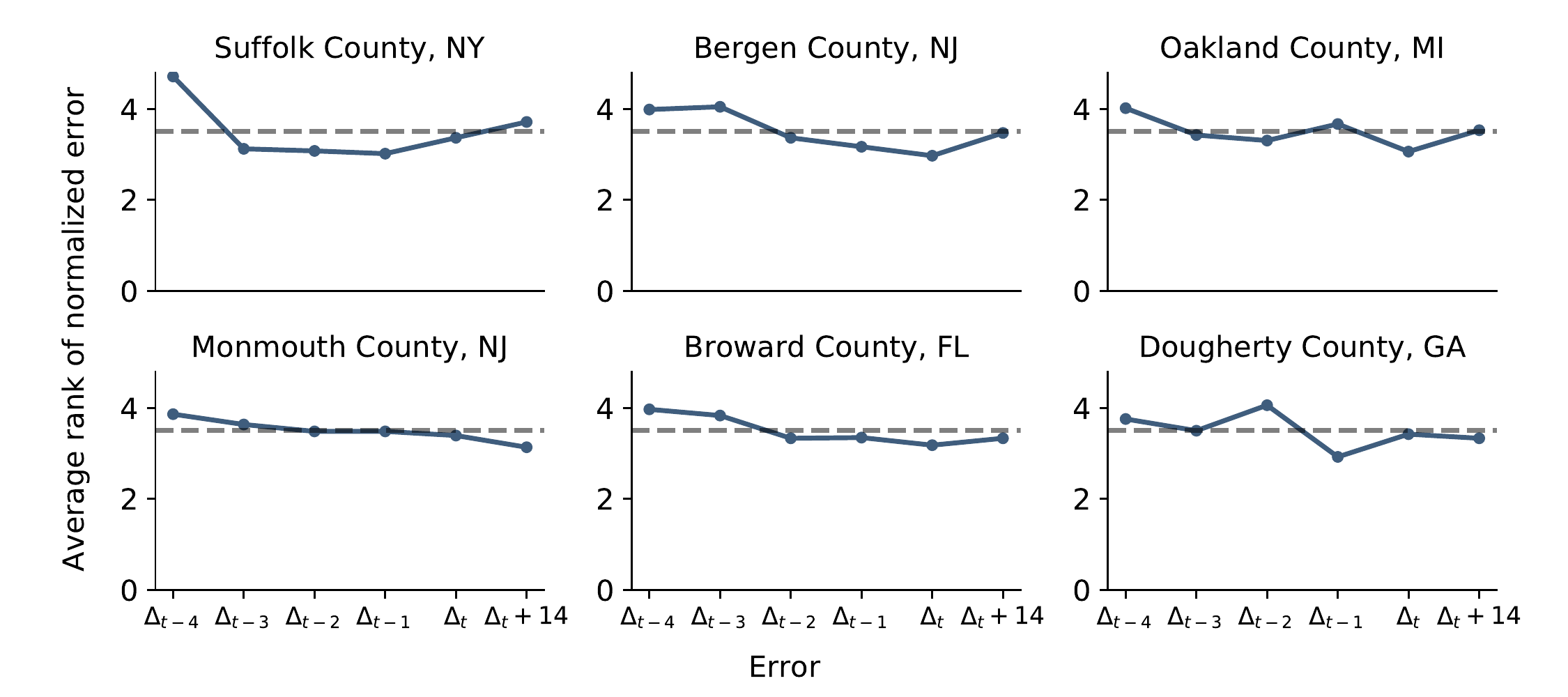} \vspace{-2mm}\\ 
        (b) Six randomly-selected counties
    \end{tabular}
    \caption[Investigating exchangeability of errors for 14-day-ahead CLEP predictions]{EDA plot for investigating exchangeability of normalized errors of \emph{14-day-ahead} CLEP predictions with its last $5$ errors made at time $t$, over the period $t =$ April 2, \ldots, Jun 6. We plot the average rank of the six errors $\{\err_{\tidx+14}, \err_{\tidx}, \err_{\tidx-1}, \ldots, \err_{\tidx-4}\}$ of our CLEP (with the expanded shared and linear predictors) for \textbf{(a)} the six worst affected counties, and \textbf{(b)} six random counties. We rank the errors $\{\err_{\tidx+14}, \err_{\tidx}, \err_{\tidx-1}, \ldots, \err_{\tidx-4}\}$ in increasing order so that the largest error has a rank of 6. If $\{\err_{\tidx+14}, \err_{\tidx}, \err_{\tidx-1}, \ldots, \err_{\tidx-4}\}$ are exchangeable for any day $t$, then the expected average rank for each of the six errors would be 3.5 (dashed black line).}
    \label{fig:average_rank_normalized_error_14_days}
\end{figure}

\bibliographystyle{abbrv}  
\begin{small}
%   \singlespacing
%   \bibliographystyle{alpha}
  \bibliography{references}

\begin{thebibliography}{10}

\bibitem{Angelopoulos2020On}
A.~N. Angelopoulos, R.~Pathak, R.~Varma, and M.~I. Jordan.
\newblock On identifying and mitigating bias in the estimation of the
  {COVID-19} case fatality rate.
\newblock {\em Harvard Data Science Review}, 6 2020.

\bibitem{apple_mobility_data}
{Apple Inc}.
\newblock {Apple Mobility Trends Reports}.
\newblock 2020.
\newblock Accessed on 05-15-2020 at
  \url{https://www.apple.com/covid19/mobility}.

\bibitem{chime}
M.~Becker and C.~Chivers.
\newblock Announcing {CHIME}, a tool for covid-19 capacity planning.
\newblock 2020.
\newblock Accessed on 04-02-2020 at
  \url{http://predictivehealthcare.pennmedicine.org/2020/03/14/accouncing-chime.html}.

\bibitem{Bregman}
D.~J. Bregman and A.~D. Langmuir.
\newblock {Farr's Law Applied to AIDS Projections}.
\newblock {\em JAMA}, 263(11):1522--1525, 03 1990.

\bibitem{plane_data}
{Bureau of Transportation Statistics}.
\newblock Airline origin and destination survey (db1b).
\newblock 2020.
\newblock Accessed on 04-20-2020 at
  \url{https://transtats.bts.gov/Databases.asp?Mode_ID=1&Mode_Desc=Aviation&Subject_ID2=0}.

\bibitem{heart_data}
{Centers for Disease Control and Prevention}.
\newblock {Interactive Atlas of Heart Disease and Stroke}.
\newblock 2018.
\newblock Accessed on 04-02-2020 at \url{http://nccd.cdc.gov/DHDSPAtlas}.

\bibitem{svi_data}
{Centers for Disease Control and Prevention}, {Agency for Toxic Substances and
  Disease Registry}, and {Geospatial Research, Analysis, and Services Program}.
\newblock {Social Vulnerability Index Database}.
\newblock 2018.
\newblock Accessed on 04-03-2020 at
  \url{https://svi.cdc.gov/data-and-tools-download.html}.

\bibitem{diabetes_data}
{Centers for Disease Control and Prevention}, {Division of Diabetes
  Translation}, and {US Diabetes Surveillance System}.
\newblock Diagnosed diabetes atlas.
\newblock 2016.
\newblock Accessed on 04-02-2020 at \url{https://www.cdc.gov/diabetes/data}.

\bibitem{chronic_data}
{Centers for Medicare \& Medicaid Services}.
\newblock {Chronic Conditions Prevalence State/County Level: All Beneficiaries
  by Age, 2007-2017}.
\newblock 2017.
\newblock Accessed on 04-02-2020 at
  \url{https://www.cms.gov/Research-Statistics-Data-and-Systems/Statistics-Trends-and-Reports/Chronic-Conditions/CC_Main}.

\bibitem{cmi_data}
{Centers for Medicares \& Medicaid Services}.
\newblock Case mix index file.
\newblock 2018.
\newblock Accessed on 04-01-2020 at
  \url{https://www.cms.gov/Medicare/Medicare-Fee-for-Service-Payment/AcuteInpatientPPS/FY2020-IPPS-Final-Rule-Home-Page-Items/FY2020-IPPS-Final-Rule-Data-Files}.

\bibitem{cms_teaching_data}
{Centers for Medicares \& Medicaid Services}.
\newblock 2020 reporting cycle: Teaching hospital list.
\newblock 2020.
\newblock Accessed on 04-01-2020 at
  \url{https://www.cms.gov/OpenPayments/Downloads/2020-Reporting-Cycle-Teaching-Hospital-List-PDF-.pdf}.

\bibitem{chiang2020hawkes}
W.-H. Chiang, X.~Liu, and G.~Mohler.
\newblock Hawkes process modeling of {COVID-19} with mobility leading
  indicators and spatial covariates.
\newblock {\em medRxiv preprint 2020.06.06.20124149}, 2020.

\bibitem{smoking_data}
{County Health Rankings \& Roadmaps}.
\newblock {County Health Rankings \& Roadmaps 2020 Measures}.
\newblock 2020.
\newblock Accessed on 04-02-2020 at
  \url{https://www.countyhealthrankings.org/explore-health-rankings/measures-data-sources/2020-measures}.

\bibitem{dh_data}
{Definitive Healthcare}.
\newblock {Definitive Healthcare: USA Hospital Beds}.
\newblock 2020.
\newblock Accessed on 04-01-2020 at
  \url{https://coronavirus-resources.esri.com/datasets/definitivehc::definitive-healthcare-usa-hospital-beds}.

\bibitem{dong2020interactive}
E.~Dong, H.~Du, and L.~Gardner.
\newblock An interactive web-based dashboard to track covid-19 in real time.
\newblock {\em The Lancet infectious diseases}, 20(5):533--534, 2020.

\bibitem{elmousalami2020day}
H.~H. Elmousalami and A.~E. Hassanien.
\newblock Day level forecasting for {C}oronavirus disease {(COVID-19)} spread:
  Analysis, modeling and recommendations.
\newblock {\em arXiv preprint arXiv:2003.07778}, 2020.

\bibitem{fanelli2020analysis}
D.~Fanelli and F.~Piazza.
\newblock Analysis and forecast of {COVID-19} spreading in {China, Italy and
  France}.
\newblock {\em Chaos, Solitons \& Fractals}, 134:109761, 2020.

\bibitem{ferguson2020impact}
N.~Ferguson, D.~Laydon, G.~Nedjati-Gilani, N.~Imai, K.~Ainslie, M.~Baguelin,
  S.~Bhatia, A.~Boonyasiri, Z.~Cucunub{\'a}, G.~Cuomo-Dannenburg, et~al.
\newblock Impact of non-pharmaceutical interventions ({NPIs}) to reduce
  {COVID}19 mortality and healthcare demand, 2020.
\newblock Accessed on 04-02-2020 at
  \url{https://www.imperial.ac.uk/media/imperial-college/medicine/sph/ide/gida-fellowships/Imperial-College-COVID19-NPI-modelling-16-03-2020.pdf}.

\bibitem{goh2020rapid}
K.~J. Goh, S.~Kalimuddin, and K.~S. Chan.
\newblock Rapid progression to acute respiratory distress syndrome: {R}eview of
  current understanding of critical illness from {COVID-19} infection.
\newblock {\em Annals of the Academy of Medicine, Singapore}, 49(1):1, 2020.

\bibitem{google_mobility_data}
{Google LLC}.
\newblock {Google {COVID-19} Community Mobility Reports}.
\newblock 2020.
\newblock Accessed on 05-15-2020 at
  \url{https://www.google.com/covid19/mobility/}.

\bibitem{guan2020comorbidity}
W.~Guan, W.~Liang, Y.~Zhao, H.~Liang, Z.~Chen, Y.~Li, X.~Liu, R.~Chen, C.~Tang,
  T.~Wang, et~al.
\newblock Comorbidity and its impact on 1590 patients with {COVID}-19 in
  {C}hina: A nationwide analysis.
\newblock {\em European Respiratory Journal}, 2020.

\bibitem{guan2020clinical}
W.~Guan, Z.~Ni, Y.~Hu, W.~Liang, C.~Ou, J.~He, L.~Liu, H.~Shan, C.~Lei, D.~S.
  Hui, et~al.
\newblock Clinical characteristics of {C}oronavirus disease 2019 in {C}hina.
\newblock {\em New England Journal of Medicine}, 2020.

\bibitem{ahrf_data}
{Health Resources and Services Administration}.
\newblock {Area Health Resources Files}.
\newblock 2019.
\newblock Accessed on 04-02-2020 at \url{https://data.hrsa.gov/data/download}.

\bibitem{hpsa_data}
{Health Resources and Services Administration}.
\newblock {Health Professional Shortage Areas - Primary Care}.
\newblock 2020.
\newblock Accessed on 04-04-2020 at \url{https://data.hrsa.gov/data/download}.

\bibitem{hifld_hospital_data}
{Homeland Infrastructure Foundation-Level Data}.
\newblock Hospitals.
\newblock 2020.
\newblock Accessed on 06-23-2020 at
  \url{https://hifld-geoplatform.opendata.arcgis.com/datasets/6ac5e325468c4cb9b905f1728d6fbf0f_0}.

\bibitem{Hsiang2020}
S.~Hsiang, D.~Allen, S.~Annan-Phan, K.~Bell, I.~Bolliger, T.~Chong,
  H.~Druckenmiller, A.~Hultgren, L.~Y. Huang, E.~Krasovich, P.~Lau, J.~Lee,
  E.~Rolf, J.~Tseng, and T.~Wu.
\newblock The effect of large-scale anti-contagion policies on the {Coronavirus
  (COVID-19)} pandemic.
\newblock {\em medRxiv}, 2020.

\bibitem{resp_data}
{Institute for Health Metrics and Evaluation}.
\newblock {United States Chronic Respiratory Disease Mortality Rates by County
  1980-2014}.
\newblock 2017.
\newblock Accessed on 04-02-2020 at
  \url{http://ghdx.healthdata.org/record/ihme-data/united-states-chronic-respiratory-disease-mortality-rates-county-1980-2014}.

\bibitem{khn_data}
{Kaiser Health News}.
\newblock {ICU Beds by County}.
\newblock 2020.
\newblock Accessed on 04-02-2020 at
  \url{https://khn.org/news/as-coronavirus-spreads-widely-millions-of-older-americans-live-in-counties-with-no-icu-beds/}.

\bibitem{killeen2020county}
B.~D. Killeen, J.~Y. Wu, K.~Shah, A.~Zapaishchykova, P.~Nikutta, A.~Tamhane,
  S.~Chakraborty, J.~Wei, T.~Gao, M.~Thies, and M.~Unberath.
\newblock A county-level dataset for informing the {United States'} response to
  {COVID-19}.
\newblock {\em arXiv preprint arXiv:2004.00756}, 2020.

\bibitem{Kucharski2020}
A.~J. Kucharski, T.~W. Russell, C.~Diamond, Y.~Liu, J.~Edmunds, S.~Funk, and
  R.~M. Eggo.
\newblock Early dynamics of transmission and control of {COVID-19}: A
  mathematical modelling study.
\newblock {\em medRxiv}, 2020.

\bibitem{marchant2020learning}
R.~Marchant, N.~I. Samia, O.~Rosen, M.~A. Tanner, and S.~Cripps.
\newblock Learning as we go: An examination of the statistical accuracy of
  covid19 daily death count predictions.
\newblock {\em arXiv preprint arXiv:2004.04734}, 2020.

\bibitem{voting_data}
{MIT Election Data and Science Lab}.
\newblock {County Presidential Election Returns 2000-2016}.
\newblock 2018.

\bibitem{IHME}
C.~J. Murray and I.~H. M.~E. {COVID-19} health service utilization~forecasting
  team.
\newblock Forecasting {COVID-19} impact on hospital bed-days, {ICU}-days,
  ventilator-days and deaths by {US} state in the next 4 months.
\newblock {\em medRxiv}, 2020.

\bibitem{exponential-growth}
S.~Nebehay and K.~Kelland.
\newblock {COVID-19} cases and deaths rising, debt relief needed for poorest
  nations: {WHO}.
\newblock {\em Reuters}, 2020-04-01.
\newblock Accessed on 04-01-2020 at
  \url{https://www.reuters.com/article/us-health-coronavirus-who/covid-19-infections-growing-exponentially-deaths-nearing-50000-who-idUSKBN21J6IL?il=0}.

\bibitem{peak2020modeling}
C.~M. Peak, R.~Kahn, Y.~H. Grad, L.~M. Childs, R.~Li, M.~Lipsitch, and C.~O.
  Buckee.
\newblock Modeling the comparative impact of individual quarantine vs. active
  monitoring of contacts for the mitigation of {COVID-19}.
\newblock {\em medRxiv}, 2020.

\bibitem{pei2020initial}
S.~Pei and J.~Shaman.
\newblock Initial simulation of {SARS-CoV2} spread and intervention effects in
  the continental {US}.
\newblock {\em medRxiv}, 2020.

\bibitem{qi2020epidem}
D.~Qi, X.~Yan, X.~Tang, J.~Peng, Q.~Yu, L.~Feng, G.~Yuan, A.~Zhang, Y.~Chen,
  J.~Yuan, X.~Huang, X.~Zhang, P.~Hu, Y.~Song, C.~Qian, Q.~Sun, D.~Wang,
  J.~Tong, and J.~Xiang.
\newblock Epidemiological and clinical features of {2019-nCoV} acute
  respiratory disease cases in {C}hongqing municipality, {C}hina: {A}
  retrospective, descriptive, multiple-center study.
\newblock {\em medRxiv}, 2020.

\bibitem{rubinson2010mechanical}
L.~Rubinson, F.~Vaughn, S.~Nelson, S.~Giordano, T.~Kallstrom, T.~Buckley,
  T.~Burney, N.~Hupert, R.~Mutter, M.~Handrigan, et~al.
\newblock Mechanical ventilators in {US} acute care hospitals.
\newblock {\em Disaster medicine and public health preparedness},
  4(3):199--206, 2010.

\bibitem{schuller_2002}
G.~D. Schuller, B.~Yu, D.~Huang, and B.~Edler.
\newblock Perceptual audio coding using adaptive pre-and post-filters and
  lossless compression.
\newblock {\em IEEE Transactions on Speech and Audio Processing},
  10(6):379--390, 2002.

\bibitem{seabold2010statsmodels}
S.~Seabold and J.~Perktold.
\newblock statsmodels: Econometric and statistical modeling with python.
\newblock In {\em 9th Python in Science Conference}, 2010.

\bibitem{shafer2008tutorial}
G.~Shafer and V.~Vovk.
\newblock A tutorial on conformal prediction.
\newblock {\em Journal of Machine Learning Research}, 9(Mar):371--421, 2008.

\bibitem{uncertainty_IHME}
{The Institute for Health Metrics and Evaluation}.
\newblock {COVID-19: What’s New for April 5, 2020}, 2020.
\newblock
  \url{http://www.healthdata.org/sites/default/files/files/Projects/COVID/Estimation_update_040520_3.pdf},
  Last accessed on 2020-04-13.

\bibitem{nytimes_data}
{The New York Times}.
\newblock {COVID-19 Data in the United States}.
\newblock \url{https://github.com/nytimes/covid-19-data}, 2020.
\newblock Accessed on 04-01-2020 at
  \url{https://github.com/nytimes/covid-19-data}.

\bibitem{adjacency_data}
{United States Census Bureau}.
\newblock {County Adjacency File}.
\newblock 2018.
\newblock Accessed on 05-15-2020 at
  \url{https://www.census.gov/geographies/reference-files/2010/geo/county-adjacency.html}.

\bibitem{poverty_data}
{United States Department of Agriculture, Economic Research Service}.
\newblock Poverty estimates for the {U.S.}, states, and counties.
\newblock 2018.
\newblock Accessed on 04-24-2020 at
  \url{https://www.ers.usda.gov/data-products/county-level-data-sets/download-data/}.

\bibitem{mortality_data}
{United States Department of Health and Human Services}, {Centers for Disease
  Control and Prevention }, and {National Center for Health Statistics}.
\newblock {Compressed Mortality File (CMF) on CDC WONDER Online Database,
  2012-2016}.
\newblock 2017.
\newblock Accessed on 04-02-2020 at
  \url{https://wonder.cdc.gov/cmf-icd10.html}.

\bibitem{usa_facts_data}
USAFacts.
\newblock {COVID-19 Deaths Data}.
\newblock 2020.
\newblock Accessed on 03-31-2020 at
  \url{https://www.reuters.com/article/us-health-coronavirus-who/covid-19-spread-map}.

\bibitem{vovk2005algorithmic}
V.~Vovk, A.~Gammerman, and G.~Shafer.
\newblock {\em Algorithmic learning in a random world}.
\newblock Springer Science \& Business Media, 2005.

\bibitem{wang2020evolving}
C.~Wang, L.~Liu, X.~Hao, H.~Guo, Q.~Wang, J.~Huang, N.~He, H.~Yu, X.~Lin,
  A.~Pan, S.~Wei, and T.~Wu.
\newblock Evolving epidemiology and impact of non-pharmaceutical interventions
  on the outbreak of {C}oronavirus disease 2019 in {W}uhan, {C}hina.
\newblock {\em medRxiv}, 2020.

\bibitem{nytimes_missing_deaths}
J.~Wu, A.~McCann, J.~Katz, and E.~Peltier.
\newblock 109,000 missing deaths: Tracking the true toll of the coronavirus
  outbreak.
\newblock {\em The New York Times}, 2020.

\bibitem{zhou2020clinical}
F.~Zhou, T.~Yu, R.~Du, G.~Fan, Y.~Liu, Z.~Liu, J.~Xiang, Y.~Wang, B.~Song,
  X.~Gu, et~al.
\newblock Clinical course and risk factors for mortality of adult inpatients
  with {COVID-19 in Wuhan, China: A} retrospective cohort study.
\newblock {\em The Lancet}, 2020.

\end{thebibliography}
\end{small}

\end{document}